\documentclass[prl,twocolumn,floatfix,superscriptaddress]{revtex4-2}
\usepackage{graphics,epsfig,amsfonts,amssymb,amsmath}
\usepackage{soul} % para tachar palabras
\usepackage[dvipsnames]{xcolor}
\usepackage{MnSymbol,wasysym}

\graphicspath{{./pics/}}

\begin{document}
\title{Doping fingerprints of spin and lattice fluctuations in moir\'{e} superlattice systems}

\author{Niklas Witt}
\email{niklas.witt@physik.uni-hamburg.de}
\affiliation{Institute of Theoretical Physics, Bremen Center for Computational Materials Science,
	and MAPEX Center for Materials and Processes, University of Bremen, Otto-Hahn-Allee 1, 28359 Bremen, Germany}
\affiliation{I. Institute of Theoretical Physics, University of Hamburg, Notkestraße 9, 22607 Hamburg, Germany}

\author{José M. Pizarro}
\email{jose.pizarro@mpsd.mpg.de}
\affiliation{Institute of Theoretical Physics, Bremen Center for Computational Materials Science,
	and MAPEX Center for Materials and Processes, University of Bremen, Otto-Hahn-Allee 1, 28359 Bremen, Germany}
\affiliation{Max Planck Institute for the Structure and Dynamics of Matter, Luruper Chaussee 149, 22671 Hamburg, Germany}

\author{Jan Berges}
\affiliation{Institute of Theoretical Physics, Bremen Center for Computational Materials Science,
	and MAPEX Center for Materials and Processes, University of Bremen, Otto-Hahn-Allee 1, 28359 Bremen, Germany}

\author{Takuya Nomoto}
\affiliation{Department of Applied Physics, The University of Tokyo, 7-3-1 Hongo, Bunkyo-ku, Tokyo 113-8656, Japan}

\author{Ryotaro Arita}
\affiliation{Department of Applied Physics, The University of Tokyo, 7-3-1 Hongo, Bunkyo-ku, Tokyo 113-8656, Japan}
\affiliation{RIKEN Center for Emergent Matter Science, 2-1 Hirosawa, Wako, Saitama 351-0198, Japan}

\author{Tim O. Wehling}
%\email{twehling@uni-bremen.de}
\affiliation{Institute of Theoretical Physics, Bremen Center for Computational Materials Science,
	and MAPEX Center for Materials and Processes, University of Bremen, Otto-Hahn-Allee 1, 28359 Bremen, Germany}
\affiliation{I. Institute of Theoretical Physics, University of Hamburg, Notkestraße 9, 22607 Hamburg, Germany}

\date{\today}
\begin{abstract}  
Twisted Van der Waals systems offer the unprecedented possibility to tune different states of correlated quantum matter with an external non-invasive electrostatic doping. The nature of the superconducting order presents a recurring open question in this context. In this work, we quantitatively assess the case of spin-fluctuation-mediated pairing for $\Gamma$-valley twisted transition metal dichalcogenide homobilayers. We self-consistently and dynamically calculate the doping dependent superconducting transition temperature $T_{\mathrm{c}}$ revealing a superconducting dome with a maximal $T_{\mathrm{c}}\approx$~0.1\,--\,1~K depending on twist angle. We compare our results with conventional phonon-mediated superconductivity and identify clear fingerprints in the doping dependence of $T_{\mathrm{c}}$, which allow experiments to distinguish between different pairing mechanisms. 
\end{abstract} 
\maketitle

\textit{Introduction.} 
Twisting layers of two-dimensional (2D) materials leads to a moir\'{e} pattern, where flat bands can emerge close to the Fermi level \cite{PhysRevB.82.121407,BisPNAS1082011,KosPRX82018}.
The associated quenching of the kinetic energy leads to strong electronic correlations, which often interplay with topology \cite{marrazzo_prediction_2018,wu_unconventional_2019,Pizarro2020}.
Amongst these effects are Mott and topological Chern insulators, and different kinds of magnetic, nematic, and superconducting ordered states \cite{CaoN5562018ins,CaoN5562018sc,Yankowitz1059,Chen2019,Sharpe605,Lu2019,PhysRevLett.123.197702,Regan2020,Tang2020,Shen2020,Stepanov2020,Saito2020,Cao2020,Wang2020,Chen2021,Polshyn2020,Nuckolls2020,Park2021,Wu2021,Xu2021,Cao264,Liu2021}. One can precisely tune between these states and change the filling of the flat bands from completely empty to filled by electrostatic doping \cite{PhysRevLett.117.116804}, which is special in the domain of correlated materials.
%These emergent states can be further tuned and studied with temperature, pressure, strain, different dielectric environments, and displacement and magnetic fields.

The nature of superconducting states in twisted 2D systems is highly controversial.
On the one hand, unconventional pairing mechanisms based on spin, orbital, and/or nematic fluctuations are regularly hypothesized \cite{PhysRevLett.121.217001,You2019,PhysRevB.100.155145,PhysRevB.103.L041103,Cao264,Fischer2022}.
The reasons are that superconductivity emerges next to a strongly correlated state \cite{CaoN5562018sc,Yankowitz1059,Wang2020,Cao264,Oh2021} and that the ratio of critical temperature $T_{\mathrm{c}}$ and Fermi temperature $T_{\mathrm{F}}$ fits within the boundary of other unconventional superconductors \cite{CaoN5562018sc,Talantsev2020,Park2021}.
On the other hand, recent experiments in magic-angle twisted bilayer graphene (MATBG) showed that the strongly correlated states and superconductivity are affected differently by the dielectric environment \cite{Stepanov2020,Saito2020,Liu2021}, which might point to a conventional origin, i.e., electron-phonon coupling.

Twisted 2D systems can be classified according to the symmetry of the low-energy Hamiltonian associated with the moir\'{e} pattern \cite{Kennes2021}. Honeycomb twisted 2D systems hold promises for hosting correlated Dirac fermions and topological $d+id$ chiral superconductivity \cite{Kuznetsova2005,BlackSchaffer2014,PhysRevLett.121.217001}. 
Examples of honeycomb systems are MATBG \cite{KosPRX82018,PhysRevX.8.031089}, twisted double bilayer graphene \cite{Lee2019,Haddadi2020,PhysRevB.102.155146}, magic-angle twisted trilayer graphene (MATTG) \cite{Carr2020,Wu2021a,PhysRevResearch.2.033357}, and twisted transition metal dichalcogenides (TMDCs) \cite{Xian2019,Xian2021,gammaTMDCsMacdonald2021}.
Most of the experimental and theoretical work has been focused on graphite-based systems. However, their complicated low-energy electronic structure makes theoretical many-body studies difficult \cite{PhysRevB.99.195455,PhysRevResearch.1.013001,PhysRevLett.125.176404,Haddadi2020,PhysRevB.102.155146,Carr2020,Wu2021a,PhysRevResearch.2.033357}.
The low-energy electronic structure of twisted TMDC homobilayers is simpler than that of twisted graphitic systems since it can be described by an effective single-orbital model (see below) and there is no topological obstruction preventing simple Wannier constructions \cite{Zou2018,PhysRevX.8.031089}, so that they are good candidates for a handshake of experiments and theoretical many-body modeling. Recently, a zero-resistance state has been reported in a twisted TMDC homobilayer \cite{Wang2020}, the nature of which remains to be understood.
%The simplest models for twisted TMDCs can be obtained for the so called $\Gamma$-valley twisted TMDCs, where the low-energy electronic structure can be modeled by a single-orbital honeycomb tight-binding model.

In this Letter we provide a quantitative study of the critical temperature $T_{\text{c}}$ due to spin-fluctuation-mediated pairing in $\Gamma$-valley twisted TMDCs in terms of doping and twisting, which we obtain dynamically by means of the fluctuation exchange approximation (FLEX) \cite{Bickers1989a,Bickers1989b}. We additionally provide a qualitative understanding of spin fluctuations versus electron-phonon coupling and propose that experimental measurements on the doping-dependent $T_{\text{c}}$ can help to unveil the nature of the superconducting states. 

\begin{figure}[h]
	\hfill
	\includegraphics[clip,width=0.48\textwidth]{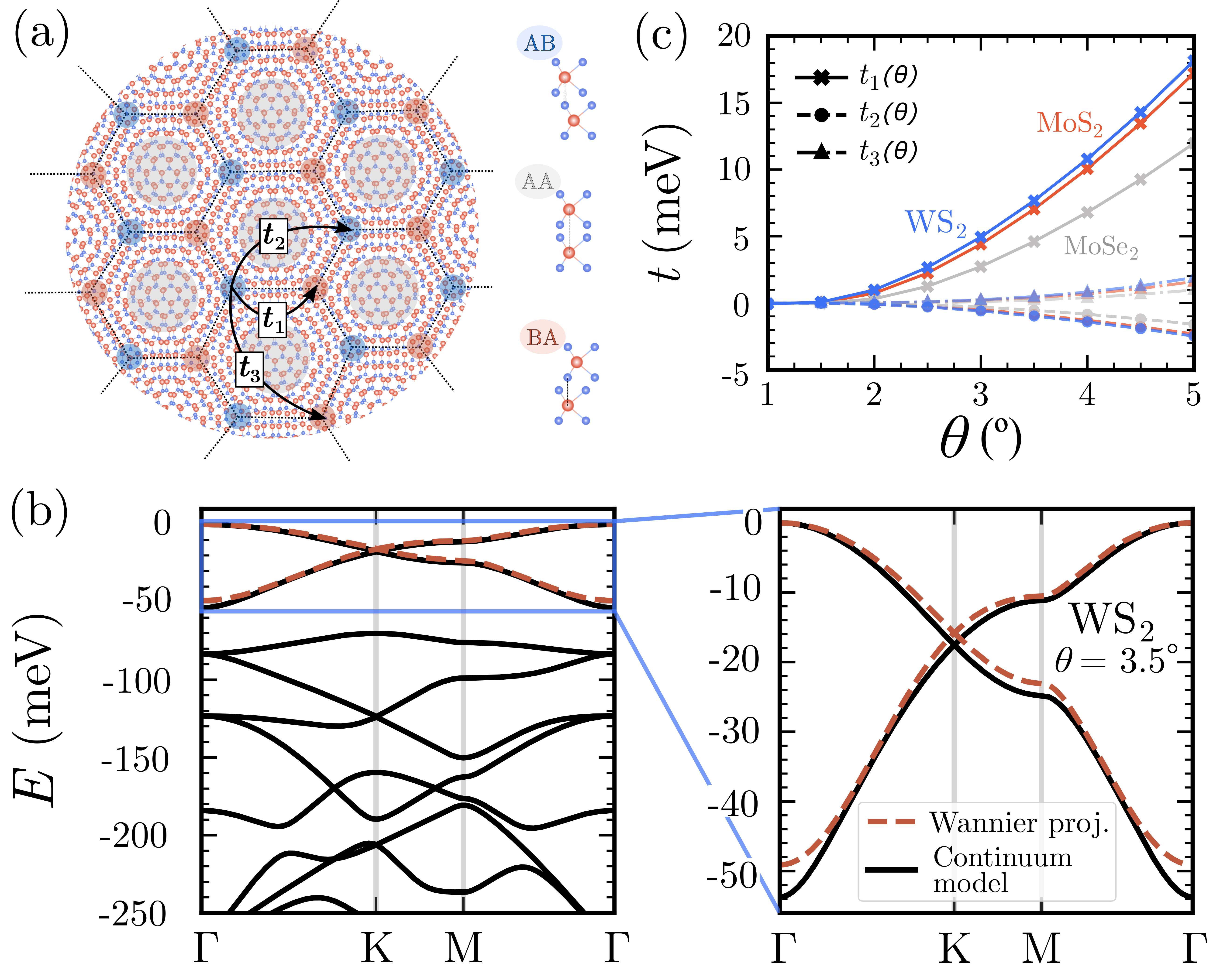}
	%\vspace{-.30 cm}
	\caption{(Color online) $\Gamma$-valley twisted TMDCs. (a) Moir\'{e} pattern of twisted TMDCs. AA (gray shaded), AB (blue shaded), and BA (red shaded) regions in the moir\'{e} pattern correspond to different stackings of the two layers as shown in the right side. Dashed black lines serve as a guide to the eye to identify the honeycomb superlattice. The most relevant hopping processes are sketched with black arrows. (b) Continuum model for WS$_2$ at a twist angle of \(\theta=3.5^{\circ}\) (black solid line) with third-nearest-neighbor hopping tight-binding model (red dashed line) of the highest valence bands. The effective honeycomb lattice is formed by the AB and BA moir\'{e} sites. Right panel shows a zoom to the flat Dirac bands. (c) Twist-angle dependence of the hopping parameters for different $\Gamma$-valley twisted TMDCs, WS$_2$, MoS$_2$, and MoSe$_2$ obtained via Wannier projection.}
	\label{FIG1} 
	\vspace{-0.3 cm}
\end{figure}

%%%%%%%%%%%%%%%%%%%%%%%%%%%%%%%%%%%%%%%%%%
%%%%%%%%%%%%%%%%%%%%%%%%%%%%%%%%%%%%%%%%%%

\textit{Band structures, Wannierization and Hartree potential effect.} 
We consider the twisting of TMDC homobilayers with respect to the untwisted ($\theta = 0^{\circ}$) situation. In Fig.~\ref{FIG1}(a) we show the emergent moir\'{e} pattern, where the AA regions form a triangular superlattice surrounded by AB and BA regions arranged in a honeycomb pattern. We focus on the so-called $\Gamma$-valley twisted TMDCs (WS$_2$, MoS$_2$, and MoSe$_2$) \cite{gammaTMDCsMacdonald2021,PhysRevB.102.201115,PhysRevB.103.155142}, in which the valence band maximum of the untwisted homobilayer is located at the center of the Brillouin zone $\Gamma$ due to the hybridization between the transition metal $d$ and chalcogen $p$ orbitals. The valence band maximum is an antibonding state energetically separated from its bonding counterpart by hundreds of meV. Also the conduction band is separated from the valence band maximum by more than an eV \cite{Rasmussen2015,Chaves2020}. Based on this observation, Angeli and MacDonald constructed a low-energy continuum model in which only the antibonding state is included \cite{gammaTMDCsMacdonald2021}. The emergent symmetry of these moir\'e valence bands is that of a 2D honeycomb lattice.

In the plane-wave basis defined by the moir\'{e} vectors $\bold{G}=m\bold{G}_1^{\mathrm{M}} + n\bold{G}_2^{\mathrm{M}}$, with integers $m,n$ and $\bold{G}_{1,2}^{\mathrm{M}}$ spanning the reciprocal lattice, the Hamiltonian of the continuum model takes the form
\begin{equation}
H = -\frac{\hbar^2 | \bold{k} + \bold{G} |^2}{2m^{*}} \delta_{\bold{G},\bold{G'}} + V_{\mathrm{M}}(\bold{G} - \bold{G'}),
\label{eq1}
\end{equation}
where $\bold{k}$ are the reciprocal vectors defined in the mini Brillouin zone, $m^{*}$ is the effective mass, and $V_{\mathrm{M}}(\bold{G})$ is the Fourier transformation of the moir\'{e} potential \cite{SM}. This Hamiltonian is expanded up to a plane-wave cutoff $G_{\mathrm{c}}=5G^{\mathrm{M}}$, where $G^{\mathrm{M}}=|\bold{G}_{1,2}^{\mathrm{M}}|$.

The low-energy electronic structure of $\Gamma$-valley twisted TMDCs shows 2D honeycomb Dirac bands for the highest valence band, see Fig.~\ref{FIG1}(b). The Dirac point can be accessed by hole doping and the Dirac bands are well isolated from higher energy bands for twist angles $1^{\circ}<\theta<5^{\circ}$. In this twist angle range, the bandwidth of the flat Dirac bands varies between $0.5$ to $100$ meV \cite{SM}.

We next construct a tight-binding Hamiltonian to describe the flat Dirac bands with one orbital per honeycomb superlattice site. Here, the AB and BA regions play the role of the A and B sublattice degrees of freedom in the honeycomb lattice. We include up to three nearest-neighbor hoppings \(t_1,t_2,t_3\) in our model which we obtain by Wannier projection \cite{SM}. The tight-binding and continuum model band structure agree very well in the twist angle range $1^{\circ}<\theta<5^{\circ}$ \cite{SM}. We observe that, when comparing amongst different $\Gamma$-valley twisted TMDCs, the transition metal does not influence the hopping amplitudes significantly, while the chalcogen atoms do. We also find dominant nearest-neighbor hopping $t_1 \gg t_2, t_3$, and that $t_1 \sim \alpha \sin^2(\theta) \approx \alpha \theta^2$  with $\alpha\approx 2$~eV/rad$^2$.

In other twisted 2D systems, such as MATBG \cite{Guinea2018,PhysRevB.100.205113,PhysRevB.102.045107,PhysRevB.102.155149} or MATTG \cite{Fischer2022}, the effect of the purely electrostatic and long-range (Hartree) potential in doped flat bands is important. Thus, we also consider its influence in our model \cite{SM}. We find that, contrary to MATBG and MATTG, the flat Dirac bands remain unaffected. Therefore, we disregard doping-dependent long-range Coulomb reconstructions on the flat bands from now on.

%%%%%%%%%%%%%%%%%%%%%%%%%%%%%%%%%%%%%%%%%%
%%%%%%%%%%%%%%%%%%%%%%%%%%%%%%%%%%%%%%%%%%

\textit{Doping- and interaction-dependent spin fluctuations.}
Since the nearest-neighbor hopping $t_1(\theta)$ dominates over $t_2$ and $t_3$ for twist angles $1^{\circ}<\theta<5^{\circ}$, we neglect $t_2$ and $t_3$ here. We discuss their influence in the Supplemental Material \cite{SM}. We study the Hubbard Hamiltonian
\begin{align}
	H_U = - \sum_{<im, jn> ,\sigma} t \left( c^{\dagger}_{im\sigma}c^{}_{jn\sigma} + \mathrm{h.c.} \right) + U\sum_{im}n_{im\uparrow}n_{im\downarrow},
	\label{eq:hubbard}
\end{align}
where the hopping amplitude $t\equiv t_1(\theta)$ sets the energy scale and $<\!\! im, jn\!\!>$ denotes that the sum is limited to neighboring lattice sites of a moir\'{e} unit cell $i,j$ and sublattice $m,n$. $c^{\dagger}_{im\sigma}$ ($c^{}_{im\sigma}$) creates (annihilates) an electron with spin $\sigma$ and $U$ is the local Coulomb repulsion between electrons on the same lattice site. In the simplified tight-binding model the system is particle-hole symmetric with respect to the Dirac point and has a logarithmically diverging density of states (DOS) at the Van Hove singularities (VHS) that are present in the $M$ points of the Brillouin zone \cite{RevModPhys.81.109}. We re-define our zero-doping level $\delta=0$ to correspond to a Fermi energy at the Dirac point, see Fig.~\ref{FIG1}(b). Then, the VHS are at $\delta=0.25$.

The Hubbard model for the honeycomb lattice has previously been studied, indicating a rich phase diagram of competing many-body instabilities \cite{Assaad2013,BlackSchaffer2014,Kiesel2012,Wang2012,Nandkishore2012,Xu2016,Raczkowski2020,Costa2021}. The emergence of spin-density waves (SDWs) and superconductivity in close proximity suggests an unconventional pairing mechanism mediated by spin fluctuations. Following this premise, we study the magnetic and superconducting excitations using FLEX \cite{SM} in the model described above \cite{Kuroki2001,Onari2002,Kuroki2010} as a representation of spin-fluctuation-mediated pairing in $\Gamma$-valley twisted TMDCs. A recently developed sparse sampling method \cite{Li2020,Witt2021} enabled us to perform the numerically demanding calculations at low temperatures.

In FLEX, the exchange of spin and charge fluctuations is treated dynamically and self-consistently with an effective electron-electron interaction of a random phase approximation (RPA)-type. Estimates of the Hubbard interaction parameter given in the Supplemental Material \cite{SM} show that $U$ is highly tunable via twist angle and the dielectric environment \cite{Roesner2016,Raja2017,Pizarro2019,Goodwin2019,Goodwin2020,Stepanov2020,Saito2020,Arora2020,Kim2020,Liu2021}. E.g. for $\theta=5^\circ$, the interaction strength is tunable in the range $4<U/t <8$ \cite{SM}. In addition, vertex corrections that are neglected in FLEX could contribute to further screening \cite{Bulut1993,Bulut1994,Zhang2009}. In what follows we treat $U$ as a free parameter.

\begin{figure}
	\centering
	\includegraphics[clip,width=0.48\textwidth]{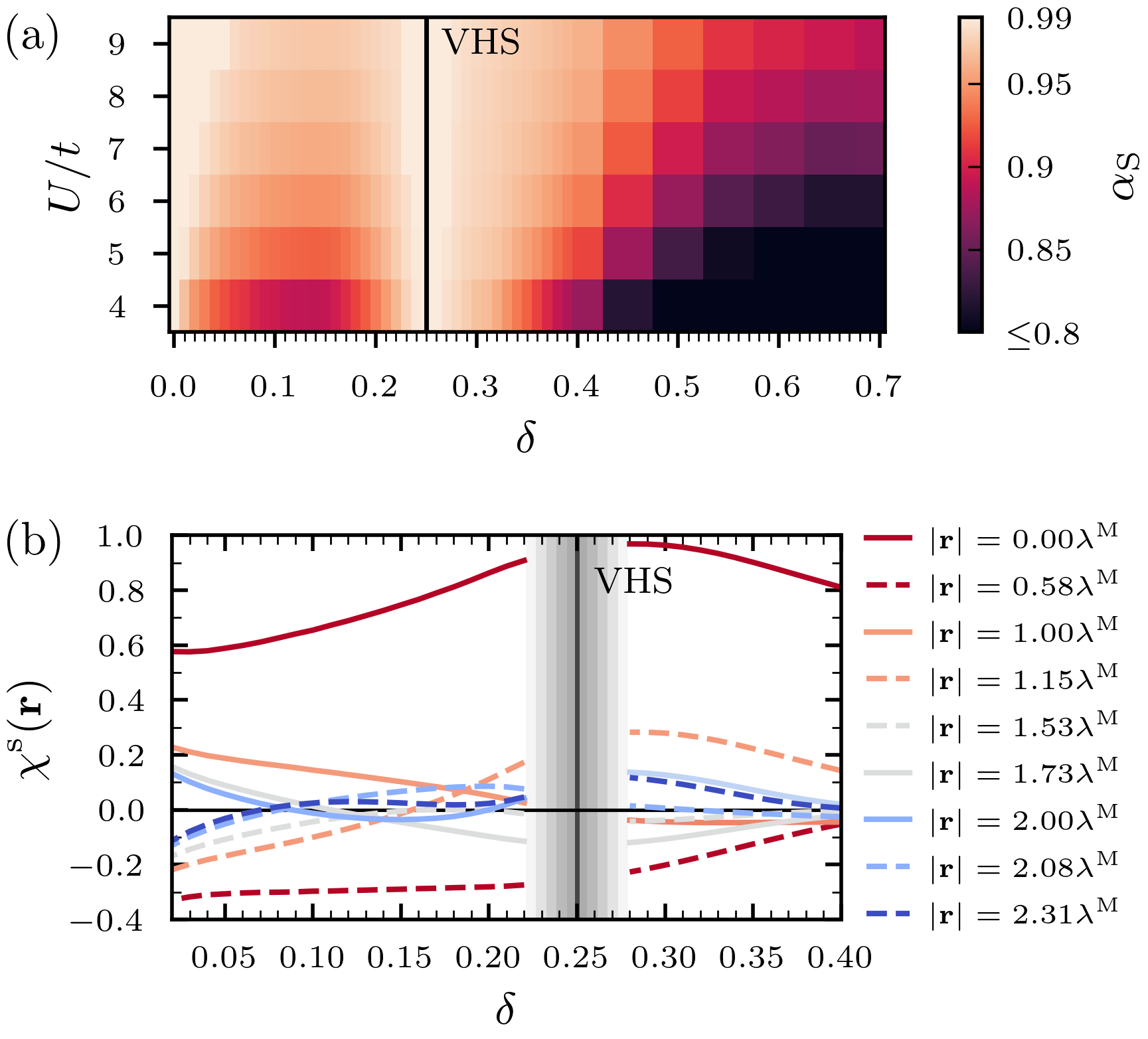}
	%\vspace{-.30 cm}
	\caption{(Color online) Spin fluctuation characteristics of $\Gamma$-valley twisted TMDCs at \(T/t=0.003\). (a) Leading Stoner enhancement factor $\alpha_{\mathrm{S}}=\mathrm{max}_{\bold{q}}\lbrace U\chi^0(\bold{q})\rbrace$ for different Coulomb interaction strengths $U/t$ and dopings $\delta$ with respect to the Dirac point as obtained from FLEX. A transition to a quasi-ordered magnetic state is assumed for $\alpha_{\mathrm{S}}\geq 0.99$. (b) Real space components of the static spin susceptibility \(\chi^{\mathrm{s}}(\mathbf{r})\) for $U/t=6$. Up to eighth nearest-neighbor components are shown with $|\mathbf{r}|$ denoting the distance between two spins in terms of the moir\'{e} unit length $\lambda^{\mathrm{M}}$. Solid (dashed) lines correspond to the AA (AB) components of \(\chi^{\mathrm{s}}\), i.e., correlations between same (different) sublattice sites. The area around the Van Hove singularities (VHS) is not accessible because of too strong fluctuations, which is marked by a gray shaded area (c.f.~panel (a)).}
	\label{FIG2} 
	%\vspace{-0.3 cm}

\end{figure}

We analyze the emergence of magnetic fluctuations by inspecting the leading Stoner enhancement factor $\alpha_{\mathrm{S}}=\mathrm{max}_{\bold{q}}\lbrace U\chi^0(\bold{q})\rbrace$ with the static irreducible susceptibility \(\chi^0(\bold{q})\), see Fig.~\ref{FIG2}(a). If \(\alpha_{\mathrm{S}}\geq0.99\), the transition to a quasi-ordered magnetic phase is assumed \cite{SM}. This situation occurs in two locations of the phase diagram: at the Dirac point ($\delta=0$) and in the vicinity of the VHS ($\delta=0.25$). Between these two points, \(\alpha_{\mathrm{S}}\) is strong but does not reach the quasi-ordering criterion. When doping beyond the VHS ($\delta\gtrsim 0.3$), the relative spin fluctuation strength rapidly decreases and the system stays paramagnetic. Increasing the interaction strength amplifies \(\alpha_{\mathrm{S}}\), but the doping dependence remains largely unaffected.

The presence of strong spin fluctuations can induce an effective electron-electron interaction with non-local attractive regions, which gives rise to superconducting pairing \cite{Moriya2003,Scalapino2012}. FLEX captures this effect with the dominant contribution to the effective interaction coming from the spin susceptibility \(\chi^{\mathrm{s}}\). Optimal pairing conditions can be inferred from its real space profile. In Fig.~\ref{FIG2}(b), we show the  doping dependence of up to eighth nearest-neighbor components of \(\chi^{\mathrm{s}}(\bold{r})\) for $U/t=6$. For doping levels in the vicinity of the Dirac point, antiferromagnetic fluctuations with respect to the sublattices A and B emerge, i.e., the AB (inter-sublattice) components have a negative sign, whereas the AA (intra-sublattice) components are positive. Upon doping, initially the longest range and successively the more short range components of \(\chi^{\mathrm{s}}\) change their sign. Hence, antiferromagnetic fluctuations are suppressed and an admixture of ferromagnetic components to \(\chi^{\mathrm{s}}\) is triggered away from the Dirac point. Beyond the VHS, fluctuations turn increasingly ferromagnetic and their relative strength weakens. Further insight into the emerging SDWs and their origin from nesting conditions can be gained by inspecting the momentum-resolved structure of \(\chi^{\mathrm{s}}\) \cite{SM}.

To investigate the dominant superconducting pairing symmetry and transition temperature \(T_{\mathrm{c}}\), we solve the linearized Eliashberg equation for different possible order parameters. In all our calculations, the degenerate singlet \(d\)-wave pairings (\(d_{xy}\),\(d_{x^2-y^2}\)) emerge as the dominant pairing symmetries \cite{SM}. This is in agreement with the antiferromagnetic fluctuations as they favor singlet-pairing symmetries. Below \(T_{\mathrm{c}}\), the order parameter forms a time-reversal symmetry broken chiral \(d+id\) pairing state \cite{Kuznetsova2005,Nandkishore2012,BlackSchaffer2014}.

\begin{figure}
	\includegraphics[clip,width=0.48\textwidth]{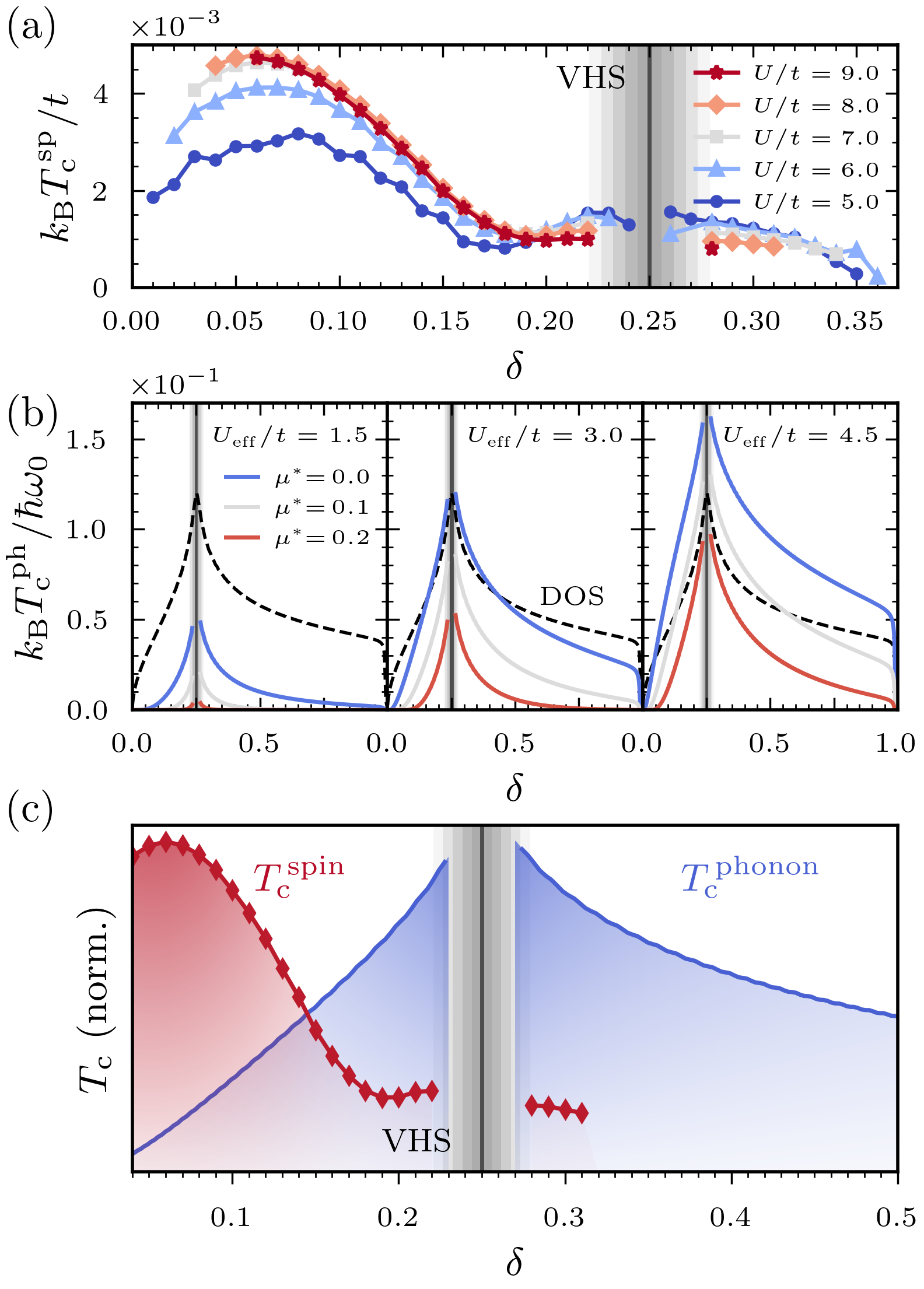}
	%\vspace{-.30 cm}
	\caption{(Color online) Doping dependence of the superconducting transition temperature $T_{\text{c}}$ in $\Gamma$-valley twisted TMDCs. (a) Phase diagram for spin-fluctuation-mediated pairing for different local Coulomb interaction strengths $U/t$ from FLEX calculations. The critical temperature $T_{\text{c}}/t$ belongs to the dominant singlet \(d\)-wave pairing symmetry.
	(b)~Phonon-mediated \(T_{\mathrm{c}}\) for an Einstein-Holstein phonon mode $\omega_0$ and for different Coulomb pseudopotentials $\mu^*=0.0$ (blue), $0.1$ (gray), and $0.2$ (red). From left to right the effective attractive interaction $U_{\mathrm{eff}}$ from electron-phonon coupling is increased. The trend of the density of states (DOS) is indicated by black dashed lines.
	(c) Comparison of doping dependent phase diagram of the maximum $T_{\text{c}}$ obtained for spin fluctuations and phonons. This phase diagram holds qualitatively for all $\Gamma$-valley twisted TMDCs.}
	\label{FIG3} 
	%\vspace{-0.3 cm}
\end{figure}

In Fig.~\ref{FIG3}(a), we show the doping dependence of \(T_{\mathrm{c}}\) for different \(U/t\). We find a superconducting dome which is characterized by a non-monotonous behavior with a maximal value \(T^{\mathrm{max}}_{\mathrm{c}}\) at an optimal doping \( \delta_{\mathrm{opt}}\). The existence of such a maximum results from the interplay of the pairing interaction pattern and the electronic DOS at the Fermi level \cite{SM}. Doping away from the Dirac point increases the DOS at the Fermi level, which supports $d$-wave pairing via antiferromagnetic spin fluctuations. As the doping level further increases, however, an increasing amount of pair-breaking ferromagnetic spin fluctuations emerges (c.f.~Fig.~\ref{FIG2}(b)). Thus, we reach a situation of optimal doping around $\delta_{\mathrm{opt}}=0.06$ and a decrease in $T_{\mathrm{c}}$ upon further doping.

We obtain increasing \(T_{\mathrm{c}}\) with increasing interaction \(U\) until the highest $T_{\mathrm{c}}$ curve for \(U/t=8\) with a maximal value of \(T_{\mathrm{c}}^{\mathrm{max}}/t=4.8\times 10^{-3}\) at \(\delta_{\mathrm{opt}}=0.06\) is reached. For larger interactions \(U/t\gtrsim 9\), the superconducting transition temperatures decrease again.

Near the VHS, possible superconducting order \cite{BlackSchaffer2014,Kiesel2012,Wang2012,Ying2018} is masked by magnetic fluctuations in FLEX, that is $\alpha_{\mathrm{S}}$ exceeds 0.99. As the spin fluctuations turn ferromagnetic towards and beyond the VHS doping, singlet-pairing emerging from antiferromagnetic spin fluctuation exchange is strongly suppressed. In addition, triplet superconductivity does not arise for any temperature $T/t > 10^{-3}$ due to the weakened fluctuation strength \cite{SM}. The material and twist-angle dependent hopping amplitudes given in Fig.~\ref{FIG1}(c) set the temperature scale. \(T_{\mathrm{c}}\) takes values on the order of 0.1 -- 1~K, which is in agreement with reports on other twisted 2D systems \cite{CaoN5562018sc,Yankowitz1059,Wang2020,Cao264}.

%%%%%%%%%%%%%%%%%%%%%%%%%%%%%%%%%%%%%%%%%%
%%%%%%%%%%%%%%%%%%%%%%%%%%%%%%%%%%%%%%%%%%

\textit{Spin fluctuations vs. electron-phonon coupling.} 
The previous discussion showed that superconductivity arising from a spin-fluctuation-mediated pairing mechanism exhibits a characteristic doping dependent transition line with a clear maximum near Dirac filling. To contrast this pairing scenario, we assess how the doping characteristics appear in the conventional case of phonon-mediated superconductivity. 

We estimate the transition temperature $T^{\mathrm{ph}}_{\mathrm{c}}$ by means of McMillan's formula \cite{PhysRev.167.331,DYNES1972615}
\begin{align}
	T^{\mathrm{ph}}_{\mathrm{c}} = \frac{\hbar\langle\omega\rangle}{1.20\,k_{\mathrm{B}}}\exp\left\lbrace\frac{-1.04(1+\lambda)}{\lambda-0.62\lambda\mu^*-\mu^*}\right\rbrace\, ,
	\label{eq:phonon_Tc}
\end{align}
where $\langle \omega \rangle$ is an effective phonon frequency, $\lambda$ denotes the effective pairing strength, and $\mu^*$ is the Tolmachev-Morel-Anderson Coulomb pseudopotential \cite{Tolmachev1961,Morel1962}. $\langle \omega \rangle$ and $\lambda$ are generally obtained from the phonon spectral function $\alpha^2 F(\omega)$. Here, we consider the limiting case of an Einstein-Holstein phonon mode, i.e., with a constant electron-phonon coupling $g$ and a constant phonon frequency $\omega_{0}$. We discuss the opposite limit of non-local Peierls coupling with dispersive phonons in the Supplemental Material \cite{SM}.

When discussing phonon-mediated superconductivity, it is simplest to do so in terms of a BCS-like effective attractive interaction $U_{\mathrm{eff}}$ such that $\lambda = U_{\mathrm{eff}} N(\delta)$ with the DOS $N(\delta)$ per spin and unit cell for a particular doping $\delta$. In the Einstein-Holstein model, we explicitly have $U_{\mathrm{eff}}=2g^2/\hbar\omega_0$ and $\langle \omega \rangle = \omega_{0}$. The exact values of $\omega_0$, $U_{\mathrm{eff}}$, and $\mu^*$ are material specific and they depend on factors like twist angle or external screening \cite{Wu2018,Lin2018,Choi2018,Lian2019,Debnath2020,Lin2021,Parzefall2021}. Twisted TMDCs display phonon modes at energies on the order of a few 10~meV as in the bulk and in addition feature moiré phonons in the range 2\,--\,5~meV \cite{Lin2021,Parzefall2021,Quan2021}. We estimate $U_{\mathrm{eff}}$ to be in the large range of 0.05\,--\,8\,$t$ \cite{SM} and typical values of $\mu^*$ are in the range $0.0$\,--\,$0.2$ \cite{Morel1962}.

The key observation is that the generic doping dependence of $T^{\mathrm{ph}}_{\mathrm{c}}$ mainly derives from the DOS. To illustrate this point, we show in Fig.~\ref{FIG3}(b) results for $T^{\mathrm{ph}}_{\mathrm{c}}$ in units of $\omega_0$ for different $U_{\mathrm{eff}}$ and $\mu^*$ together with the DOS. We tune $U_{\mathrm{eff}}$ to yield weak to intermediate coupling strengths ($\lambda \lesssim 1$). Increasing $\mu^*$ suppresses $T^{\mathrm{ph}}_{\mathrm{c}}$,  while increasing $U_{\mathrm{eff}}$ has the opposite effect. The quantitative details may vary, but the qualitative shape of the $T^{\mathrm{ph}}_{\mathrm{c}}$ curve is unaffected in both cases, mainly following $N(\delta)$. Our findings for non-local coupling \cite{SM} support the robustness of the doping dependence of $T^{\,\mathrm{ph}}_{\mathrm{c}}$: A peaked structure emerges around the VHS and extends over the whole range of dopings $\delta\in[0,1]$, i.e., in particular also beyond the VHS in the region of $\delta > 0.25$. The relevant temperature scale is set by $\omega_0$ with $T_{\mathrm{c}}^{\mathrm{ph}}$ taking values on the order of 0.1\,--\,10~K. Note that we excluded the immediate region around the VHS in our discussion since the competition of different instabilities complicates the determination of the doping dependence \cite{Rice1975,Jiang2014,Berges2020}.

A direct comparison of the doping dependent superconducting phase diagram obtained for the different pairing mechanisms, spin fluctuations and phonons, is given in Fig.~\ref{FIG3}(c). We use the normalized results of Fig.~\ref{FIG3}(a) for \(U/t=8\) and Fig.~\ref{FIG3}(b) for $U_{\mathrm{eff}}/t = 3$ and $\mu^*=0.0$. Each pairing mechanism shows unique fingerprints for which we identify two key differences. Firstly, phonon-mediated superconductivity shows a clear increase towards the VHS, whereas for spin-fluctuation-mediated pairing a global maximum appears close to the Dirac point at $\delta_{\mathrm{opt}}$. Secondly, phonon-induced superconductivity persists over a wider doping range and is closely linked to the DOS, while spin-fluctuation-mediated superconductivity is confined to a narrow doping region near an antiferromagnetic instability, which diminishes rapidly after the VHS due to the emergence of pair-breaking ferromagnetic fluctuations.

%%%%%%%%%%%%%%%%%%%%%%%%%%%%%%%%%%%%%%%%%%
%%%%%%%%%%%%%%%%%%%%%%%%%%%%%%%%%%%%%%%%%%

\textit{Summary and outlook.} 
We have shown that the superconducting response to doping in $\Gamma$-valley twisted TMDCs depends decisively on the quantum nature of the pairing fluctuations. Superconducting pairing mechanisms and their experimental determination present a major open problem in twisted 2D systems. Thus, the question is: Are there simple experimental ways to discern different pairing mechanisms? 

Our analysis demonstrates that different pairing mechanisms can be distinguished by simple doping-dependent transport experiments of \(T_{\mathrm{c}}\). Fingerprints unique to the particular microscopic mechanism can be found with respect to the  the doping levels of maximal \(T_{\mathrm{c}}\) or the doping extent over which superconductivity persists.

This possibility has not been explored in other unconventional superconductors \cite{KeiN5182015} because of the difficulties of performing systematic doping-dependent studies. The consideration of multiple local and non-local electron-phonon coupling profiles \cite{SM} indicates that the distinct doping dependence between $T_{\mathrm{c}}^{\mathrm{sp}}$ and $T_{\mathrm{c}}^{\mathrm{ph}}$ is generic. Hence, our conclusions are not only valid for the $\Gamma$-valley twisted TMDCs, but they can help to elucidate pairing mechanisms in other twisted 2D Van der Waals materials, such as MATBG or MATTG.

% Acknowledgements
\textit{Acknowledgements.} 
We acknowledge support and funding by the Deutsche Forschungsgemeinschaft (DFG) via RTG 2247 (QM$^3$) (project number 286518848), via the priority program SPP 2244 (project number 422707584), via EXC 2077 (University Allowance, University of Bremen, project number 390741603), and via the Cluster of Excellence `CUI: Advanced Imaging of Matter' -- EXC 2056 (project number 390715994). Funding from the European Commission via the Graphene Flagship Core Project 3 (grant agreement ID: 881603) and computing time at the HLRN facilities (Berlin and G\"ottingen) is acknowledged. This work was supported by a Grant-in-Aid for Scientific Research (No.~19H05825) by MEXT and by JST PRESTO (No.~JPMJPR20L7), Japan.

\bibliography{twistFLEX_bib}

\begin{thebibliography}{106}%
\makeatletter
\providecommand \@ifxundefined [1]{%
 \@ifx{#1\undefined}
}%
\providecommand \@ifnum [1]{%
 \ifnum #1\expandafter \@firstoftwo
 \else \expandafter \@secondoftwo
 \fi
}%
\providecommand \@ifx [1]{%
 \ifx #1\expandafter \@firstoftwo
 \else \expandafter \@secondoftwo
 \fi
}%
\providecommand \natexlab [1]{#1}%
\providecommand \enquote  [1]{``#1''}%
\providecommand \bibnamefont  [1]{#1}%
\providecommand \bibfnamefont [1]{#1}%
\providecommand \citenamefont [1]{#1}%
\providecommand \href@noop [0]{\@secondoftwo}%
\providecommand \href [0]{\begingroup \@sanitize@url \@href}%
\providecommand \@href[1]{\@@startlink{#1}\@@href}%
\providecommand \@@href[1]{\endgroup#1\@@endlink}%
\providecommand \@sanitize@url [0]{\catcode `\\12\catcode `\$12\catcode
  `\&12\catcode `\#12\catcode `\^12\catcode `\_12\catcode `\%12\relax}%
\providecommand \@@startlink[1]{}%
\providecommand \@@endlink[0]{}%
\providecommand \url  [0]{\begingroup\@sanitize@url \@url }%
\providecommand \@url [1]{\endgroup\@href {#1}{\urlprefix }}%
\providecommand \urlprefix  [0]{URL }%
\providecommand \Eprint [0]{\href }%
\providecommand \doibase [0]{https://doi.org/}%
\providecommand \selectlanguage [0]{\@gobble}%
\providecommand \bibinfo  [0]{\@secondoftwo}%
\providecommand \bibfield  [0]{\@secondoftwo}%
\providecommand \translation [1]{[#1]}%
\providecommand \BibitemOpen [0]{}%
\providecommand \bibitemStop [0]{}%
\providecommand \bibitemNoStop [0]{.\EOS\space}%
\providecommand \EOS [0]{\spacefactor3000\relax}%
\providecommand \BibitemShut  [1]{\csname bibitem#1\endcsname}%
\let\auto@bib@innerbib\@empty
%</preamble>
\bibitem [{\citenamefont {Su\'arez~Morell}\ \emph {et~al.}(2010)\citenamefont
  {Su\'arez~Morell}, \citenamefont {Correa}, \citenamefont {Vargas},
  \citenamefont {Pacheco},\ and\ \citenamefont
  {Barticevic}}]{PhysRevB.82.121407}%
  \BibitemOpen
  \bibfield  {author} {\bibinfo {author} {\bibfnamefont {E.}~\bibnamefont
  {Su\'arez~Morell}}, \bibinfo {author} {\bibfnamefont {J.~D.}\ \bibnamefont
  {Correa}}, \bibinfo {author} {\bibfnamefont {P.}~\bibnamefont {Vargas}},
  \bibinfo {author} {\bibfnamefont {M.}~\bibnamefont {Pacheco}},\ and\ \bibinfo
  {author} {\bibfnamefont {Z.}~\bibnamefont {Barticevic}},\ }\bibfield  {title}
  {\bibinfo {title} {Flat bands in slightly twisted bilayer graphene:
  {T}ight-binding calculations},\ }\href
  {https://doi.org/10.1103/PhysRevB.82.121407} {\bibfield  {journal} {\bibinfo
  {journal} {Phys. Rev. B}\ }\textbf {\bibinfo {volume} {82}},\ \bibinfo
  {pages} {121407} (\bibinfo {year} {2010})}\BibitemShut {NoStop}%
\bibitem [{\citenamefont {Bistritzer}\ and\ \citenamefont
  {MacDonald}(2011)}]{BisPNAS1082011}%
  \BibitemOpen
  \bibfield  {author} {\bibinfo {author} {\bibfnamefont {R.}~\bibnamefont
  {Bistritzer}}\ and\ \bibinfo {author} {\bibfnamefont {A.~H.}\ \bibnamefont
  {MacDonald}},\ }\bibfield  {title} {\bibinfo {title} {Moir{\'e} bands in
  twisted double-layer graphene},\ }\href
  {https://doi.org/10.1073/pnas.1108174108} {\bibfield  {journal} {\bibinfo
  {journal} {Proc. Natl. Acad. Sci.}\ }\textbf {\bibinfo {volume} {108}},\
  \bibinfo {pages} {12233} (\bibinfo {year} {2011})}\BibitemShut {NoStop}%
\bibitem [{\citenamefont {Koshino}\ \emph {et~al.}(2018)\citenamefont
  {Koshino}, \citenamefont {Yuan}, \citenamefont {Koretsune}, \citenamefont
  {Ochi}, \citenamefont {Kuroki},\ and\ \citenamefont {Fu}}]{KosPRX82018}%
  \BibitemOpen
  \bibfield  {author} {\bibinfo {author} {\bibfnamefont {M.}~\bibnamefont
  {Koshino}}, \bibinfo {author} {\bibfnamefont {N.~F.~Q.}\ \bibnamefont
  {Yuan}}, \bibinfo {author} {\bibfnamefont {T.}~\bibnamefont {Koretsune}},
  \bibinfo {author} {\bibfnamefont {M.}~\bibnamefont {Ochi}}, \bibinfo {author}
  {\bibfnamefont {K.}~\bibnamefont {Kuroki}},\ and\ \bibinfo {author}
  {\bibfnamefont {L.}~\bibnamefont {Fu}},\ }\bibfield  {title} {\bibinfo
  {title} {{Maximally Localized Wannier Orbitals and the Extended Hubbard Model
  for Twisted Bilayer Graphene}},\ }\href
  {https://doi.org/10.1103/PhysRevX.8.031087} {\bibfield  {journal} {\bibinfo
  {journal} {Phys. Rev. X}\ }\textbf {\bibinfo {volume} {8}},\ \bibinfo {pages}
  {031087} (\bibinfo {year} {2018})}\BibitemShut {NoStop}%
\bibitem [{\citenamefont {Marrazzo}\ \emph {et~al.}(2018)\citenamefont
  {Marrazzo}, \citenamefont {Gibertini}, \citenamefont {Campi}, \citenamefont
  {Mounet},\ and\ \citenamefont {Marzari}}]{marrazzo_prediction_2018}%
  \BibitemOpen
  \bibfield  {author} {\bibinfo {author} {\bibfnamefont {A.}~\bibnamefont
  {Marrazzo}}, \bibinfo {author} {\bibfnamefont {M.}~\bibnamefont {Gibertini}},
  \bibinfo {author} {\bibfnamefont {D.}~\bibnamefont {Campi}}, \bibinfo
  {author} {\bibfnamefont {N.}~\bibnamefont {Mounet}},\ and\ \bibinfo {author}
  {\bibfnamefont {N.}~\bibnamefont {Marzari}},\ }\bibfield  {title} {\bibinfo
  {title} {Prediction of a {Large}-{Gap} and {Switchable} {Kane}-{Mele}
  {Quantum} {Spin} {Hall} {Insulator}},\ }\href
  {https://doi.org/10.1103/PhysRevLett.120.117701} {\bibfield  {journal}
  {\bibinfo  {journal} {Phys. Rev. Lett.}\ }\textbf {\bibinfo {volume} {120}},\
  \bibinfo {pages} {117701} (\bibinfo {year} {2018})}\BibitemShut {NoStop}%
\bibitem [{\citenamefont {Wu}\ \emph {et~al.}(2019)\citenamefont {Wu},
  \citenamefont {Fink}, \citenamefont {Hanke}, \citenamefont {Thomale},\ and\
  \citenamefont {Di~Sante}}]{wu_unconventional_2019}%
  \BibitemOpen
  \bibfield  {author} {\bibinfo {author} {\bibfnamefont {X.}~\bibnamefont
  {Wu}}, \bibinfo {author} {\bibfnamefont {M.}~\bibnamefont {Fink}}, \bibinfo
  {author} {\bibfnamefont {W.}~\bibnamefont {Hanke}}, \bibinfo {author}
  {\bibfnamefont {R.}~\bibnamefont {Thomale}},\ and\ \bibinfo {author}
  {\bibfnamefont {D.}~\bibnamefont {Di~Sante}},\ }\bibfield  {title} {\bibinfo
  {title} {Unconventional superconductivity in a doped quantum spin {Hall}
  insulator},\ }\href {https://doi.org/10.1103/PhysRevB.100.041117} {\bibfield
  {journal} {\bibinfo  {journal} {Phys. Rev. B}\ }\textbf {\bibinfo {volume}
  {100}},\ \bibinfo {pages} {041117} (\bibinfo {year} {2019})}\BibitemShut
  {NoStop}%
\bibitem [{\citenamefont {Pizarro}\ \emph {et~al.}(2020)\citenamefont
  {Pizarro}, \citenamefont {Adler}, \citenamefont {Zantout}, \citenamefont
  {Mertz}, \citenamefont {Barone}, \citenamefont {Valent{\'i}}, \citenamefont
  {Sangiovanni},\ and\ \citenamefont {Wehling}}]{Pizarro2020}%
  \BibitemOpen
  \bibfield  {author} {\bibinfo {author} {\bibfnamefont {J.~M.}\ \bibnamefont
  {Pizarro}}, \bibinfo {author} {\bibfnamefont {S.}~\bibnamefont {Adler}},
  \bibinfo {author} {\bibfnamefont {K.}~\bibnamefont {Zantout}}, \bibinfo
  {author} {\bibfnamefont {T.}~\bibnamefont {Mertz}}, \bibinfo {author}
  {\bibfnamefont {P.}~\bibnamefont {Barone}}, \bibinfo {author} {\bibfnamefont
  {R.}~\bibnamefont {Valent{\'i}}}, \bibinfo {author} {\bibfnamefont
  {G.}~\bibnamefont {Sangiovanni}},\ and\ \bibinfo {author} {\bibfnamefont
  {T.~O.}\ \bibnamefont {Wehling}},\ }\bibfield  {title} {\bibinfo {title}
  {Deconfinement of {M}ott localized electrons into topological and
  spin--orbit-coupled {D}irac fermions},\ }\href
  {https://doi.org/10.1038/s41535-020-00277-3} {\bibfield  {journal} {\bibinfo
  {journal} {npj Quantum Mater.}\ }\textbf {\bibinfo {volume} {5}},\ \bibinfo
  {pages} {79} (\bibinfo {year} {2020})}\BibitemShut {NoStop}%
\bibitem [{\citenamefont {Cao}\ \emph {et~al.}(2018{\natexlab{a}})\citenamefont
  {Cao}, \citenamefont {Fatemi}, \citenamefont {Demir}, \citenamefont {Fang},
  \citenamefont {Tomarken}, \citenamefont {Luo}, \citenamefont
  {Sanchez-Yamagishi}, \citenamefont {Watanabe}, \citenamefont {Taniguchi},
  \citenamefont {Kaxiras}, \citenamefont {Ashoori},\ and\ \citenamefont
  {Jarillo-Herrero}}]{CaoN5562018ins}%
  \BibitemOpen
  \bibfield  {author} {\bibinfo {author} {\bibfnamefont {Y.}~\bibnamefont
  {Cao}}, \bibinfo {author} {\bibfnamefont {V.}~\bibnamefont {Fatemi}},
  \bibinfo {author} {\bibfnamefont {A.}~\bibnamefont {Demir}}, \bibinfo
  {author} {\bibfnamefont {S.}~\bibnamefont {Fang}}, \bibinfo {author}
  {\bibfnamefont {S.~L.}\ \bibnamefont {Tomarken}}, \bibinfo {author}
  {\bibfnamefont {J.~Y.}\ \bibnamefont {Luo}}, \bibinfo {author} {\bibfnamefont
  {J.~D.}\ \bibnamefont {Sanchez-Yamagishi}}, \bibinfo {author} {\bibfnamefont
  {K.}~\bibnamefont {Watanabe}}, \bibinfo {author} {\bibfnamefont
  {T.}~\bibnamefont {Taniguchi}}, \bibinfo {author} {\bibfnamefont
  {E.}~\bibnamefont {Kaxiras}}, \bibinfo {author} {\bibfnamefont {R.~C.}\
  \bibnamefont {Ashoori}},\ and\ \bibinfo {author} {\bibfnamefont
  {P.}~\bibnamefont {Jarillo-Herrero}},\ }\bibfield  {title} {\bibinfo {title}
  {Correlated insulator behaviour at half-filling in magic-angle graphene
  superlattices},\ }\href {https://doi.org/10.1038/nature26154} {\bibfield
  {journal} {\bibinfo  {journal} {Nature}\ }\textbf {\bibinfo {volume} {556}},\
  \bibinfo {pages} {80} (\bibinfo {year} {2018}{\natexlab{a}})}\BibitemShut
  {NoStop}%
\bibitem [{\citenamefont {Cao}\ \emph {et~al.}(2018{\natexlab{b}})\citenamefont
  {Cao}, \citenamefont {Fatemi}, \citenamefont {Fang}, \citenamefont
  {Watanabe}, \citenamefont {Taniguchi}, \citenamefont {Kaxiras},\ and\
  \citenamefont {Jarillo-Herrero}}]{CaoN5562018sc}%
  \BibitemOpen
  \bibfield  {author} {\bibinfo {author} {\bibfnamefont {Y.}~\bibnamefont
  {Cao}}, \bibinfo {author} {\bibfnamefont {V.}~\bibnamefont {Fatemi}},
  \bibinfo {author} {\bibfnamefont {S.}~\bibnamefont {Fang}}, \bibinfo {author}
  {\bibfnamefont {K.}~\bibnamefont {Watanabe}}, \bibinfo {author}
  {\bibfnamefont {T.}~\bibnamefont {Taniguchi}}, \bibinfo {author}
  {\bibfnamefont {E.}~\bibnamefont {Kaxiras}},\ and\ \bibinfo {author}
  {\bibfnamefont {P.}~\bibnamefont {Jarillo-Herrero}},\ }\bibfield  {title}
  {\bibinfo {title} {Unconventional superconductivity in magic-angle graphene
  superlattices},\ }\href {https://doi.org/10.1038/nature26160} {\bibfield
  {journal} {\bibinfo  {journal} {Nature}\ }\textbf {\bibinfo {volume} {556}},\
  \bibinfo {pages} {43} (\bibinfo {year} {2018}{\natexlab{b}})}\BibitemShut
  {NoStop}%
\bibitem [{\citenamefont {Yankowitz}\ \emph {et~al.}(2019)\citenamefont
  {Yankowitz}, \citenamefont {Chen}, \citenamefont {Polshyn}, \citenamefont
  {Zhang}, \citenamefont {Watanabe}, \citenamefont {Taniguchi}, \citenamefont
  {Graf}, \citenamefont {Young},\ and\ \citenamefont {Dean}}]{Yankowitz1059}%
  \BibitemOpen
  \bibfield  {author} {\bibinfo {author} {\bibfnamefont {M.}~\bibnamefont
  {Yankowitz}}, \bibinfo {author} {\bibfnamefont {S.}~\bibnamefont {Chen}},
  \bibinfo {author} {\bibfnamefont {H.}~\bibnamefont {Polshyn}}, \bibinfo
  {author} {\bibfnamefont {Y.}~\bibnamefont {Zhang}}, \bibinfo {author}
  {\bibfnamefont {K.}~\bibnamefont {Watanabe}}, \bibinfo {author}
  {\bibfnamefont {T.}~\bibnamefont {Taniguchi}}, \bibinfo {author}
  {\bibfnamefont {D.}~\bibnamefont {Graf}}, \bibinfo {author} {\bibfnamefont
  {A.~F.}\ \bibnamefont {Young}},\ and\ \bibinfo {author} {\bibfnamefont
  {C.~R.}\ \bibnamefont {Dean}},\ }\bibfield  {title} {\bibinfo {title} {Tuning
  superconductivity in twisted bilayer graphene},\ }\href
  {https://doi.org/10.1126/science.aav1910} {\bibfield  {journal} {\bibinfo
  {journal} {Science}\ }\textbf {\bibinfo {volume} {363}},\ \bibinfo {pages}
  {1059} (\bibinfo {year} {2019})}\BibitemShut {NoStop}%
\bibitem [{\citenamefont {Chen}\ \emph {et~al.}(2019)\citenamefont {Chen},
  \citenamefont {Jiang}, \citenamefont {Wu}, \citenamefont {Lyu}, \citenamefont
  {Li}, \citenamefont {Chittari}, \citenamefont {Watanabe}, \citenamefont
  {Taniguchi}, \citenamefont {Shi}, \citenamefont {Jung}, \citenamefont
  {Zhang},\ and\ \citenamefont {Wang}}]{Chen2019}%
  \BibitemOpen
  \bibfield  {author} {\bibinfo {author} {\bibfnamefont {G.}~\bibnamefont
  {Chen}}, \bibinfo {author} {\bibfnamefont {L.}~\bibnamefont {Jiang}},
  \bibinfo {author} {\bibfnamefont {S.}~\bibnamefont {Wu}}, \bibinfo {author}
  {\bibfnamefont {B.}~\bibnamefont {Lyu}}, \bibinfo {author} {\bibfnamefont
  {H.}~\bibnamefont {Li}}, \bibinfo {author} {\bibfnamefont {B.~L.}\
  \bibnamefont {Chittari}}, \bibinfo {author} {\bibfnamefont {K.}~\bibnamefont
  {Watanabe}}, \bibinfo {author} {\bibfnamefont {T.}~\bibnamefont {Taniguchi}},
  \bibinfo {author} {\bibfnamefont {Z.}~\bibnamefont {Shi}}, \bibinfo {author}
  {\bibfnamefont {J.}~\bibnamefont {Jung}}, \bibinfo {author} {\bibfnamefont
  {Y.}~\bibnamefont {Zhang}},\ and\ \bibinfo {author} {\bibfnamefont
  {F.}~\bibnamefont {Wang}},\ }\bibfield  {title} {\bibinfo {title} {Evidence
  of a gate-tunable {M}ott insulator in a trilayer graphene moir{\'e}
  superlattice},\ }\href {https://doi.org/10.1038/s41567-019-0631-4} {\bibfield
   {journal} {\bibinfo  {journal} {Nat. Phys.}\ }\textbf {\bibinfo {volume}
  {15}},\ \bibinfo {pages} {237} (\bibinfo {year} {2019})}\BibitemShut
  {NoStop}%
\bibitem [{\citenamefont {Sharpe}\ \emph {et~al.}(2019)\citenamefont {Sharpe},
  \citenamefont {Fox}, \citenamefont {Barnard}, \citenamefont {Finney},
  \citenamefont {Watanabe}, \citenamefont {Taniguchi}, \citenamefont
  {Kastner},\ and\ \citenamefont {Goldhaber-Gordon}}]{Sharpe605}%
  \BibitemOpen
  \bibfield  {author} {\bibinfo {author} {\bibfnamefont {A.~L.}\ \bibnamefont
  {Sharpe}}, \bibinfo {author} {\bibfnamefont {E.~J.}\ \bibnamefont {Fox}},
  \bibinfo {author} {\bibfnamefont {A.~W.}\ \bibnamefont {Barnard}}, \bibinfo
  {author} {\bibfnamefont {J.}~\bibnamefont {Finney}}, \bibinfo {author}
  {\bibfnamefont {K.}~\bibnamefont {Watanabe}}, \bibinfo {author}
  {\bibfnamefont {T.}~\bibnamefont {Taniguchi}}, \bibinfo {author}
  {\bibfnamefont {M.~A.}\ \bibnamefont {Kastner}},\ and\ \bibinfo {author}
  {\bibfnamefont {D.}~\bibnamefont {Goldhaber-Gordon}},\ }\bibfield  {title}
  {\bibinfo {title} {Emergent ferromagnetism near three-quarters filling in
  twisted bilayer graphene},\ }\href {https://doi.org/10.1126/science.aaw3780}
  {\bibfield  {journal} {\bibinfo  {journal} {Science}\ }\textbf {\bibinfo
  {volume} {365}},\ \bibinfo {pages} {605} (\bibinfo {year}
  {2019})}\BibitemShut {NoStop}%
\bibitem [{\citenamefont {Lu}\ \emph {et~al.}(2019)\citenamefont {Lu},
  \citenamefont {Stepanov}, \citenamefont {Yang}, \citenamefont {Xie},
  \citenamefont {Aamir}, \citenamefont {Das}, \citenamefont {Urgell},
  \citenamefont {Watanabe}, \citenamefont {Taniguchi}, \citenamefont {Zhang},
  \citenamefont {Bachtold}, \citenamefont {MacDonald},\ and\ \citenamefont
  {Efetov}}]{Lu2019}%
  \BibitemOpen
  \bibfield  {author} {\bibinfo {author} {\bibfnamefont {X.}~\bibnamefont
  {Lu}}, \bibinfo {author} {\bibfnamefont {P.}~\bibnamefont {Stepanov}},
  \bibinfo {author} {\bibfnamefont {W.}~\bibnamefont {Yang}}, \bibinfo {author}
  {\bibfnamefont {M.}~\bibnamefont {Xie}}, \bibinfo {author} {\bibfnamefont
  {M.~A.}\ \bibnamefont {Aamir}}, \bibinfo {author} {\bibfnamefont
  {I.}~\bibnamefont {Das}}, \bibinfo {author} {\bibfnamefont {C.}~\bibnamefont
  {Urgell}}, \bibinfo {author} {\bibfnamefont {K.}~\bibnamefont {Watanabe}},
  \bibinfo {author} {\bibfnamefont {T.}~\bibnamefont {Taniguchi}}, \bibinfo
  {author} {\bibfnamefont {G.}~\bibnamefont {Zhang}}, \bibinfo {author}
  {\bibfnamefont {A.}~\bibnamefont {Bachtold}}, \bibinfo {author}
  {\bibfnamefont {A.~H.}\ \bibnamefont {MacDonald}},\ and\ \bibinfo {author}
  {\bibfnamefont {D.~K.}\ \bibnamefont {Efetov}},\ }\bibfield  {title}
  {\bibinfo {title} {Superconductors, orbital magnets and correlated states in
  magic-angle bilayer graphene},\ }\href
  {https://doi.org/10.1038/s41586-019-1695-0} {\bibfield  {journal} {\bibinfo
  {journal} {Nature}\ }\textbf {\bibinfo {volume} {574}},\ \bibinfo {pages}
  {653} (\bibinfo {year} {2019})}\BibitemShut {NoStop}%
\bibitem [{\citenamefont {Burg}\ \emph {et~al.}(2019)\citenamefont {Burg},
  \citenamefont {Zhu}, \citenamefont {Taniguchi}, \citenamefont {Watanabe},
  \citenamefont {MacDonald},\ and\ \citenamefont
  {Tutuc}}]{PhysRevLett.123.197702}%
  \BibitemOpen
  \bibfield  {author} {\bibinfo {author} {\bibfnamefont {G.~W.}\ \bibnamefont
  {Burg}}, \bibinfo {author} {\bibfnamefont {J.}~\bibnamefont {Zhu}}, \bibinfo
  {author} {\bibfnamefont {T.}~\bibnamefont {Taniguchi}}, \bibinfo {author}
  {\bibfnamefont {K.}~\bibnamefont {Watanabe}}, \bibinfo {author}
  {\bibfnamefont {A.~H.}\ \bibnamefont {MacDonald}},\ and\ \bibinfo {author}
  {\bibfnamefont {E.}~\bibnamefont {Tutuc}},\ }\bibfield  {title} {\bibinfo
  {title} {Correlated {I}nsulating {S}tates in {T}wisted {D}ouble {B}ilayer
  {G}raphene},\ }\href {https://doi.org/10.1103/PhysRevLett.123.197702}
  {\bibfield  {journal} {\bibinfo  {journal} {Phys. Rev. Lett.}\ }\textbf
  {\bibinfo {volume} {123}},\ \bibinfo {pages} {197702} (\bibinfo {year}
  {2019})}\BibitemShut {NoStop}%
\bibitem [{\citenamefont {Regan}\ \emph {et~al.}(2020)\citenamefont {Regan},
  \citenamefont {Wang}, \citenamefont {Jin}, \citenamefont {Bakti~Utama},
  \citenamefont {Gao}, \citenamefont {Wei}, \citenamefont {Zhao}, \citenamefont
  {Zhao}, \citenamefont {Zhang}, \citenamefont {Yumigeta}, \citenamefont
  {Blei}, \citenamefont {Carlstr{\"o}m}, \citenamefont {Watanabe},
  \citenamefont {Taniguchi}, \citenamefont {Tongay}, \citenamefont {Crommie},
  \citenamefont {Zettl},\ and\ \citenamefont {Wang}}]{Regan2020}%
  \BibitemOpen
  \bibfield  {author} {\bibinfo {author} {\bibfnamefont {E.~C.}\ \bibnamefont
  {Regan}}, \bibinfo {author} {\bibfnamefont {D.}~\bibnamefont {Wang}},
  \bibinfo {author} {\bibfnamefont {C.}~\bibnamefont {Jin}}, \bibinfo {author}
  {\bibfnamefont {M.~I.}\ \bibnamefont {Bakti~Utama}}, \bibinfo {author}
  {\bibfnamefont {B.}~\bibnamefont {Gao}}, \bibinfo {author} {\bibfnamefont
  {X.}~\bibnamefont {Wei}}, \bibinfo {author} {\bibfnamefont {S.}~\bibnamefont
  {Zhao}}, \bibinfo {author} {\bibfnamefont {W.}~\bibnamefont {Zhao}}, \bibinfo
  {author} {\bibfnamefont {Z.}~\bibnamefont {Zhang}}, \bibinfo {author}
  {\bibfnamefont {K.}~\bibnamefont {Yumigeta}}, \bibinfo {author}
  {\bibfnamefont {M.}~\bibnamefont {Blei}}, \bibinfo {author} {\bibfnamefont
  {J.~D.}\ \bibnamefont {Carlstr{\"o}m}}, \bibinfo {author} {\bibfnamefont
  {K.}~\bibnamefont {Watanabe}}, \bibinfo {author} {\bibfnamefont
  {T.}~\bibnamefont {Taniguchi}}, \bibinfo {author} {\bibfnamefont
  {S.}~\bibnamefont {Tongay}}, \bibinfo {author} {\bibfnamefont
  {M.}~\bibnamefont {Crommie}}, \bibinfo {author} {\bibfnamefont
  {A.}~\bibnamefont {Zettl}},\ and\ \bibinfo {author} {\bibfnamefont
  {F.}~\bibnamefont {Wang}},\ }\bibfield  {title} {\bibinfo {title} {Mott and
  generalized {W}igner crystal states in {WS}e{$_2$}/{WS}{$_2$} moir{\'e}
  superlattices},\ }\href {https://doi.org/10.1038/s41586-020-2092-4}
  {\bibfield  {journal} {\bibinfo  {journal} {Nature}\ }\textbf {\bibinfo
  {volume} {579}},\ \bibinfo {pages} {359} (\bibinfo {year}
  {2020})}\BibitemShut {NoStop}%
\bibitem [{\citenamefont {Tang}\ \emph {et~al.}(2020)\citenamefont {Tang},
  \citenamefont {Li}, \citenamefont {Li}, \citenamefont {Xu}, \citenamefont
  {Liu}, \citenamefont {Barmak}, \citenamefont {Watanabe}, \citenamefont
  {Taniguchi}, \citenamefont {MacDonald}, \citenamefont {Shan},\ and\
  \citenamefont {Mak}}]{Tang2020}%
  \BibitemOpen
  \bibfield  {author} {\bibinfo {author} {\bibfnamefont {Y.}~\bibnamefont
  {Tang}}, \bibinfo {author} {\bibfnamefont {L.}~\bibnamefont {Li}}, \bibinfo
  {author} {\bibfnamefont {T.}~\bibnamefont {Li}}, \bibinfo {author}
  {\bibfnamefont {Y.}~\bibnamefont {Xu}}, \bibinfo {author} {\bibfnamefont
  {S.}~\bibnamefont {Liu}}, \bibinfo {author} {\bibfnamefont {K.}~\bibnamefont
  {Barmak}}, \bibinfo {author} {\bibfnamefont {K.}~\bibnamefont {Watanabe}},
  \bibinfo {author} {\bibfnamefont {T.}~\bibnamefont {Taniguchi}}, \bibinfo
  {author} {\bibfnamefont {A.~H.}\ \bibnamefont {MacDonald}}, \bibinfo {author}
  {\bibfnamefont {J.}~\bibnamefont {Shan}},\ and\ \bibinfo {author}
  {\bibfnamefont {K.~F.}\ \bibnamefont {Mak}},\ }\bibfield  {title} {\bibinfo
  {title} {Simulation of {H}ubbard model physics in {WS}e{$_2$}/{WS}{$_2$}
  moir{\'e} superlattices},\ }\href {https://doi.org/10.1038/s41586-020-2085-3}
  {\bibfield  {journal} {\bibinfo  {journal} {Nature}\ }\textbf {\bibinfo
  {volume} {579}},\ \bibinfo {pages} {353} (\bibinfo {year}
  {2020})}\BibitemShut {NoStop}%
\bibitem [{\citenamefont {Shen}\ \emph {et~al.}(2020)\citenamefont {Shen},
  \citenamefont {Chu}, \citenamefont {Wu}, \citenamefont {Li}, \citenamefont
  {Wang}, \citenamefont {Zhao}, \citenamefont {Tang}, \citenamefont {Liu},
  \citenamefont {Tian}, \citenamefont {Watanabe}, \citenamefont {Taniguchi},
  \citenamefont {Yang}, \citenamefont {Meng}, \citenamefont {Shi},
  \citenamefont {Yazyev},\ and\ \citenamefont {Zhang}}]{Shen2020}%
  \BibitemOpen
  \bibfield  {author} {\bibinfo {author} {\bibfnamefont {C.}~\bibnamefont
  {Shen}}, \bibinfo {author} {\bibfnamefont {Y.}~\bibnamefont {Chu}}, \bibinfo
  {author} {\bibfnamefont {Q.}~\bibnamefont {Wu}}, \bibinfo {author}
  {\bibfnamefont {N.}~\bibnamefont {Li}}, \bibinfo {author} {\bibfnamefont
  {S.}~\bibnamefont {Wang}}, \bibinfo {author} {\bibfnamefont {Y.}~\bibnamefont
  {Zhao}}, \bibinfo {author} {\bibfnamefont {J.}~\bibnamefont {Tang}}, \bibinfo
  {author} {\bibfnamefont {J.}~\bibnamefont {Liu}}, \bibinfo {author}
  {\bibfnamefont {J.}~\bibnamefont {Tian}}, \bibinfo {author} {\bibfnamefont
  {K.}~\bibnamefont {Watanabe}}, \bibinfo {author} {\bibfnamefont
  {T.}~\bibnamefont {Taniguchi}}, \bibinfo {author} {\bibfnamefont
  {R.}~\bibnamefont {Yang}}, \bibinfo {author} {\bibfnamefont {Z.~Y.}\
  \bibnamefont {Meng}}, \bibinfo {author} {\bibfnamefont {D.}~\bibnamefont
  {Shi}}, \bibinfo {author} {\bibfnamefont {O.~V.}\ \bibnamefont {Yazyev}},\
  and\ \bibinfo {author} {\bibfnamefont {G.}~\bibnamefont {Zhang}},\ }\bibfield
   {title} {\bibinfo {title} {Correlated states in twisted double bilayer
  graphene},\ }\href {https://doi.org/10.1038/s41567-020-0825-9} {\bibfield
  {journal} {\bibinfo  {journal} {Nat. Phys.}\ }\textbf {\bibinfo {volume}
  {16}},\ \bibinfo {pages} {520} (\bibinfo {year} {2020})}\BibitemShut
  {NoStop}%
\bibitem [{\citenamefont {Stepanov}\ \emph {et~al.}(2020)\citenamefont
  {Stepanov}, \citenamefont {Das}, \citenamefont {Lu}, \citenamefont
  {Fahimniya}, \citenamefont {Watanabe}, \citenamefont {Taniguchi},
  \citenamefont {Koppens}, \citenamefont {Lischner}, \citenamefont {Levitov},\
  and\ \citenamefont {Efetov}}]{Stepanov2020}%
  \BibitemOpen
  \bibfield  {author} {\bibinfo {author} {\bibfnamefont {P.}~\bibnamefont
  {Stepanov}}, \bibinfo {author} {\bibfnamefont {I.}~\bibnamefont {Das}},
  \bibinfo {author} {\bibfnamefont {X.}~\bibnamefont {Lu}}, \bibinfo {author}
  {\bibfnamefont {A.}~\bibnamefont {Fahimniya}}, \bibinfo {author}
  {\bibfnamefont {K.}~\bibnamefont {Watanabe}}, \bibinfo {author}
  {\bibfnamefont {T.}~\bibnamefont {Taniguchi}}, \bibinfo {author}
  {\bibfnamefont {F.~H.~L.}\ \bibnamefont {Koppens}}, \bibinfo {author}
  {\bibfnamefont {J.}~\bibnamefont {Lischner}}, \bibinfo {author}
  {\bibfnamefont {L.}~\bibnamefont {Levitov}},\ and\ \bibinfo {author}
  {\bibfnamefont {D.~K.}\ \bibnamefont {Efetov}},\ }\bibfield  {title}
  {\bibinfo {title} {Untying the insulating and superconducting orders in
  magic-angle graphene},\ }\href {https://doi.org/10.1038/s41586-020-2459-6}
  {\bibfield  {journal} {\bibinfo  {journal} {Nature}\ }\textbf {\bibinfo
  {volume} {583}},\ \bibinfo {pages} {375} (\bibinfo {year}
  {2020})}\BibitemShut {NoStop}%
\bibitem [{\citenamefont {Saito}\ \emph {et~al.}(2020)\citenamefont {Saito},
  \citenamefont {Ge}, \citenamefont {Watanabe}, \citenamefont {Taniguchi},\
  and\ \citenamefont {Young}}]{Saito2020}%
  \BibitemOpen
  \bibfield  {author} {\bibinfo {author} {\bibfnamefont {Y.}~\bibnamefont
  {Saito}}, \bibinfo {author} {\bibfnamefont {J.}~\bibnamefont {Ge}}, \bibinfo
  {author} {\bibfnamefont {K.}~\bibnamefont {Watanabe}}, \bibinfo {author}
  {\bibfnamefont {T.}~\bibnamefont {Taniguchi}},\ and\ \bibinfo {author}
  {\bibfnamefont {A.~F.}\ \bibnamefont {Young}},\ }\bibfield  {title} {\bibinfo
  {title} {Independent superconductors and correlated insulators in twisted
  bilayer graphene},\ }\href {https://doi.org/10.1038/s41567-020-0928-3}
  {\bibfield  {journal} {\bibinfo  {journal} {Nat. Phys.}\ }\textbf {\bibinfo
  {volume} {16}},\ \bibinfo {pages} {926} (\bibinfo {year} {2020})}\BibitemShut
  {NoStop}%
\bibitem [{\citenamefont {Cao}\ \emph {et~al.}(2020)\citenamefont {Cao},
  \citenamefont {Rodan-Legrain}, \citenamefont {Rubies-Bigorda}, \citenamefont
  {Park}, \citenamefont {Watanabe}, \citenamefont {Taniguchi},\ and\
  \citenamefont {Jarillo-Herrero}}]{Cao2020}%
  \BibitemOpen
  \bibfield  {author} {\bibinfo {author} {\bibfnamefont {Y.}~\bibnamefont
  {Cao}}, \bibinfo {author} {\bibfnamefont {D.}~\bibnamefont {Rodan-Legrain}},
  \bibinfo {author} {\bibfnamefont {O.}~\bibnamefont {Rubies-Bigorda}},
  \bibinfo {author} {\bibfnamefont {J.~M.}\ \bibnamefont {Park}}, \bibinfo
  {author} {\bibfnamefont {K.}~\bibnamefont {Watanabe}}, \bibinfo {author}
  {\bibfnamefont {T.}~\bibnamefont {Taniguchi}},\ and\ \bibinfo {author}
  {\bibfnamefont {P.}~\bibnamefont {Jarillo-Herrero}},\ }\bibfield  {title}
  {\bibinfo {title} {Tunable correlated states and spin-polarized phases in
  twisted bilayer--bilayer graphene},\ }\href
  {https://doi.org/10.1038/s41586-020-2260-6} {\bibfield  {journal} {\bibinfo
  {journal} {Nature}\ }\textbf {\bibinfo {volume} {583}},\ \bibinfo {pages}
  {215} (\bibinfo {year} {2020})}\BibitemShut {NoStop}%
\bibitem [{\citenamefont {Wang}\ \emph {et~al.}(2020)\citenamefont {Wang},
  \citenamefont {Shih}, \citenamefont {Ghiotto}, \citenamefont {Xian},
  \citenamefont {Rhodes}, \citenamefont {Tan}, \citenamefont {Claassen},
  \citenamefont {Kennes}, \citenamefont {Bai}, \citenamefont {Kim},
  \citenamefont {Watanabe}, \citenamefont {Taniguchi}, \citenamefont {Zhu},
  \citenamefont {Hone}, \citenamefont {Rubio}, \citenamefont {Pasupathy},\ and\
  \citenamefont {Dean}}]{Wang2020}%
  \BibitemOpen
  \bibfield  {author} {\bibinfo {author} {\bibfnamefont {L.}~\bibnamefont
  {Wang}}, \bibinfo {author} {\bibfnamefont {E.-M.}\ \bibnamefont {Shih}},
  \bibinfo {author} {\bibfnamefont {A.}~\bibnamefont {Ghiotto}}, \bibinfo
  {author} {\bibfnamefont {L.}~\bibnamefont {Xian}}, \bibinfo {author}
  {\bibfnamefont {D.~A.}\ \bibnamefont {Rhodes}}, \bibinfo {author}
  {\bibfnamefont {C.}~\bibnamefont {Tan}}, \bibinfo {author} {\bibfnamefont
  {M.}~\bibnamefont {Claassen}}, \bibinfo {author} {\bibfnamefont {D.~M.}\
  \bibnamefont {Kennes}}, \bibinfo {author} {\bibfnamefont {Y.}~\bibnamefont
  {Bai}}, \bibinfo {author} {\bibfnamefont {B.}~\bibnamefont {Kim}}, \bibinfo
  {author} {\bibfnamefont {K.}~\bibnamefont {Watanabe}}, \bibinfo {author}
  {\bibfnamefont {T.}~\bibnamefont {Taniguchi}}, \bibinfo {author}
  {\bibfnamefont {X.}~\bibnamefont {Zhu}}, \bibinfo {author} {\bibfnamefont
  {J.}~\bibnamefont {Hone}}, \bibinfo {author} {\bibfnamefont {A.}~\bibnamefont
  {Rubio}}, \bibinfo {author} {\bibfnamefont {A.~N.}\ \bibnamefont
  {Pasupathy}},\ and\ \bibinfo {author} {\bibfnamefont {C.~R.}\ \bibnamefont
  {Dean}},\ }\bibfield  {title} {\bibinfo {title} {Correlated electronic phases
  in twisted bilayer transition metal dichalcogenides},\ }\href
  {https://doi.org/10.1038/s41563-020-0708-6} {\bibfield  {journal} {\bibinfo
  {journal} {Nat. Mater.}\ }\textbf {\bibinfo {volume} {19}},\ \bibinfo {pages}
  {861} (\bibinfo {year} {2020})}\BibitemShut {NoStop}%
\bibitem [{\citenamefont {Chen}\ \emph {et~al.}(2021)\citenamefont {Chen},
  \citenamefont {He}, \citenamefont {Zhang}, \citenamefont {Hsieh},
  \citenamefont {Fei}, \citenamefont {Watanabe}, \citenamefont {Taniguchi},
  \citenamefont {Cobden}, \citenamefont {Xu}, \citenamefont {Dean},\ and\
  \citenamefont {Yankowitz}}]{Chen2021}%
  \BibitemOpen
  \bibfield  {author} {\bibinfo {author} {\bibfnamefont {S.}~\bibnamefont
  {Chen}}, \bibinfo {author} {\bibfnamefont {M.}~\bibnamefont {He}}, \bibinfo
  {author} {\bibfnamefont {Y.-H.}\ \bibnamefont {Zhang}}, \bibinfo {author}
  {\bibfnamefont {V.}~\bibnamefont {Hsieh}}, \bibinfo {author} {\bibfnamefont
  {Z.}~\bibnamefont {Fei}}, \bibinfo {author} {\bibfnamefont {K.}~\bibnamefont
  {Watanabe}}, \bibinfo {author} {\bibfnamefont {T.}~\bibnamefont {Taniguchi}},
  \bibinfo {author} {\bibfnamefont {D.~H.}\ \bibnamefont {Cobden}}, \bibinfo
  {author} {\bibfnamefont {X.}~\bibnamefont {Xu}}, \bibinfo {author}
  {\bibfnamefont {C.~R.}\ \bibnamefont {Dean}},\ and\ \bibinfo {author}
  {\bibfnamefont {M.}~\bibnamefont {Yankowitz}},\ }\bibfield  {title} {\bibinfo
  {title} {Electrically tunable correlated and topological states in twisted
  monolayer--bilayer graphene},\ }\href
  {https://doi.org/10.1038/s41567-020-01062-6} {\bibfield  {journal} {\bibinfo
  {journal} {Nat. Phys.}\ }\textbf {\bibinfo {volume} {17}},\ \bibinfo {pages}
  {374} (\bibinfo {year} {2021})}\BibitemShut {NoStop}%
\bibitem [{\citenamefont {Polshyn}\ \emph {et~al.}(2020)\citenamefont
  {Polshyn}, \citenamefont {Zhu}, \citenamefont {Kumar}, \citenamefont {Zhang},
  \citenamefont {Yang}, \citenamefont {Tschirhart}, \citenamefont {Serlin},
  \citenamefont {Watanabe}, \citenamefont {Taniguchi}, \citenamefont
  {MacDonald},\ and\ \citenamefont {Young}}]{Polshyn2020}%
  \BibitemOpen
  \bibfield  {author} {\bibinfo {author} {\bibfnamefont {H.}~\bibnamefont
  {Polshyn}}, \bibinfo {author} {\bibfnamefont {J.}~\bibnamefont {Zhu}},
  \bibinfo {author} {\bibfnamefont {M.~A.}\ \bibnamefont {Kumar}}, \bibinfo
  {author} {\bibfnamefont {Y.}~\bibnamefont {Zhang}}, \bibinfo {author}
  {\bibfnamefont {F.}~\bibnamefont {Yang}}, \bibinfo {author} {\bibfnamefont
  {C.~L.}\ \bibnamefont {Tschirhart}}, \bibinfo {author} {\bibfnamefont
  {M.}~\bibnamefont {Serlin}}, \bibinfo {author} {\bibfnamefont
  {K.}~\bibnamefont {Watanabe}}, \bibinfo {author} {\bibfnamefont
  {T.}~\bibnamefont {Taniguchi}}, \bibinfo {author} {\bibfnamefont {A.~H.}\
  \bibnamefont {MacDonald}},\ and\ \bibinfo {author} {\bibfnamefont {A.~F.}\
  \bibnamefont {Young}},\ }\bibfield  {title} {\bibinfo {title} {Electrical
  switching of magnetic order in an orbital {C}hern insulator},\ }\href
  {https://doi.org/10.1038/s41586-020-2963-8} {\bibfield  {journal} {\bibinfo
  {journal} {Nature}\ }\textbf {\bibinfo {volume} {588}},\ \bibinfo {pages}
  {66} (\bibinfo {year} {2020})}\BibitemShut {NoStop}%
\bibitem [{\citenamefont {Nuckolls}\ \emph {et~al.}(2020)\citenamefont
  {Nuckolls}, \citenamefont {Oh}, \citenamefont {Wong}, \citenamefont {Lian},
  \citenamefont {Watanabe}, \citenamefont {Taniguchi}, \citenamefont
  {Bernevig},\ and\ \citenamefont {Yazdani}}]{Nuckolls2020}%
  \BibitemOpen
  \bibfield  {author} {\bibinfo {author} {\bibfnamefont {K.~P.}\ \bibnamefont
  {Nuckolls}}, \bibinfo {author} {\bibfnamefont {M.}~\bibnamefont {Oh}},
  \bibinfo {author} {\bibfnamefont {D.}~\bibnamefont {Wong}}, \bibinfo {author}
  {\bibfnamefont {B.}~\bibnamefont {Lian}}, \bibinfo {author} {\bibfnamefont
  {K.}~\bibnamefont {Watanabe}}, \bibinfo {author} {\bibfnamefont
  {T.}~\bibnamefont {Taniguchi}}, \bibinfo {author} {\bibfnamefont {B.~A.}\
  \bibnamefont {Bernevig}},\ and\ \bibinfo {author} {\bibfnamefont
  {A.}~\bibnamefont {Yazdani}},\ }\bibfield  {title} {\bibinfo {title}
  {Strongly correlated {C}hern insulators in magic-angle twisted bilayer
  graphene},\ }\href {https://doi.org/10.1038/s41586-020-3028-8} {\bibfield
  {journal} {\bibinfo  {journal} {Nature}\ }\textbf {\bibinfo {volume} {588}},\
  \bibinfo {pages} {610} (\bibinfo {year} {2020})}\BibitemShut {NoStop}%
\bibitem [{\citenamefont {Park}\ \emph {et~al.}(2021)\citenamefont {Park},
  \citenamefont {Cao}, \citenamefont {Watanabe}, \citenamefont {Taniguchi},\
  and\ \citenamefont {Jarillo-Herrero}}]{Park2021}%
  \BibitemOpen
  \bibfield  {author} {\bibinfo {author} {\bibfnamefont {J.~M.}\ \bibnamefont
  {Park}}, \bibinfo {author} {\bibfnamefont {Y.}~\bibnamefont {Cao}}, \bibinfo
  {author} {\bibfnamefont {K.}~\bibnamefont {Watanabe}}, \bibinfo {author}
  {\bibfnamefont {T.}~\bibnamefont {Taniguchi}},\ and\ \bibinfo {author}
  {\bibfnamefont {P.}~\bibnamefont {Jarillo-Herrero}},\ }\bibfield  {title}
  {\bibinfo {title} {Tunable strongly coupled superconductivity in magic-angle
  twisted trilayer graphene},\ }\href
  {https://doi.org/10.1038/s41586-021-03192-0} {\bibfield  {journal} {\bibinfo
  {journal} {Nature}\ }\textbf {\bibinfo {volume} {590}},\ \bibinfo {pages}
  {249} (\bibinfo {year} {2021})}\BibitemShut {NoStop}%
\bibitem [{\citenamefont {Wu}\ \emph {et~al.}(2021{\natexlab{a}})\citenamefont
  {Wu}, \citenamefont {Zhang}, \citenamefont {Watanabe}, \citenamefont
  {Taniguchi},\ and\ \citenamefont {Andrei}}]{Wu2021}%
  \BibitemOpen
  \bibfield  {author} {\bibinfo {author} {\bibfnamefont {S.}~\bibnamefont
  {Wu}}, \bibinfo {author} {\bibfnamefont {Z.}~\bibnamefont {Zhang}}, \bibinfo
  {author} {\bibfnamefont {K.}~\bibnamefont {Watanabe}}, \bibinfo {author}
  {\bibfnamefont {T.}~\bibnamefont {Taniguchi}},\ and\ \bibinfo {author}
  {\bibfnamefont {E.~Y.}\ \bibnamefont {Andrei}},\ }\bibfield  {title}
  {\bibinfo {title} {Chern insulators, van {H}ove singularities and topological
  flat bands in magic-angle twisted bilayer graphene},\ }\href
  {https://doi.org/10.1038/s41563-020-00911-2} {\bibfield  {journal} {\bibinfo
  {journal} {Nat. Mater.}\ }\textbf {\bibinfo {volume} {20}},\ \bibinfo {pages}
  {488} (\bibinfo {year} {2021}{\natexlab{a}})}\BibitemShut {NoStop}%
\bibitem [{\citenamefont {Xu}\ \emph {et~al.}(2021)\citenamefont {Xu},
  \citenamefont {Al~Ezzi}, \citenamefont {Balakrishnan}, \citenamefont
  {Garcia-Ruiz}, \citenamefont {Tsim}, \citenamefont {Mullan}, \citenamefont
  {Barrier}, \citenamefont {Xin}, \citenamefont {Piot}, \citenamefont
  {Taniguchi}, \citenamefont {Watanabe}, \citenamefont {Carvalho},
  \citenamefont {Mishchenko}, \citenamefont {Geim}, \citenamefont {Fal'ko},
  \citenamefont {Adam}, \citenamefont {Neto}, \citenamefont {Novoselov},\ and\
  \citenamefont {Shi}}]{Xu2021}%
  \BibitemOpen
  \bibfield  {author} {\bibinfo {author} {\bibfnamefont {S.}~\bibnamefont
  {Xu}}, \bibinfo {author} {\bibfnamefont {M.~M.}\ \bibnamefont {Al~Ezzi}},
  \bibinfo {author} {\bibfnamefont {N.}~\bibnamefont {Balakrishnan}}, \bibinfo
  {author} {\bibfnamefont {A.}~\bibnamefont {Garcia-Ruiz}}, \bibinfo {author}
  {\bibfnamefont {B.}~\bibnamefont {Tsim}}, \bibinfo {author} {\bibfnamefont
  {C.}~\bibnamefont {Mullan}}, \bibinfo {author} {\bibfnamefont
  {J.}~\bibnamefont {Barrier}}, \bibinfo {author} {\bibfnamefont
  {N.}~\bibnamefont {Xin}}, \bibinfo {author} {\bibfnamefont {B.~A.}\
  \bibnamefont {Piot}}, \bibinfo {author} {\bibfnamefont {T.}~\bibnamefont
  {Taniguchi}}, \bibinfo {author} {\bibfnamefont {K.}~\bibnamefont {Watanabe}},
  \bibinfo {author} {\bibfnamefont {A.}~\bibnamefont {Carvalho}}, \bibinfo
  {author} {\bibfnamefont {A.}~\bibnamefont {Mishchenko}}, \bibinfo {author}
  {\bibfnamefont {A.~K.}\ \bibnamefont {Geim}}, \bibinfo {author}
  {\bibfnamefont {V.~I.}\ \bibnamefont {Fal'ko}}, \bibinfo {author}
  {\bibfnamefont {S.}~\bibnamefont {Adam}}, \bibinfo {author} {\bibfnamefont
  {A.~H.~C.}\ \bibnamefont {Neto}}, \bibinfo {author} {\bibfnamefont {K.~S.}\
  \bibnamefont {Novoselov}},\ and\ \bibinfo {author} {\bibfnamefont
  {Y.}~\bibnamefont {Shi}},\ }\bibfield  {title} {\bibinfo {title} {Tunable van
  {H}ove singularities and correlated states in twisted monolayer--bilayer
  graphene},\ }\href
  {https://doi.org/https://doi.org/10.1038/s41567-021-01172-9} {\bibfield
  {journal} {\bibinfo  {journal} {Nat. Phys.}\ }\textbf {\bibinfo {volume}
  {17}},\ \bibinfo {pages} {619} (\bibinfo {year} {2021})}\BibitemShut
  {NoStop}%
\bibitem [{\citenamefont {Cao}\ \emph {et~al.}(2021)\citenamefont {Cao},
  \citenamefont {Rodan-Legrain}, \citenamefont {Park}, \citenamefont {Yuan},
  \citenamefont {Watanabe}, \citenamefont {Taniguchi}, \citenamefont
  {Fernandes}, \citenamefont {Fu},\ and\ \citenamefont
  {Jarillo-Herrero}}]{Cao264}%
  \BibitemOpen
  \bibfield  {author} {\bibinfo {author} {\bibfnamefont {Y.}~\bibnamefont
  {Cao}}, \bibinfo {author} {\bibfnamefont {D.}~\bibnamefont {Rodan-Legrain}},
  \bibinfo {author} {\bibfnamefont {J.~M.}\ \bibnamefont {Park}}, \bibinfo
  {author} {\bibfnamefont {N.~F.~Q.}\ \bibnamefont {Yuan}}, \bibinfo {author}
  {\bibfnamefont {K.}~\bibnamefont {Watanabe}}, \bibinfo {author}
  {\bibfnamefont {T.}~\bibnamefont {Taniguchi}}, \bibinfo {author}
  {\bibfnamefont {R.~M.}\ \bibnamefont {Fernandes}}, \bibinfo {author}
  {\bibfnamefont {L.}~\bibnamefont {Fu}},\ and\ \bibinfo {author}
  {\bibfnamefont {P.}~\bibnamefont {Jarillo-Herrero}},\ }\bibfield  {title}
  {\bibinfo {title} {Nematicity and competing orders in superconducting
  magic-angle graphene},\ }\href {https://doi.org/10.1126/science.abc2836}
  {\bibfield  {journal} {\bibinfo  {journal} {Science}\ }\textbf {\bibinfo
  {volume} {372}},\ \bibinfo {pages} {264} (\bibinfo {year}
  {2021})}\BibitemShut {NoStop}%
\bibitem [{\citenamefont {Liu}\ \emph {et~al.}(2021)\citenamefont {Liu},
  \citenamefont {Wang}, \citenamefont {Watanabe}, \citenamefont {Taniguchi},
  \citenamefont {Vafek},\ and\ \citenamefont {Li}}]{Liu2021}%
  \BibitemOpen
  \bibfield  {author} {\bibinfo {author} {\bibfnamefont {X.}~\bibnamefont
  {Liu}}, \bibinfo {author} {\bibfnamefont {Z.}~\bibnamefont {Wang}}, \bibinfo
  {author} {\bibfnamefont {K.}~\bibnamefont {Watanabe}}, \bibinfo {author}
  {\bibfnamefont {T.}~\bibnamefont {Taniguchi}}, \bibinfo {author}
  {\bibfnamefont {O.}~\bibnamefont {Vafek}},\ and\ \bibinfo {author}
  {\bibfnamefont {J.~I.~A.}\ \bibnamefont {Li}},\ }\bibfield  {title} {\bibinfo
  {title} {Tuning electron correlation in magic-angle twisted bilayer graphene
  using {C}oulomb screening},\ }\href {https://doi.org/10.1126/science.abb8754}
  {\bibfield  {journal} {\bibinfo  {journal} {Science}\ }\textbf {\bibinfo
  {volume} {371}},\ \bibinfo {pages} {1261} (\bibinfo {year}
  {2021})}\BibitemShut {NoStop}%
\bibitem [{\citenamefont {Cao}\ \emph {et~al.}(2016)\citenamefont {Cao},
  \citenamefont {Luo}, \citenamefont {Fatemi}, \citenamefont {Fang},
  \citenamefont {Sanchez-Yamagishi}, \citenamefont {Watanabe}, \citenamefont
  {Taniguchi}, \citenamefont {Kaxiras},\ and\ \citenamefont
  {Jarillo-Herrero}}]{PhysRevLett.117.116804}%
  \BibitemOpen
  \bibfield  {author} {\bibinfo {author} {\bibfnamefont {Y.}~\bibnamefont
  {Cao}}, \bibinfo {author} {\bibfnamefont {J.~Y.}\ \bibnamefont {Luo}},
  \bibinfo {author} {\bibfnamefont {V.}~\bibnamefont {Fatemi}}, \bibinfo
  {author} {\bibfnamefont {S.}~\bibnamefont {Fang}}, \bibinfo {author}
  {\bibfnamefont {J.~D.}\ \bibnamefont {Sanchez-Yamagishi}}, \bibinfo {author}
  {\bibfnamefont {K.}~\bibnamefont {Watanabe}}, \bibinfo {author}
  {\bibfnamefont {T.}~\bibnamefont {Taniguchi}}, \bibinfo {author}
  {\bibfnamefont {E.}~\bibnamefont {Kaxiras}},\ and\ \bibinfo {author}
  {\bibfnamefont {P.}~\bibnamefont {Jarillo-Herrero}},\ }\bibfield  {title}
  {\bibinfo {title} {{Superlattice-Induced Insulating States and
  Valley-Protected Orbits in Twisted Bilayer Graphene}},\ }\href
  {https://doi.org/10.1103/PhysRevLett.117.116804} {\bibfield  {journal}
  {\bibinfo  {journal} {Phys. Rev. Lett.}\ }\textbf {\bibinfo {volume} {117}},\
  \bibinfo {pages} {116804} (\bibinfo {year} {2016})}\BibitemShut {NoStop}%
\bibitem [{\citenamefont {Liu}\ \emph {et~al.}(2018)\citenamefont {Liu},
  \citenamefont {Zhang}, \citenamefont {Chen},\ and\ \citenamefont
  {Yang}}]{PhysRevLett.121.217001}%
  \BibitemOpen
  \bibfield  {author} {\bibinfo {author} {\bibfnamefont {C.-C.}\ \bibnamefont
  {Liu}}, \bibinfo {author} {\bibfnamefont {L.-D.}\ \bibnamefont {Zhang}},
  \bibinfo {author} {\bibfnamefont {W.-Q.}\ \bibnamefont {Chen}},\ and\
  \bibinfo {author} {\bibfnamefont {F.}~\bibnamefont {Yang}},\ }\bibfield
  {title} {\bibinfo {title} {Chiral {S}pin {D}ensity {W}ave and $d+id$
  {S}uperconductivity in the {M}agic-{A}ngle-{T}wisted {B}ilayer {G}raphene},\
  }\href {https://doi.org/10.1103/PhysRevLett.121.217001} {\bibfield  {journal}
  {\bibinfo  {journal} {Phys. Rev. Lett.}\ }\textbf {\bibinfo {volume} {121}},\
  \bibinfo {pages} {217001} (\bibinfo {year} {2018})}\BibitemShut {NoStop}%
\bibitem [{\citenamefont {You}\ and\ \citenamefont
  {Vishwanath}(2019)}]{You2019}%
  \BibitemOpen
  \bibfield  {author} {\bibinfo {author} {\bibfnamefont {Y.-Z.}\ \bibnamefont
  {You}}\ and\ \bibinfo {author} {\bibfnamefont {A.}~\bibnamefont
  {Vishwanath}},\ }\bibfield  {title} {\bibinfo {title} {Superconductivity from
  valley fluctuations and approximate {SO}(4) symmetry in a weak coupling
  theory of twisted bilayer graphene},\ }\href
  {https://doi.org/10.1038/s41535-019-0153-4} {\bibfield  {journal} {\bibinfo
  {journal} {npj Quantum Mater.}\ }\textbf {\bibinfo {volume} {4}},\ \bibinfo
  {pages} {16} (\bibinfo {year} {2019})}\BibitemShut {NoStop}%
\bibitem [{\citenamefont {Klebl}\ and\ \citenamefont
  {Honerkamp}(2019)}]{PhysRevB.100.155145}%
  \BibitemOpen
  \bibfield  {author} {\bibinfo {author} {\bibfnamefont {L.}~\bibnamefont
  {Klebl}}\ and\ \bibinfo {author} {\bibfnamefont {C.}~\bibnamefont
  {Honerkamp}},\ }\bibfield  {title} {\bibinfo {title} {Inherited and
  flatband-induced ordering in twisted graphene bilayers},\ }\href
  {https://doi.org/10.1103/PhysRevB.100.155145} {\bibfield  {journal} {\bibinfo
   {journal} {Phys. Rev. B}\ }\textbf {\bibinfo {volume} {100}},\ \bibinfo
  {pages} {155145} (\bibinfo {year} {2019})}\BibitemShut {NoStop}%
\bibitem [{\citenamefont {Fischer}\ \emph {et~al.}(2021)\citenamefont
  {Fischer}, \citenamefont {Klebl}, \citenamefont {Honerkamp},\ and\
  \citenamefont {Kennes}}]{PhysRevB.103.L041103}%
  \BibitemOpen
  \bibfield  {author} {\bibinfo {author} {\bibfnamefont {A.}~\bibnamefont
  {Fischer}}, \bibinfo {author} {\bibfnamefont {L.}~\bibnamefont {Klebl}},
  \bibinfo {author} {\bibfnamefont {C.}~\bibnamefont {Honerkamp}},\ and\
  \bibinfo {author} {\bibfnamefont {D.~M.}\ \bibnamefont {Kennes}},\ }\bibfield
   {title} {\bibinfo {title} {Spin-fluctuation-induced pairing in twisted
  bilayer graphene},\ }\href {https://doi.org/10.1103/PhysRevB.103.L041103}
  {\bibfield  {journal} {\bibinfo  {journal} {Phys. Rev. B}\ }\textbf {\bibinfo
  {volume} {103}},\ \bibinfo {pages} {L041103} (\bibinfo {year}
  {2021})}\BibitemShut {NoStop}%
\bibitem [{\citenamefont {Fischer}\ \emph {et~al.}(2022)\citenamefont
  {Fischer}, \citenamefont {Goodwin}, \citenamefont {Mostofi}, \citenamefont
  {Lischner}, \citenamefont {Kennes},\ and\ \citenamefont
  {Klebl}}]{Fischer2022}%
  \BibitemOpen
  \bibfield  {author} {\bibinfo {author} {\bibfnamefont {A.}~\bibnamefont
  {Fischer}}, \bibinfo {author} {\bibfnamefont {Z.~A.~H.}\ \bibnamefont
  {Goodwin}}, \bibinfo {author} {\bibfnamefont {A.~A.}\ \bibnamefont
  {Mostofi}}, \bibinfo {author} {\bibfnamefont {J.}~\bibnamefont {Lischner}},
  \bibinfo {author} {\bibfnamefont {D.~M.}\ \bibnamefont {Kennes}},\ and\
  \bibinfo {author} {\bibfnamefont {L.}~\bibnamefont {Klebl}},\ }\bibfield
  {title} {\bibinfo {title} {Unconventional superconductivity in magic-angle
  twisted trilayer graphene},\ }\href
  {https://doi.org/10.1038/s41535-021-00410-w} {\bibfield  {journal} {\bibinfo
  {journal} {npj Quantum Mater.}\ }\textbf {\bibinfo {volume} {7}},\ \bibinfo
  {pages} {5} (\bibinfo {year} {2022})}\BibitemShut {NoStop}%
\bibitem [{\citenamefont {Oh}\ \emph {et~al.}(2021)\citenamefont {Oh},
  \citenamefont {Nuckolls}, \citenamefont {Wong}, \citenamefont {Lee},
  \citenamefont {Liu}, \citenamefont {Watanabe}, \citenamefont {Taniguchi},\
  and\ \citenamefont {Yazdani}}]{Oh2021}%
  \BibitemOpen
  \bibfield  {author} {\bibinfo {author} {\bibfnamefont {M.}~\bibnamefont
  {Oh}}, \bibinfo {author} {\bibfnamefont {K.~P.}\ \bibnamefont {Nuckolls}},
  \bibinfo {author} {\bibfnamefont {D.}~\bibnamefont {Wong}}, \bibinfo {author}
  {\bibfnamefont {R.~L.}\ \bibnamefont {Lee}}, \bibinfo {author} {\bibfnamefont
  {X.}~\bibnamefont {Liu}}, \bibinfo {author} {\bibfnamefont {K.}~\bibnamefont
  {Watanabe}}, \bibinfo {author} {\bibfnamefont {T.}~\bibnamefont
  {Taniguchi}},\ and\ \bibinfo {author} {\bibfnamefont {A.}~\bibnamefont
  {Yazdani}},\ }\bibfield  {title} {\bibinfo {title} {Evidence for
  unconventional superconductivity in twisted bilayer graphene},\ }\href
  {https://doi.org/10.1038/s41586-021-04121-x} {\bibfield  {journal} {\bibinfo
  {journal} {Nature}\ }\textbf {\bibinfo {volume} {600}},\ \bibinfo {pages}
  {240} (\bibinfo {year} {2021})}\BibitemShut {NoStop}%
\bibitem [{\citenamefont {Talantsev}\ \emph {et~al.}(2020)\citenamefont
  {Talantsev}, \citenamefont {Mataira},\ and\ \citenamefont
  {Crump}}]{Talantsev2020}%
  \BibitemOpen
  \bibfield  {author} {\bibinfo {author} {\bibfnamefont {E.~F.}\ \bibnamefont
  {Talantsev}}, \bibinfo {author} {\bibfnamefont {R.~C.}\ \bibnamefont
  {Mataira}},\ and\ \bibinfo {author} {\bibfnamefont {W.~P.}\ \bibnamefont
  {Crump}},\ }\bibfield  {title} {\bibinfo {title} {{C}lassifying
  superconductivity in {M}oir{\'e} graphene superlattices},\ }\href
  {https://doi.org/10.1038/s41598-019-57055-w} {\bibfield  {journal} {\bibinfo
  {journal} {Sci. Rep.}\ }\textbf {\bibinfo {volume} {10}},\ \bibinfo {pages}
  {212} (\bibinfo {year} {2020})}\BibitemShut {NoStop}%
\bibitem [{\citenamefont {Kennes}\ \emph {et~al.}(2021)\citenamefont {Kennes},
  \citenamefont {Claassen}, \citenamefont {Xian}, \citenamefont {Georges},
  \citenamefont {Millis}, \citenamefont {Hone}, \citenamefont {Dean},
  \citenamefont {Basov}, \citenamefont {Pasupathy},\ and\ \citenamefont
  {Rubio}}]{Kennes2021}%
  \BibitemOpen
  \bibfield  {author} {\bibinfo {author} {\bibfnamefont {D.~M.}\ \bibnamefont
  {Kennes}}, \bibinfo {author} {\bibfnamefont {M.}~\bibnamefont {Claassen}},
  \bibinfo {author} {\bibfnamefont {L.}~\bibnamefont {Xian}}, \bibinfo {author}
  {\bibfnamefont {A.}~\bibnamefont {Georges}}, \bibinfo {author} {\bibfnamefont
  {A.~J.}\ \bibnamefont {Millis}}, \bibinfo {author} {\bibfnamefont
  {J.}~\bibnamefont {Hone}}, \bibinfo {author} {\bibfnamefont {C.~R.}\
  \bibnamefont {Dean}}, \bibinfo {author} {\bibfnamefont {D.~N.}\ \bibnamefont
  {Basov}}, \bibinfo {author} {\bibfnamefont {A.~N.}\ \bibnamefont
  {Pasupathy}},\ and\ \bibinfo {author} {\bibfnamefont {A.}~\bibnamefont
  {Rubio}},\ }\bibfield  {title} {\bibinfo {title} {Moir{\'e} heterostructures
  as a condensed-matter quantum simulator},\ }\href
  {https://doi.org/10.1038/s41567-020-01154-3} {\bibfield  {journal} {\bibinfo
  {journal} {Nat. Phys.}\ }\textbf {\bibinfo {volume} {17}},\ \bibinfo {pages}
  {155} (\bibinfo {year} {2021})}\BibitemShut {NoStop}%
\bibitem [{\citenamefont {Kuznetsova}\ and\ \citenamefont
  {Barzykin}(2005)}]{Kuznetsova2005}%
  \BibitemOpen
  \bibfield  {author} {\bibinfo {author} {\bibfnamefont {Z.}~\bibnamefont
  {Kuznetsova}}\ and\ \bibinfo {author} {\bibfnamefont {V.}~\bibnamefont
  {Barzykin}},\ }\bibfield  {title} {\bibinfo {title} {Pairing state in
  multicomponent superconductors},\ }\href
  {https://doi.org/10.1209/epl/i2005-10242-8} {\bibfield  {journal} {\bibinfo
  {journal} {Europhys. Lett.}\ }\textbf {\bibinfo {volume} {72}},\ \bibinfo
  {pages} {437} (\bibinfo {year} {2005})}\BibitemShut {NoStop}%
\bibitem [{\citenamefont {Black-Schaffer}\ and\ \citenamefont
  {Honerkamp}(2014)}]{BlackSchaffer2014}%
  \BibitemOpen
  \bibfield  {author} {\bibinfo {author} {\bibfnamefont {A.~M.}\ \bibnamefont
  {Black-Schaffer}}\ and\ \bibinfo {author} {\bibfnamefont {C.}~\bibnamefont
  {Honerkamp}},\ }\bibfield  {title} {\bibinfo {title} {Chiral {$d$}-wave
  superconductivity in doped graphene},\ }\href
  {https://doi.org/10.1088/0953-8984/26/42/423201} {\bibfield  {journal}
  {\bibinfo  {journal} {J. Phys.: Condens. Matter}\ }\textbf {\bibinfo {volume}
  {26}},\ \bibinfo {pages} {423201} (\bibinfo {year} {2014})},\ \Eprint
  {https://arxiv.org/abs/1406.0101} {arxiv:1406.0101} \BibitemShut {NoStop}%
\bibitem [{\citenamefont {Po}\ \emph {et~al.}(2018)\citenamefont {Po},
  \citenamefont {Zou}, \citenamefont {Vishwanath},\ and\ \citenamefont
  {Senthil}}]{PhysRevX.8.031089}%
  \BibitemOpen
  \bibfield  {author} {\bibinfo {author} {\bibfnamefont {H.~C.}\ \bibnamefont
  {Po}}, \bibinfo {author} {\bibfnamefont {L.}~\bibnamefont {Zou}}, \bibinfo
  {author} {\bibfnamefont {A.}~\bibnamefont {Vishwanath}},\ and\ \bibinfo
  {author} {\bibfnamefont {T.}~\bibnamefont {Senthil}},\ }\bibfield  {title}
  {\bibinfo {title} {{Origin of {M}ott Insulating Behavior and
  Superconductivity in Twisted Bilayer Graphene}},\ }\href
  {https://doi.org/10.1103/PhysRevX.8.031089} {\bibfield  {journal} {\bibinfo
  {journal} {Phys. Rev. X}\ }\textbf {\bibinfo {volume} {8}},\ \bibinfo {pages}
  {031089} (\bibinfo {year} {2018})}\BibitemShut {NoStop}%
\bibitem [{\citenamefont {Lee}\ \emph {et~al.}(2019)\citenamefont {Lee},
  \citenamefont {Khalaf}, \citenamefont {Liu}, \citenamefont {Liu},
  \citenamefont {Hao}, \citenamefont {Kim},\ and\ \citenamefont
  {Vishwanath}}]{Lee2019}%
  \BibitemOpen
  \bibfield  {author} {\bibinfo {author} {\bibfnamefont {J.~Y.}\ \bibnamefont
  {Lee}}, \bibinfo {author} {\bibfnamefont {E.}~\bibnamefont {Khalaf}},
  \bibinfo {author} {\bibfnamefont {S.}~\bibnamefont {Liu}}, \bibinfo {author}
  {\bibfnamefont {X.}~\bibnamefont {Liu}}, \bibinfo {author} {\bibfnamefont
  {Z.}~\bibnamefont {Hao}}, \bibinfo {author} {\bibfnamefont {P.}~\bibnamefont
  {Kim}},\ and\ \bibinfo {author} {\bibfnamefont {A.}~\bibnamefont
  {Vishwanath}},\ }\bibfield  {title} {\bibinfo {title} {Theory of correlated
  insulating behaviour and spin-triplet superconductivity in twisted double
  bilayer graphene},\ }\href {https://doi.org/10.1038/s41467-019-12981-1}
  {\bibfield  {journal} {\bibinfo  {journal} {Nat Commun.}\ }\textbf {\bibinfo
  {volume} {10}},\ \bibinfo {pages} {5333} (\bibinfo {year}
  {2019})}\BibitemShut {NoStop}%
\bibitem [{\citenamefont {Haddadi}\ \emph {et~al.}(2020)\citenamefont
  {Haddadi}, \citenamefont {Wu}, \citenamefont {Kruchkov},\ and\ \citenamefont
  {Yazyev}}]{Haddadi2020}%
  \BibitemOpen
  \bibfield  {author} {\bibinfo {author} {\bibfnamefont {F.}~\bibnamefont
  {Haddadi}}, \bibinfo {author} {\bibfnamefont {Q.}~\bibnamefont {Wu}},
  \bibinfo {author} {\bibfnamefont {A.~J.}\ \bibnamefont {Kruchkov}},\ and\
  \bibinfo {author} {\bibfnamefont {O.~V.}\ \bibnamefont {Yazyev}},\ }\bibfield
   {title} {\bibinfo {title} {Moir{\'e} {F}lat {B}ands in {T}wisted {D}ouble
  {B}ilayer {G}raphene},\ }\href {https://doi.org/10.1021/acs.nanolett.9b05117}
  {\bibfield  {journal} {\bibinfo  {journal} {Nano Lett.}\ }\textbf {\bibinfo
  {volume} {20}},\ \bibinfo {pages} {2410} (\bibinfo {year}
  {2020})}\BibitemShut {NoStop}%
\bibitem [{\citenamefont {Liang}\ \emph {et~al.}(2020)\citenamefont {Liang},
  \citenamefont {Goodwin}, \citenamefont {Vitale}, \citenamefont {Corsetti},
  \citenamefont {Mostofi},\ and\ \citenamefont
  {Lischner}}]{PhysRevB.102.155146}%
  \BibitemOpen
  \bibfield  {author} {\bibinfo {author} {\bibfnamefont {X.}~\bibnamefont
  {Liang}}, \bibinfo {author} {\bibfnamefont {Z.~A.~H.}\ \bibnamefont
  {Goodwin}}, \bibinfo {author} {\bibfnamefont {V.}~\bibnamefont {Vitale}},
  \bibinfo {author} {\bibfnamefont {F.}~\bibnamefont {Corsetti}}, \bibinfo
  {author} {\bibfnamefont {A.~A.}\ \bibnamefont {Mostofi}},\ and\ \bibinfo
  {author} {\bibfnamefont {J.}~\bibnamefont {Lischner}},\ }\bibfield  {title}
  {\bibinfo {title} {Effect of bilayer stacking on the atomic and electronic
  structure of twisted double bilayer graphene},\ }\href
  {https://doi.org/10.1103/PhysRevB.102.155146} {\bibfield  {journal} {\bibinfo
   {journal} {Phys. Rev. B}\ }\textbf {\bibinfo {volume} {102}},\ \bibinfo
  {pages} {155146} (\bibinfo {year} {2020})}\BibitemShut {NoStop}%
\bibitem [{\citenamefont {Carr}\ \emph {et~al.}(2020)\citenamefont {Carr},
  \citenamefont {Li}, \citenamefont {Zhu}, \citenamefont {Kaxiras},
  \citenamefont {Sachdev},\ and\ \citenamefont {Kruchkov}}]{Carr2020}%
  \BibitemOpen
  \bibfield  {author} {\bibinfo {author} {\bibfnamefont {S.}~\bibnamefont
  {Carr}}, \bibinfo {author} {\bibfnamefont {C.}~\bibnamefont {Li}}, \bibinfo
  {author} {\bibfnamefont {Z.}~\bibnamefont {Zhu}}, \bibinfo {author}
  {\bibfnamefont {E.}~\bibnamefont {Kaxiras}}, \bibinfo {author} {\bibfnamefont
  {S.}~\bibnamefont {Sachdev}},\ and\ \bibinfo {author} {\bibfnamefont
  {A.}~\bibnamefont {Kruchkov}},\ }\bibfield  {title} {\bibinfo {title}
  {Ultraheavy and {U}ltrarelativistic {D}irac {Q}uasiparticles in {S}andwiched
  {G}raphenes},\ }\href {https://doi.org/10.1021/acs.nanolett.9b04979}
  {\bibfield  {journal} {\bibinfo  {journal} {Nano Lett.}\ }\textbf {\bibinfo
  {volume} {20}},\ \bibinfo {pages} {3030} (\bibinfo {year}
  {2020})}\BibitemShut {NoStop}%
\bibitem [{\citenamefont {Wu}\ \emph {et~al.}(2021{\natexlab{b}})\citenamefont
  {Wu}, \citenamefont {Zhan},\ and\ \citenamefont {Yuan}}]{Wu2021a}%
  \BibitemOpen
  \bibfield  {author} {\bibinfo {author} {\bibfnamefont {Z.}~\bibnamefont
  {Wu}}, \bibinfo {author} {\bibfnamefont {Z.}~\bibnamefont {Zhan}},\ and\
  \bibinfo {author} {\bibfnamefont {S.}~\bibnamefont {Yuan}},\ }\bibfield
  {title} {\bibinfo {title} {Lattice relaxation, mirror symmetry and magnetic
  field effects on ultraflat bands in twisted trilayer graphene},\ }\href
  {https://doi.org/10.1007/s11433-020-1690-4} {\bibfield  {journal} {\bibinfo
  {journal} {Sci. China Phys. Mech. Astron.}\ }\textbf {\bibinfo {volume}
  {64}},\ \bibinfo {pages} {267811} (\bibinfo {year} {2021}{\natexlab{b}})},\
  \Eprint {https://arxiv.org/abs/2012.13741} {2012.13741} \BibitemShut
  {NoStop}%
\bibitem [{\citenamefont {Lopez-Bezanilla}\ and\ \citenamefont
  {Lado}(2020)}]{PhysRevResearch.2.033357}%
  \BibitemOpen
  \bibfield  {author} {\bibinfo {author} {\bibfnamefont {A.}~\bibnamefont
  {Lopez-Bezanilla}}\ and\ \bibinfo {author} {\bibfnamefont {J.~L.}\
  \bibnamefont {Lado}},\ }\bibfield  {title} {\bibinfo {title} {Electrical band
  flattening, valley flux, and superconductivity in twisted trilayer
  graphene},\ }\href {https://doi.org/10.1103/PhysRevResearch.2.033357}
  {\bibfield  {journal} {\bibinfo  {journal} {Phys. Rev. Research}\ }\textbf
  {\bibinfo {volume} {2}},\ \bibinfo {pages} {033357} (\bibinfo {year}
  {2020})}\BibitemShut {NoStop}%
\bibitem [{\citenamefont {Xian}\ \emph {et~al.}(2019)\citenamefont {Xian},
  \citenamefont {Kennes}, \citenamefont {Tancogne-Dejean}, \citenamefont
  {Altarelli},\ and\ \citenamefont {Rubio}}]{Xian2019}%
  \BibitemOpen
  \bibfield  {author} {\bibinfo {author} {\bibfnamefont {L.}~\bibnamefont
  {Xian}}, \bibinfo {author} {\bibfnamefont {D.~M.}\ \bibnamefont {Kennes}},
  \bibinfo {author} {\bibfnamefont {N.}~\bibnamefont {Tancogne-Dejean}},
  \bibinfo {author} {\bibfnamefont {M.}~\bibnamefont {Altarelli}},\ and\
  \bibinfo {author} {\bibfnamefont {A.}~\bibnamefont {Rubio}},\ }\bibfield
  {title} {\bibinfo {title} {Multiflat {B}ands and {S}trong {C}orrelations in
  {T}wisted {B}ilayer {B}oron {N}itride: {D}oping-{I}nduced {C}orrelated
  {I}nsulator and {S}uperconductor},\ }\href
  {https://doi.org/10.1021/acs.nanolett.9b00986} {\bibfield  {journal}
  {\bibinfo  {journal} {Nano Lett.}\ }\textbf {\bibinfo {volume} {19}},\
  \bibinfo {pages} {4934} (\bibinfo {year} {2019})}\BibitemShut {NoStop}%
\bibitem [{\citenamefont {Xian}\ \emph {et~al.}(2021)\citenamefont {Xian},
  \citenamefont {Claassen}, \citenamefont {Kiese}, \citenamefont {Scherer},
  \citenamefont {Trebst}, \citenamefont {Kennes},\ and\ \citenamefont
  {Rubio}}]{Xian2021}%
  \BibitemOpen
  \bibfield  {author} {\bibinfo {author} {\bibfnamefont {L.}~\bibnamefont
  {Xian}}, \bibinfo {author} {\bibfnamefont {M.}~\bibnamefont {Claassen}},
  \bibinfo {author} {\bibfnamefont {D.}~\bibnamefont {Kiese}}, \bibinfo
  {author} {\bibfnamefont {M.~M.}\ \bibnamefont {Scherer}}, \bibinfo {author}
  {\bibfnamefont {S.}~\bibnamefont {Trebst}}, \bibinfo {author} {\bibfnamefont
  {D.~M.}\ \bibnamefont {Kennes}},\ and\ \bibinfo {author} {\bibfnamefont
  {A.}~\bibnamefont {Rubio}},\ }\bibfield  {title} {\bibinfo {title}
  {Realization of nearly dispersionless bands with strong orbital anisotropy
  from destructive interference in twisted bilayer {MoS}{$_2$}},\ }\href
  {https://doi.org/10.1038/s41467-021-25922-8} {\bibfield  {journal} {\bibinfo
  {journal} {Nat. Comm.}\ }\textbf {\bibinfo {volume} {12}},\ \bibinfo {pages}
  {5644} (\bibinfo {year} {2021})},\ \Eprint {https://arxiv.org/abs/2004.02964}
  {2004.02964} \BibitemShut {NoStop}%
\bibitem [{\citenamefont {{Angeli}}\ and\ \citenamefont
  {{MacDonald}}(2021)}]{gammaTMDCsMacdonald2021}%
  \BibitemOpen
  \bibfield  {author} {\bibinfo {author} {\bibfnamefont {M.}~\bibnamefont
  {{Angeli}}}\ and\ \bibinfo {author} {\bibfnamefont {A.~H.}\ \bibnamefont
  {{MacDonald}}},\ }\bibfield  {title} {\bibinfo {title}
  {{{\ensuremath{\Gamma}} valley transition metal dichalcogenide moir{\'e}
  bands}},\ }\href {https://doi.org/10.1073/pnas.2021826118} {\bibfield
  {journal} {\bibinfo  {journal} {Proc. Natl. Acad. Sci.}\ }\textbf {\bibinfo
  {volume} {118}},\ \bibinfo {pages} {e2021826118} (\bibinfo {year} {2021})},\
  \Eprint {https://arxiv.org/abs/2008.01735} {arXiv:2008.01735} \BibitemShut
  {NoStop}%
\bibitem [{\citenamefont {Po}\ \emph {et~al.}(2019)\citenamefont {Po},
  \citenamefont {Zou}, \citenamefont {Senthil},\ and\ \citenamefont
  {Vishwanath}}]{PhysRevB.99.195455}%
  \BibitemOpen
  \bibfield  {author} {\bibinfo {author} {\bibfnamefont {H.~C.}\ \bibnamefont
  {Po}}, \bibinfo {author} {\bibfnamefont {L.}~\bibnamefont {Zou}}, \bibinfo
  {author} {\bibfnamefont {T.}~\bibnamefont {Senthil}},\ and\ \bibinfo {author}
  {\bibfnamefont {A.}~\bibnamefont {Vishwanath}},\ }\bibfield  {title}
  {\bibinfo {title} {Faithful tight-binding models and fragile topology of
  magic-angle bilayer graphene},\ }\href
  {https://doi.org/10.1103/PhysRevB.99.195455} {\bibfield  {journal} {\bibinfo
  {journal} {Phys. Rev. B}\ }\textbf {\bibinfo {volume} {99}},\ \bibinfo
  {pages} {195455} (\bibinfo {year} {2019})}\BibitemShut {NoStop}%
\bibitem [{\citenamefont {Carr}\ \emph {et~al.}(2019)\citenamefont {Carr},
  \citenamefont {Fang}, \citenamefont {Zhu},\ and\ \citenamefont
  {Kaxiras}}]{PhysRevResearch.1.013001}%
  \BibitemOpen
  \bibfield  {author} {\bibinfo {author} {\bibfnamefont {S.}~\bibnamefont
  {Carr}}, \bibinfo {author} {\bibfnamefont {S.}~\bibnamefont {Fang}}, \bibinfo
  {author} {\bibfnamefont {Z.}~\bibnamefont {Zhu}},\ and\ \bibinfo {author}
  {\bibfnamefont {E.}~\bibnamefont {Kaxiras}},\ }\bibfield  {title} {\bibinfo
  {title} {Exact continuum model for low-energy electronic states of twisted
  bilayer graphene},\ }\href {https://doi.org/10.1103/PhysRevResearch.1.013001}
  {\bibfield  {journal} {\bibinfo  {journal} {Phys. Rev. Research}\ }\textbf
  {\bibinfo {volume} {1}},\ \bibinfo {pages} {013001} (\bibinfo {year}
  {2019})}\BibitemShut {NoStop}%
\bibitem [{\citenamefont {Phong}\ and\ \citenamefont
  {Mele}(2020)}]{PhysRevLett.125.176404}%
  \BibitemOpen
  \bibfield  {author} {\bibinfo {author} {\bibfnamefont {V.~o.~T.}\
  \bibnamefont {Phong}}\ and\ \bibinfo {author} {\bibfnamefont {E.~J.}\
  \bibnamefont {Mele}},\ }\bibfield  {title} {\bibinfo {title} {Obstruction and
  {I}nterference in {L}ow-{E}nergy {M}odels for {T}wisted {B}ilayer
  {G}raphene},\ }\href {https://doi.org/10.1103/PhysRevLett.125.176404}
  {\bibfield  {journal} {\bibinfo  {journal} {Phys. Rev. Lett.}\ }\textbf
  {\bibinfo {volume} {125}},\ \bibinfo {pages} {176404} (\bibinfo {year}
  {2020})}\BibitemShut {NoStop}%
\bibitem [{\citenamefont {Zou}\ \emph {et~al.}(2018)\citenamefont {Zou},
  \citenamefont {Po}, \citenamefont {Vishwanath},\ and\ \citenamefont
  {Senthil}}]{Zou2018}%
  \BibitemOpen
  \bibfield  {author} {\bibinfo {author} {\bibfnamefont {L.}~\bibnamefont
  {Zou}}, \bibinfo {author} {\bibfnamefont {H.~C.}\ \bibnamefont {Po}},
  \bibinfo {author} {\bibfnamefont {A.}~\bibnamefont {Vishwanath}},\ and\
  \bibinfo {author} {\bibfnamefont {T.}~\bibnamefont {Senthil}},\ }\bibfield
  {title} {\bibinfo {title} {Band structure of twisted bilayer graphene:
  Emergent symmetries, commensurate approximants, and {W}annier obstructions},\
  }\href {https://doi.org/10.1103/physrevb.98.085435} {\bibfield  {journal}
  {\bibinfo  {journal} {Phys. Rev. B}\ }\textbf {\bibinfo {volume} {98}},\
  \bibinfo {pages} {085435} (\bibinfo {year} {2018})}\BibitemShut {NoStop}%
\bibitem [{\citenamefont {Bickers}\ \emph {et~al.}(1989)\citenamefont
  {Bickers}, \citenamefont {Scalapino},\ and\ \citenamefont
  {White}}]{Bickers1989a}%
  \BibitemOpen
  \bibfield  {author} {\bibinfo {author} {\bibfnamefont {N.~E.}\ \bibnamefont
  {Bickers}}, \bibinfo {author} {\bibfnamefont {D.~J.}\ \bibnamefont
  {Scalapino}},\ and\ \bibinfo {author} {\bibfnamefont {S.~R.}\ \bibnamefont
  {White}},\ }\bibfield  {title} {\bibinfo {title} {Conserving {A}pproximations
  for {S}trongly {C}orrelated {E}lectron {S}ystems: {B}ethe-{S}alpeter
  {E}quation and {D}ynamics for the {T}wo-{D}imensional {H}ubbard {M}odel},\
  }\href {https://doi.org/10.1103/physrevlett.62.961} {\bibfield  {journal}
  {\bibinfo  {journal} {Phys. Rev. Lett.}\ }\textbf {\bibinfo {volume} {62}},\
  \bibinfo {pages} {961} (\bibinfo {year} {1989})}\BibitemShut {NoStop}%
\bibitem [{\citenamefont {Bickers}\ and\ \citenamefont
  {Scalapino}(1989)}]{Bickers1989b}%
  \BibitemOpen
  \bibfield  {author} {\bibinfo {author} {\bibfnamefont {N.~E.}\ \bibnamefont
  {Bickers}}\ and\ \bibinfo {author} {\bibfnamefont {D.~J.}\ \bibnamefont
  {Scalapino}},\ }\bibfield  {title} {\bibinfo {title} {Conserving
  approximations for strongly fluctuating electron systems. {I}. {F}ormalism
  and calculational approach},\ }\href
  {https://doi.org/10.1016/0003-4916(89)90359-x} {\bibfield  {journal}
  {\bibinfo  {journal} {Ann. Phys. (NY)}\ }\textbf {\bibinfo {volume} {193}},\
  \bibinfo {pages} {206} (\bibinfo {year} {1989})}\BibitemShut {NoStop}%
\bibitem [{\citenamefont {Zhang}\ \emph {et~al.}(2020)\citenamefont {Zhang},
  \citenamefont {Yuan},\ and\ \citenamefont {Fu}}]{PhysRevB.102.201115}%
  \BibitemOpen
  \bibfield  {author} {\bibinfo {author} {\bibfnamefont {Y.}~\bibnamefont
  {Zhang}}, \bibinfo {author} {\bibfnamefont {N.~F.~Q.}\ \bibnamefont {Yuan}},\
  and\ \bibinfo {author} {\bibfnamefont {L.}~\bibnamefont {Fu}},\ }\bibfield
  {title} {\bibinfo {title} {Moir\'e quantum chemistry: Charge transfer in
  transition metal dichalcogenide superlattices},\ }\href
  {https://doi.org/10.1103/PhysRevB.102.201115} {\bibfield  {journal} {\bibinfo
   {journal} {Phys. Rev. B}\ }\textbf {\bibinfo {volume} {102}},\ \bibinfo
  {pages} {201115} (\bibinfo {year} {2020})}\BibitemShut {NoStop}%
\bibitem [{\citenamefont {Zhang}\ \emph {et~al.}(2021)\citenamefont {Zhang},
  \citenamefont {Liu},\ and\ \citenamefont {Fu}}]{PhysRevB.103.155142}%
  \BibitemOpen
  \bibfield  {author} {\bibinfo {author} {\bibfnamefont {Y.}~\bibnamefont
  {Zhang}}, \bibinfo {author} {\bibfnamefont {T.}~\bibnamefont {Liu}},\ and\
  \bibinfo {author} {\bibfnamefont {L.}~\bibnamefont {Fu}},\ }\bibfield
  {title} {\bibinfo {title} {Electronic structures, charge transfer, and charge
  order in twisted transition metal dichalcogenide bilayers},\ }\href
  {https://doi.org/10.1103/PhysRevB.103.155142} {\bibfield  {journal} {\bibinfo
   {journal} {Phys. Rev. B}\ }\textbf {\bibinfo {volume} {103}},\ \bibinfo
  {pages} {155142} (\bibinfo {year} {2021})}\BibitemShut {NoStop}%
\bibitem [{\citenamefont {Rasmussen}\ and\ \citenamefont
  {Thygesen}(2015)}]{Rasmussen2015}%
  \BibitemOpen
  \bibfield  {author} {\bibinfo {author} {\bibfnamefont {F.~A.}\ \bibnamefont
  {Rasmussen}}\ and\ \bibinfo {author} {\bibfnamefont {K.~S.}\ \bibnamefont
  {Thygesen}},\ }\bibfield  {title} {\bibinfo {title} {{Computational 2D
  Materials Database: Electronic Structure of Transition-Metal Dichalcogenides
  and Oxides}},\ }\href {https://doi.org/10.1021/acs.jpcc.5b02950} {\bibfield
  {journal} {\bibinfo  {journal} {J. Phys. Chem. C}\ }\textbf {\bibinfo
  {volume} {119}},\ \bibinfo {pages} {13169} (\bibinfo {year}
  {2015})}\BibitemShut {NoStop}%
\bibitem [{\citenamefont {Chaves}\ \emph {et~al.}(2020)\citenamefont {Chaves},
  \citenamefont {Azadani}, \citenamefont {Alsalman}, \citenamefont {da~Costa},
  \citenamefont {Frisenda}, \citenamefont {Chaves}, \citenamefont {Song},
  \citenamefont {Kim}, \citenamefont {He}, \citenamefont {Zhou}, \citenamefont
  {Castellanos-Gomez}, \citenamefont {Peeters}, \citenamefont {Liu},
  \citenamefont {Hinkle}, \citenamefont {Oh}, \citenamefont {Ye}, \citenamefont
  {Koester}, \citenamefont {Lee}, \citenamefont {Avouris}, \citenamefont
  {Wang},\ and\ \citenamefont {Low}}]{Chaves2020}%
  \BibitemOpen
  \bibfield  {author} {\bibinfo {author} {\bibfnamefont {A.}~\bibnamefont
  {Chaves}}, \bibinfo {author} {\bibfnamefont {J.~G.}\ \bibnamefont {Azadani}},
  \bibinfo {author} {\bibfnamefont {H.}~\bibnamefont {Alsalman}}, \bibinfo
  {author} {\bibfnamefont {D.~R.}\ \bibnamefont {da~Costa}}, \bibinfo {author}
  {\bibfnamefont {R.}~\bibnamefont {Frisenda}}, \bibinfo {author}
  {\bibfnamefont {A.~J.}\ \bibnamefont {Chaves}}, \bibinfo {author}
  {\bibfnamefont {S.~H.}\ \bibnamefont {Song}}, \bibinfo {author}
  {\bibfnamefont {Y.~D.}\ \bibnamefont {Kim}}, \bibinfo {author} {\bibfnamefont
  {D.}~\bibnamefont {He}}, \bibinfo {author} {\bibfnamefont {J.}~\bibnamefont
  {Zhou}}, \bibinfo {author} {\bibfnamefont {A.}~\bibnamefont
  {Castellanos-Gomez}}, \bibinfo {author} {\bibfnamefont {F.~M.}\ \bibnamefont
  {Peeters}}, \bibinfo {author} {\bibfnamefont {Z.}~\bibnamefont {Liu}},
  \bibinfo {author} {\bibfnamefont {C.~L.}\ \bibnamefont {Hinkle}}, \bibinfo
  {author} {\bibfnamefont {S.-H.}\ \bibnamefont {Oh}}, \bibinfo {author}
  {\bibfnamefont {P.~D.}\ \bibnamefont {Ye}}, \bibinfo {author} {\bibfnamefont
  {S.~J.}\ \bibnamefont {Koester}}, \bibinfo {author} {\bibfnamefont {Y.~H.}\
  \bibnamefont {Lee}}, \bibinfo {author} {\bibfnamefont {P.}~\bibnamefont
  {Avouris}}, \bibinfo {author} {\bibfnamefont {X.}~\bibnamefont {Wang}},\ and\
  \bibinfo {author} {\bibfnamefont {T.}~\bibnamefont {Low}},\ }\bibfield
  {title} {\bibinfo {title} {Bandgap engineering of two-dimensional
  semiconductor materials},\ }\href
  {https://doi.org/10.1038/s41699-020-00162-4} {\bibfield  {journal} {\bibinfo
  {journal} {npj 2D Mater. Appl.}\ }\textbf {\bibinfo {volume} {4}},\ \bibinfo
  {pages} {29} (\bibinfo {year} {2020})}\BibitemShut {NoStop}%
\bibitem [{SM()}]{SM}%
  \BibitemOpen
  \href@noop {} {}\bibinfo {note} {See Supplemental Material for details on the
  low-energy continuum model of the electronic bands in $\Gamma$-valley twisted
  TMDCs, for the construction of the Wannier orbitals in the 2D honeycomb
  moir\'e pattern, for a study of the electrostatic and long-range Hartree
  potential in the flat Dirac bands, for an estimation of the on-site and
  nearest-neighbor Coulomb interaction, for numerical details on the FLEX
  calculations, for a discussion on the nature of magnetic ordering, for an
  in-depth discussion of the spin susceptibility structure, for the temperature
  dependence of the $d$- and $f$-wave superconducting eigenvalues and order
  parameters, for a discussion of the influence of band asymmetry as induced by
  long-range hopping, for a study of superconductivity arising from non-local
  dispersive phonons, and for an estimation of the phonon-induced attractive
  interaction $U_{\mathrm{eff}}$.}\BibitemShut {Stop}%
\bibitem [{\citenamefont {Guinea}\ and\ \citenamefont
  {Walet}(2018)}]{Guinea2018}%
  \BibitemOpen
  \bibfield  {author} {\bibinfo {author} {\bibfnamefont {F.}~\bibnamefont
  {Guinea}}\ and\ \bibinfo {author} {\bibfnamefont {N.~R.}\ \bibnamefont
  {Walet}},\ }\bibfield  {title} {\bibinfo {title} {Electrostatic effects, band
  distortions, and superconductivity in twisted graphene bilayers},\ }\href
  {https://doi.org/10.1073/pnas.1810947115} {\bibfield  {journal} {\bibinfo
  {journal} {Proc. Natl. Acad. Sci.}\ }\textbf {\bibinfo {volume} {115}},\
  \bibinfo {pages} {13174} (\bibinfo {year} {2018})}\BibitemShut {NoStop}%
\bibitem [{\citenamefont {Cea}\ \emph {et~al.}(2019)\citenamefont {Cea},
  \citenamefont {Walet},\ and\ \citenamefont {Guinea}}]{PhysRevB.100.205113}%
  \BibitemOpen
  \bibfield  {author} {\bibinfo {author} {\bibfnamefont {T.}~\bibnamefont
  {Cea}}, \bibinfo {author} {\bibfnamefont {N.~R.}\ \bibnamefont {Walet}},\
  and\ \bibinfo {author} {\bibfnamefont {F.}~\bibnamefont {Guinea}},\
  }\bibfield  {title} {\bibinfo {title} {Electronic band structure and pinning
  of {F}ermi energy to {V}an {H}ove singularities in twisted bilayer graphene:
  {A} self-consistent approach},\ }\href
  {https://doi.org/10.1103/PhysRevB.100.205113} {\bibfield  {journal} {\bibinfo
   {journal} {Phys. Rev. B}\ }\textbf {\bibinfo {volume} {100}},\ \bibinfo
  {pages} {205113} (\bibinfo {year} {2019})}\BibitemShut {NoStop}%
\bibitem [{\citenamefont {Cea}\ and\ \citenamefont
  {Guinea}(2020)}]{PhysRevB.102.045107}%
  \BibitemOpen
  \bibfield  {author} {\bibinfo {author} {\bibfnamefont {T.}~\bibnamefont
  {Cea}}\ and\ \bibinfo {author} {\bibfnamefont {F.}~\bibnamefont {Guinea}},\
  }\bibfield  {title} {\bibinfo {title} {Band structure and insulating states
  driven by {C}oulomb interaction in twisted bilayer graphene},\ }\href
  {https://doi.org/10.1103/PhysRevB.102.045107} {\bibfield  {journal} {\bibinfo
   {journal} {Phys. Rev. B}\ }\textbf {\bibinfo {volume} {102}},\ \bibinfo
  {pages} {045107} (\bibinfo {year} {2020})}\BibitemShut {NoStop}%
\bibitem [{\citenamefont {Calder\'on}\ and\ \citenamefont
  {Bascones}(2020)}]{PhysRevB.102.155149}%
  \BibitemOpen
  \bibfield  {author} {\bibinfo {author} {\bibfnamefont {M.~J.}\ \bibnamefont
  {Calder\'on}}\ and\ \bibinfo {author} {\bibfnamefont {E.}~\bibnamefont
  {Bascones}},\ }\bibfield  {title} {\bibinfo {title} {Interactions in the
  8-orbital model for twisted bilayer graphene},\ }\href
  {https://doi.org/10.1103/PhysRevB.102.155149} {\bibfield  {journal} {\bibinfo
   {journal} {Phys. Rev. B}\ }\textbf {\bibinfo {volume} {102}},\ \bibinfo
  {pages} {155149} (\bibinfo {year} {2020})}\BibitemShut {NoStop}%
\bibitem [{\citenamefont {Castro~Neto}\ \emph {et~al.}(2009)\citenamefont
  {Castro~Neto}, \citenamefont {Guinea}, \citenamefont {Peres}, \citenamefont
  {Novoselov},\ and\ \citenamefont {Geim}}]{RevModPhys.81.109}%
  \BibitemOpen
  \bibfield  {author} {\bibinfo {author} {\bibfnamefont {A.~H.}\ \bibnamefont
  {Castro~Neto}}, \bibinfo {author} {\bibfnamefont {F.}~\bibnamefont {Guinea}},
  \bibinfo {author} {\bibfnamefont {N.~M.~R.}\ \bibnamefont {Peres}}, \bibinfo
  {author} {\bibfnamefont {K.~S.}\ \bibnamefont {Novoselov}},\ and\ \bibinfo
  {author} {\bibfnamefont {A.~K.}\ \bibnamefont {Geim}},\ }\bibfield  {title}
  {\bibinfo {title} {The electronic properties of graphene},\ }\href
  {https://doi.org/10.1103/RevModPhys.81.109} {\bibfield  {journal} {\bibinfo
  {journal} {Rev. Mod. Phys.}\ }\textbf {\bibinfo {volume} {81}},\ \bibinfo
  {pages} {109} (\bibinfo {year} {2009})}\BibitemShut {NoStop}%
\bibitem [{\citenamefont {Assaad}\ and\ \citenamefont
  {Herbut}(2013)}]{Assaad2013}%
  \BibitemOpen
  \bibfield  {author} {\bibinfo {author} {\bibfnamefont {F.~F.}\ \bibnamefont
  {Assaad}}\ and\ \bibinfo {author} {\bibfnamefont {I.~F.}\ \bibnamefont
  {Herbut}},\ }\bibfield  {title} {\bibinfo {title} {Pinning the {O}rder: {T}he
  {N}ature of {Q}uantum {C}riticality in the {H}ubbard {M}odel on {H}oneycomb
  {L}attice},\ }\href {https://doi.org/10.1103/physrevx.3.031010} {\bibfield
  {journal} {\bibinfo  {journal} {Phys. Rev. X}\ }\textbf {\bibinfo {volume}
  {3}},\ \bibinfo {pages} {031010} (\bibinfo {year} {2013})}\BibitemShut
  {NoStop}%
\bibitem [{\citenamefont {Kiesel}\ \emph {et~al.}(2012)\citenamefont {Kiesel},
  \citenamefont {Platt}, \citenamefont {Hanke}, \citenamefont {Abanin},\ and\
  \citenamefont {Thomale}}]{Kiesel2012}%
  \BibitemOpen
  \bibfield  {author} {\bibinfo {author} {\bibfnamefont {M.~L.}\ \bibnamefont
  {Kiesel}}, \bibinfo {author} {\bibfnamefont {C.}~\bibnamefont {Platt}},
  \bibinfo {author} {\bibfnamefont {W.}~\bibnamefont {Hanke}}, \bibinfo
  {author} {\bibfnamefont {D.~A.}\ \bibnamefont {Abanin}},\ and\ \bibinfo
  {author} {\bibfnamefont {R.}~\bibnamefont {Thomale}},\ }\bibfield  {title}
  {\bibinfo {title} {Competing many-body instabilities and unconventional
  superconductivity in graphene},\ }\href
  {https://doi.org/10.1103/physrevb.86.020507} {\bibfield  {journal} {\bibinfo
  {journal} {Phys. Rev. B}\ }\textbf {\bibinfo {volume} {86}},\ \bibinfo
  {pages} {020507} (\bibinfo {year} {2012})}\BibitemShut {NoStop}%
\bibitem [{\citenamefont {Wang}\ \emph {et~al.}(2012)\citenamefont {Wang},
  \citenamefont {Xiang}, \citenamefont {Wang}, \citenamefont {Wang},
  \citenamefont {Yang},\ and\ \citenamefont {Lee}}]{Wang2012}%
  \BibitemOpen
  \bibfield  {author} {\bibinfo {author} {\bibfnamefont {W.-S.}\ \bibnamefont
  {Wang}}, \bibinfo {author} {\bibfnamefont {Y.-Y.}\ \bibnamefont {Xiang}},
  \bibinfo {author} {\bibfnamefont {Q.-H.}\ \bibnamefont {Wang}}, \bibinfo
  {author} {\bibfnamefont {F.}~\bibnamefont {Wang}}, \bibinfo {author}
  {\bibfnamefont {F.}~\bibnamefont {Yang}},\ and\ \bibinfo {author}
  {\bibfnamefont {D.-H.}\ \bibnamefont {Lee}},\ }\bibfield  {title} {\bibinfo
  {title} {Functional renormalization group and variational {M}onte {C}arlo
  studies of the electronic instabilities in graphene near {$\frac{1}{4}$}
  doping},\ }\href {https://doi.org/10.1103/physrevb.85.035414} {\bibfield
  {journal} {\bibinfo  {journal} {Phys. Rev. B}\ }\textbf {\bibinfo {volume}
  {85}},\ \bibinfo {pages} {035414} (\bibinfo {year} {2012})}\BibitemShut
  {NoStop}%
\bibitem [{\citenamefont {Nandkishore}\ \emph {et~al.}(2012)\citenamefont
  {Nandkishore}, \citenamefont {Levitov},\ and\ \citenamefont
  {Chubukov}}]{Nandkishore2012}%
  \BibitemOpen
  \bibfield  {author} {\bibinfo {author} {\bibfnamefont {R.}~\bibnamefont
  {Nandkishore}}, \bibinfo {author} {\bibfnamefont {L.~S.}\ \bibnamefont
  {Levitov}},\ and\ \bibinfo {author} {\bibfnamefont {A.~V.}\ \bibnamefont
  {Chubukov}},\ }\bibfield  {title} {\bibinfo {title} {Chiral superconductivity
  from repulsive interactions in doped graphene},\ }\href
  {https://doi.org/10.1038/nphys2208} {\bibfield  {journal} {\bibinfo
  {journal} {Nat. Phys.}\ }\textbf {\bibinfo {volume} {8}},\ \bibinfo {pages}
  {158} (\bibinfo {year} {2012})}\BibitemShut {NoStop}%
\bibitem [{\citenamefont {Xu}\ \emph {et~al.}(2016)\citenamefont {Xu},
  \citenamefont {Wessel},\ and\ \citenamefont {Meng}}]{Xu2016}%
  \BibitemOpen
  \bibfield  {author} {\bibinfo {author} {\bibfnamefont {X.~Y.}\ \bibnamefont
  {Xu}}, \bibinfo {author} {\bibfnamefont {S.}~\bibnamefont {Wessel}},\ and\
  \bibinfo {author} {\bibfnamefont {Z.~Y.}\ \bibnamefont {Meng}},\ }\bibfield
  {title} {\bibinfo {title} {Competing pairing channels in the doped honeycomb
  lattice {H}ubbard model},\ }\href
  {https://doi.org/10.1103/physrevb.94.115105} {\bibfield  {journal} {\bibinfo
  {journal} {Phys. Rev. B}\ }\textbf {\bibinfo {volume} {94}},\ \bibinfo
  {pages} {115105} (\bibinfo {year} {2016})}\BibitemShut {NoStop}%
\bibitem [{\citenamefont {Raczkowski}\ \emph {et~al.}(2020)\citenamefont
  {Raczkowski}, \citenamefont {Peters}, \citenamefont {Ph{\`{u}}ng},
  \citenamefont {Takemori}, \citenamefont {Assaad}, \citenamefont {Honecker},\
  and\ \citenamefont {Vahedi}}]{Raczkowski2020}%
  \BibitemOpen
  \bibfield  {author} {\bibinfo {author} {\bibfnamefont {M.}~\bibnamefont
  {Raczkowski}}, \bibinfo {author} {\bibfnamefont {R.}~\bibnamefont {Peters}},
  \bibinfo {author} {\bibfnamefont {T.~T.}\ \bibnamefont {Ph{\`{u}}ng}},
  \bibinfo {author} {\bibfnamefont {N.}~\bibnamefont {Takemori}}, \bibinfo
  {author} {\bibfnamefont {F.~F.}\ \bibnamefont {Assaad}}, \bibinfo {author}
  {\bibfnamefont {A.}~\bibnamefont {Honecker}},\ and\ \bibinfo {author}
  {\bibfnamefont {J.}~\bibnamefont {Vahedi}},\ }\bibfield  {title} {\bibinfo
  {title} {Hubbard model on the honeycomb lattice: {F}rom static and dynamical
  mean-field theories to lattice quantum {M}onte {C}arlo simulations},\ }\href
  {https://doi.org/10.1103/physrevb.101.125103} {\bibfield  {journal} {\bibinfo
   {journal} {Phys. Rev. B}\ }\textbf {\bibinfo {volume} {101}},\ \bibinfo
  {pages} {125103} (\bibinfo {year} {2020})}\BibitemShut {NoStop}%
\bibitem [{\citenamefont {Costa}\ \emph {et~al.}(2021)\citenamefont {Costa},
  \citenamefont {Seki},\ and\ \citenamefont {Sorella}}]{Costa2021}%
  \BibitemOpen
  \bibfield  {author} {\bibinfo {author} {\bibfnamefont {N.~C.}\ \bibnamefont
  {Costa}}, \bibinfo {author} {\bibfnamefont {K.}~\bibnamefont {Seki}},\ and\
  \bibinfo {author} {\bibfnamefont {S.}~\bibnamefont {Sorella}},\ }\bibfield
  {title} {\bibinfo {title} {Magnetism and {C}harge {O}rder in the {H}oneycomb
  {L}attice},\ }\href {https://doi.org/10.1103/physrevlett.126.107205}
  {\bibfield  {journal} {\bibinfo  {journal} {Phys. Rev. Lett.}\ }\textbf
  {\bibinfo {volume} {126}},\ \bibinfo {pages} {107205} (\bibinfo {year}
  {2021})}\BibitemShut {NoStop}%
\bibitem [{\citenamefont {Kuroki}\ and\ \citenamefont
  {Arita}(2001)}]{Kuroki2001}%
  \BibitemOpen
  \bibfield  {author} {\bibinfo {author} {\bibfnamefont {K.}~\bibnamefont
  {Kuroki}}\ and\ \bibinfo {author} {\bibfnamefont {R.}~\bibnamefont {Arita}},\
  }\bibfield  {title} {\bibinfo {title} {Spin-triplet superconductivity in
  repulsive {H}ubbard models with disconnected {F}ermi surfaces: {A} case study
  on triangular and honeycomb lattices},\ }\href
  {https://doi.org/10.1103/physrevb.63.174507} {\bibfield  {journal} {\bibinfo
  {journal} {Phys. Rev. B}\ }\textbf {\bibinfo {volume} {63}},\ \bibinfo
  {pages} {174507} (\bibinfo {year} {2001})}\BibitemShut {NoStop}%
\bibitem [{\citenamefont {Onari}\ \emph {et~al.}(2002)\citenamefont {Onari},
  \citenamefont {Kuroki}, \citenamefont {Arita},\ and\ \citenamefont
  {Aoki}}]{Onari2002}%
  \BibitemOpen
  \bibfield  {author} {\bibinfo {author} {\bibfnamefont {S.}~\bibnamefont
  {Onari}}, \bibinfo {author} {\bibfnamefont {K.}~\bibnamefont {Kuroki}},
  \bibinfo {author} {\bibfnamefont {R.}~\bibnamefont {Arita}},\ and\ \bibinfo
  {author} {\bibfnamefont {H.}~\bibnamefont {Aoki}},\ }\bibfield  {title}
  {\bibinfo {title} {Superconductivity induced by interband nesting in the
  three-dimensional honeycomb lattice},\ }\href
  {https://doi.org/10.1103/physrevb.65.184525} {\bibfield  {journal} {\bibinfo
  {journal} {Phys. Rev. B}\ }\textbf {\bibinfo {volume} {65}},\ \bibinfo
  {pages} {184525} (\bibinfo {year} {2002})}\BibitemShut {NoStop}%
\bibitem [{\citenamefont {Kuroki}(2010)}]{Kuroki2010}%
  \BibitemOpen
  \bibfield  {author} {\bibinfo {author} {\bibfnamefont {K.}~\bibnamefont
  {Kuroki}},\ }\bibfield  {title} {\bibinfo {title} {Spin-fluctuation-mediated
  {$d+id^{\prime}$} pairing mechanism in doped
  {$\beta$}-{MNCl}({M}={H}f,{Z}r)superconductors},\ }\href
  {https://doi.org/10.1103/physrevb.81.104502} {\bibfield  {journal} {\bibinfo
  {journal} {Phys. Rev. B}\ }\textbf {\bibinfo {volume} {81}},\ \bibinfo
  {pages} {104502} (\bibinfo {year} {2010})}\BibitemShut {NoStop}%
\bibitem [{\citenamefont {Li}\ \emph {et~al.}(2020)\citenamefont {Li},
  \citenamefont {Wallerberger}, \citenamefont {Chikano}, \citenamefont {Yeh},
  \citenamefont {Gull},\ and\ \citenamefont {Shinaoka}}]{Li2020}%
  \BibitemOpen
  \bibfield  {author} {\bibinfo {author} {\bibfnamefont {J.}~\bibnamefont
  {Li}}, \bibinfo {author} {\bibfnamefont {M.}~\bibnamefont {Wallerberger}},
  \bibinfo {author} {\bibfnamefont {N.}~\bibnamefont {Chikano}}, \bibinfo
  {author} {\bibfnamefont {C.-N.}\ \bibnamefont {Yeh}}, \bibinfo {author}
  {\bibfnamefont {E.}~\bibnamefont {Gull}},\ and\ \bibinfo {author}
  {\bibfnamefont {H.}~\bibnamefont {Shinaoka}},\ }\bibfield  {title} {\bibinfo
  {title} {Sparse sampling approach to efficient ab initio calculations at
  finite temperature},\ }\href {https://doi.org/10.1103/physrevb.101.035144}
  {\bibfield  {journal} {\bibinfo  {journal} {Phys. Rev. B}\ }\textbf {\bibinfo
  {volume} {101}},\ \bibinfo {pages} {035144} (\bibinfo {year} {2020})},\
  \Eprint {https://arxiv.org/abs/1908.07575} {arxiv:1908.07575} \BibitemShut
  {NoStop}%
\bibitem [{\citenamefont {Witt}\ \emph {et~al.}(2021)\citenamefont {Witt},
  \citenamefont {van Loon}, \citenamefont {Nomoto}, \citenamefont {Arita},\
  and\ \citenamefont {Wehling}}]{Witt2021}%
  \BibitemOpen
  \bibfield  {author} {\bibinfo {author} {\bibfnamefont {N.}~\bibnamefont
  {Witt}}, \bibinfo {author} {\bibfnamefont {E.~G. C.~P.}\ \bibnamefont {van
  Loon}}, \bibinfo {author} {\bibfnamefont {T.}~\bibnamefont {Nomoto}},
  \bibinfo {author} {\bibfnamefont {R.}~\bibnamefont {Arita}},\ and\ \bibinfo
  {author} {\bibfnamefont {T.~O.}\ \bibnamefont {Wehling}},\ }\bibfield
  {title} {\bibinfo {title} {Efficient fluctuation-exchange approach to
  low-temperature spin fluctuations and superconductivity: {F}rom the {H}ubbard
  model to {N}a{$_x$}{C}o{O}{$_2$}{$\cdotp$}{$y$}{H}{$_2$}{O}},\ }\href
  {https://doi.org/10.1103/physrevb.103.205148} {\bibfield  {journal} {\bibinfo
   {journal} {Phys. Rev. B}\ }\textbf {\bibinfo {volume} {103}},\ \bibinfo
  {pages} {205148} (\bibinfo {year} {2021})},\ \Eprint
  {https://arxiv.org/abs/2012.04562} {arXiv:2012.04562} \BibitemShut {NoStop}%
\bibitem [{\citenamefont {Rösner}\ \emph {et~al.}(2016)\citenamefont
  {Rösner}, \citenamefont {Steinke}, \citenamefont {Lorke}, \citenamefont
  {Gies}, \citenamefont {Jahnke},\ and\ \citenamefont {Wehling}}]{Roesner2016}%
  \BibitemOpen
  \bibfield  {author} {\bibinfo {author} {\bibfnamefont {M.}~\bibnamefont
  {Rösner}}, \bibinfo {author} {\bibfnamefont {C.}~\bibnamefont {Steinke}},
  \bibinfo {author} {\bibfnamefont {M.}~\bibnamefont {Lorke}}, \bibinfo
  {author} {\bibfnamefont {C.}~\bibnamefont {Gies}}, \bibinfo {author}
  {\bibfnamefont {F.}~\bibnamefont {Jahnke}},\ and\ \bibinfo {author}
  {\bibfnamefont {T.~O.}\ \bibnamefont {Wehling}},\ }\bibfield  {title}
  {\bibinfo {title} {Two-{D}imensional {H}eterojunctions from {N}onlocal
  {M}anipulations of the {I}nteractions},\ }\href
  {https://doi.org/10.1021/acs.nanolett.5b05009} {\bibfield  {journal}
  {\bibinfo  {journal} {Nano Lett.}\ }\textbf {\bibinfo {volume} {16}},\
  \bibinfo {pages} {2322} (\bibinfo {year} {2016})}\BibitemShut {NoStop}%
\bibitem [{\citenamefont {Raja}\ \emph {et~al.}(2017)\citenamefont {Raja},
  \citenamefont {Chaves}, \citenamefont {Yu}, \citenamefont {Arefe},
  \citenamefont {Hill}, \citenamefont {Rigosi}, \citenamefont {Berkelbach},
  \citenamefont {Nagler}, \citenamefont {Schüller}, \citenamefont {Korn},
  \citenamefont {Nuckolls}, \citenamefont {Hone}, \citenamefont {Brus},
  \citenamefont {Heinz}, \citenamefont {Reichman},\ and\ \citenamefont
  {Chernikov}}]{Raja2017}%
  \BibitemOpen
  \bibfield  {author} {\bibinfo {author} {\bibfnamefont {A.}~\bibnamefont
  {Raja}}, \bibinfo {author} {\bibfnamefont {A.}~\bibnamefont {Chaves}},
  \bibinfo {author} {\bibfnamefont {J.}~\bibnamefont {Yu}}, \bibinfo {author}
  {\bibfnamefont {G.}~\bibnamefont {Arefe}}, \bibinfo {author} {\bibfnamefont
  {H.~M.}\ \bibnamefont {Hill}}, \bibinfo {author} {\bibfnamefont {A.~F.}\
  \bibnamefont {Rigosi}}, \bibinfo {author} {\bibfnamefont {T.~C.}\
  \bibnamefont {Berkelbach}}, \bibinfo {author} {\bibfnamefont
  {P.}~\bibnamefont {Nagler}}, \bibinfo {author} {\bibfnamefont
  {C.}~\bibnamefont {Schüller}}, \bibinfo {author} {\bibfnamefont
  {T.}~\bibnamefont {Korn}}, \bibinfo {author} {\bibfnamefont {C.}~\bibnamefont
  {Nuckolls}}, \bibinfo {author} {\bibfnamefont {J.}~\bibnamefont {Hone}},
  \bibinfo {author} {\bibfnamefont {L.~E.}\ \bibnamefont {Brus}}, \bibinfo
  {author} {\bibfnamefont {T.~F.}\ \bibnamefont {Heinz}}, \bibinfo {author}
  {\bibfnamefont {D.~R.}\ \bibnamefont {Reichman}},\ and\ \bibinfo {author}
  {\bibfnamefont {A.}~\bibnamefont {Chernikov}},\ }\bibfield  {title} {\bibinfo
  {title} {Coulomb engineering of the bandgap and excitons in two-dimensional
  materials},\ }\href {https://doi.org/10.1038/ncomms15251} {\bibfield
  {journal} {\bibinfo  {journal} {Nat Commun.}\ }\textbf {\bibinfo {volume}
  {8}},\ \bibinfo {pages} {15251} (\bibinfo {year} {2017})}\BibitemShut
  {NoStop}%
\bibitem [{\citenamefont {Pizarro}\ \emph {et~al.}(2019)\citenamefont
  {Pizarro}, \citenamefont {Rösner}, \citenamefont {Thomale}, \citenamefont
  {Valent{\'{\i}}},\ and\ \citenamefont {Wehling}}]{Pizarro2019}%
  \BibitemOpen
  \bibfield  {author} {\bibinfo {author} {\bibfnamefont {J.~M.}\ \bibnamefont
  {Pizarro}}, \bibinfo {author} {\bibfnamefont {M.}~\bibnamefont {Rösner}},
  \bibinfo {author} {\bibfnamefont {R.}~\bibnamefont {Thomale}}, \bibinfo
  {author} {\bibfnamefont {R.}~\bibnamefont {Valent{\'{\i}}}},\ and\ \bibinfo
  {author} {\bibfnamefont {T.~O.}\ \bibnamefont {Wehling}},\ }\bibfield
  {title} {\bibinfo {title} {Internal screening and dielectric engineering in
  magic-angle twisted bilayer graphene},\ }\href
  {https://doi.org/10.1103/physrevb.100.161102} {\bibfield  {journal} {\bibinfo
   {journal} {Phys. Rev. B}\ }\textbf {\bibinfo {volume} {100}},\ \bibinfo
  {pages} {161102} (\bibinfo {year} {2019})}\BibitemShut {NoStop}%
\bibitem [{\citenamefont {Goodwin}\ \emph {et~al.}(2019)\citenamefont
  {Goodwin}, \citenamefont {Corsetti}, \citenamefont {Mostofi},\ and\
  \citenamefont {Lischner}}]{Goodwin2019}%
  \BibitemOpen
  \bibfield  {author} {\bibinfo {author} {\bibfnamefont {Z.~A.~H.}\
  \bibnamefont {Goodwin}}, \bibinfo {author} {\bibfnamefont {F.}~\bibnamefont
  {Corsetti}}, \bibinfo {author} {\bibfnamefont {A.~A.}\ \bibnamefont
  {Mostofi}},\ and\ \bibinfo {author} {\bibfnamefont {J.}~\bibnamefont
  {Lischner}},\ }\bibfield  {title} {\bibinfo {title} {Twist-angle sensitivity
  of electron correlations in moir{\'{e}} graphene bilayers},\ }\href
  {https://doi.org/10.1103/physrevb.100.121106} {\bibfield  {journal} {\bibinfo
   {journal} {Phys. Rev. B}\ }\textbf {\bibinfo {volume} {100}},\ \bibinfo
  {pages} {121106} (\bibinfo {year} {2019})}\BibitemShut {NoStop}%
\bibitem [{\citenamefont {Goodwin}\ \emph {et~al.}(2020)\citenamefont
  {Goodwin}, \citenamefont {Vitale}, \citenamefont {Corsetti}, \citenamefont
  {Efetov}, \citenamefont {Mostofi},\ and\ \citenamefont
  {Lischner}}]{Goodwin2020}%
  \BibitemOpen
  \bibfield  {author} {\bibinfo {author} {\bibfnamefont {Z.~A.~H.}\
  \bibnamefont {Goodwin}}, \bibinfo {author} {\bibfnamefont {V.}~\bibnamefont
  {Vitale}}, \bibinfo {author} {\bibfnamefont {F.}~\bibnamefont {Corsetti}},
  \bibinfo {author} {\bibfnamefont {D.~K.}\ \bibnamefont {Efetov}}, \bibinfo
  {author} {\bibfnamefont {A.~A.}\ \bibnamefont {Mostofi}},\ and\ \bibinfo
  {author} {\bibfnamefont {J.}~\bibnamefont {Lischner}},\ }\bibfield  {title}
  {\bibinfo {title} {Critical role of device geometry for the phase diagram of
  twisted bilayer graphene},\ }\href
  {https://doi.org/10.1103/physrevb.101.165110} {\bibfield  {journal} {\bibinfo
   {journal} {Phys. Rev. B}\ }\textbf {\bibinfo {volume} {101}},\ \bibinfo
  {pages} {165110} (\bibinfo {year} {2020})}\BibitemShut {NoStop}%
\bibitem [{\citenamefont {Arora}\ \emph {et~al.}(2020)\citenamefont {Arora},
  \citenamefont {Polski}, \citenamefont {Zhang}, \citenamefont {Thomson},
  \citenamefont {Choi}, \citenamefont {Kim}, \citenamefont {Lin}, \citenamefont
  {Wilson}, \citenamefont {Xu}, \citenamefont {Chu}, \citenamefont {Watanabe},
  \citenamefont {Taniguchi}, \citenamefont {Alicea},\ and\ \citenamefont
  {Nadj-Perge}}]{Arora2020}%
  \BibitemOpen
  \bibfield  {author} {\bibinfo {author} {\bibfnamefont {H.~S.}\ \bibnamefont
  {Arora}}, \bibinfo {author} {\bibfnamefont {R.}~\bibnamefont {Polski}},
  \bibinfo {author} {\bibfnamefont {Y.}~\bibnamefont {Zhang}}, \bibinfo
  {author} {\bibfnamefont {A.}~\bibnamefont {Thomson}}, \bibinfo {author}
  {\bibfnamefont {Y.}~\bibnamefont {Choi}}, \bibinfo {author} {\bibfnamefont
  {H.}~\bibnamefont {Kim}}, \bibinfo {author} {\bibfnamefont {Z.}~\bibnamefont
  {Lin}}, \bibinfo {author} {\bibfnamefont {I.~Z.}\ \bibnamefont {Wilson}},
  \bibinfo {author} {\bibfnamefont {X.}~\bibnamefont {Xu}}, \bibinfo {author}
  {\bibfnamefont {J.-H.}\ \bibnamefont {Chu}}, \bibinfo {author} {\bibfnamefont
  {K.}~\bibnamefont {Watanabe}}, \bibinfo {author} {\bibfnamefont
  {T.}~\bibnamefont {Taniguchi}}, \bibinfo {author} {\bibfnamefont
  {J.}~\bibnamefont {Alicea}},\ and\ \bibinfo {author} {\bibfnamefont
  {S.}~\bibnamefont {Nadj-Perge}},\ }\bibfield  {title} {\bibinfo {title}
  {Superconductivity in metallic twisted bilayer graphene stabilized by
  {WSe}{$_2$}},\ }\href {https://doi.org/10.1038/s41586-020-2473-8} {\bibfield
  {journal} {\bibinfo  {journal} {Nature}\ }\textbf {\bibinfo {volume} {583}},\
  \bibinfo {pages} {379} (\bibinfo {year} {2020})}\BibitemShut {NoStop}%
\bibitem [{\citenamefont {Kim}\ \emph {et~al.}(2020)\citenamefont {Kim},
  \citenamefont {Xu}, \citenamefont {Berdyugin}, \citenamefont {Principi},
  \citenamefont {Slizovskiy}, \citenamefont {Xin}, \citenamefont
  {Kumaravadivel}, \citenamefont {Kuang}, \citenamefont {Hamer}, \citenamefont
  {Kumar}, \citenamefont {Gorbachev}, \citenamefont {Watanabe}, \citenamefont
  {Taniguchi}, \citenamefont {Grigorieva}, \citenamefont {Fal'ko},
  \citenamefont {Polini},\ and\ \citenamefont {Geim}}]{Kim2020}%
  \BibitemOpen
  \bibfield  {author} {\bibinfo {author} {\bibfnamefont {M.}~\bibnamefont
  {Kim}}, \bibinfo {author} {\bibfnamefont {S.~G.}\ \bibnamefont {Xu}},
  \bibinfo {author} {\bibfnamefont {A.~I.}\ \bibnamefont {Berdyugin}}, \bibinfo
  {author} {\bibfnamefont {A.}~\bibnamefont {Principi}}, \bibinfo {author}
  {\bibfnamefont {S.}~\bibnamefont {Slizovskiy}}, \bibinfo {author}
  {\bibfnamefont {N.}~\bibnamefont {Xin}}, \bibinfo {author} {\bibfnamefont
  {P.}~\bibnamefont {Kumaravadivel}}, \bibinfo {author} {\bibfnamefont
  {W.}~\bibnamefont {Kuang}}, \bibinfo {author} {\bibfnamefont
  {M.}~\bibnamefont {Hamer}}, \bibinfo {author} {\bibfnamefont {R.~K.}\
  \bibnamefont {Kumar}}, \bibinfo {author} {\bibfnamefont {R.~V.}\ \bibnamefont
  {Gorbachev}}, \bibinfo {author} {\bibfnamefont {K.}~\bibnamefont {Watanabe}},
  \bibinfo {author} {\bibfnamefont {T.}~\bibnamefont {Taniguchi}}, \bibinfo
  {author} {\bibfnamefont {I.~V.}\ \bibnamefont {Grigorieva}}, \bibinfo
  {author} {\bibfnamefont {V.~I.}\ \bibnamefont {Fal'ko}}, \bibinfo {author}
  {\bibfnamefont {M.}~\bibnamefont {Polini}},\ and\ \bibinfo {author}
  {\bibfnamefont {A.~K.}\ \bibnamefont {Geim}},\ }\bibfield  {title} {\bibinfo
  {title} {Control of electron-electron interaction in graphene by proximity
  screening},\ }\href {https://doi.org/10.1038/s41467-020-15829-1} {\bibfield
  {journal} {\bibinfo  {journal} {Nat Commun.}\ }\textbf {\bibinfo {volume}
  {11}},\ \bibinfo {pages} {2339} (\bibinfo {year} {2020})}\BibitemShut
  {NoStop}%
\bibitem [{\citenamefont {Bulut}\ \emph {et~al.}(1993)\citenamefont {Bulut},
  \citenamefont {Scalapino},\ and\ \citenamefont {White}}]{Bulut1993}%
  \BibitemOpen
  \bibfield  {author} {\bibinfo {author} {\bibfnamefont {N.}~\bibnamefont
  {Bulut}}, \bibinfo {author} {\bibfnamefont {D.~J.}\ \bibnamefont
  {Scalapino}},\ and\ \bibinfo {author} {\bibfnamefont {S.~R.}\ \bibnamefont
  {White}},\ }\bibfield  {title} {\bibinfo {title} {Comparison of {M}onte
  {C}arlo and diagrammatic calculations for the two-dimensional {H}ubbard
  model},\ }\href {https://doi.org/10.1103/physrevb.47.2742} {\bibfield
  {journal} {\bibinfo  {journal} {Phys. Rev. B}\ }\textbf {\bibinfo {volume}
  {47}},\ \bibinfo {pages} {2742} (\bibinfo {year} {1993})}\BibitemShut
  {NoStop}%
\bibitem [{\citenamefont {Bulut}\ \emph {et~al.}(1994)\citenamefont {Bulut},
  \citenamefont {Scalapino},\ and\ \citenamefont {White}}]{Bulut1994}%
  \BibitemOpen
  \bibfield  {author} {\bibinfo {author} {\bibfnamefont {N.}~\bibnamefont
  {Bulut}}, \bibinfo {author} {\bibfnamefont {D.~J.}\ \bibnamefont
  {Scalapino}},\ and\ \bibinfo {author} {\bibfnamefont {S.~R.}\ \bibnamefont
  {White}},\ }\bibfield  {title} {\bibinfo {title} {Effective electron-electron
  interaction in the two-dimensional {H}ubbard model},\ }\href
  {https://doi.org/10.1103/physrevb.50.9623} {\bibfield  {journal} {\bibinfo
  {journal} {Phys. Rev. B}\ }\textbf {\bibinfo {volume} {50}},\ \bibinfo
  {pages} {9623} (\bibinfo {year} {1994})}\BibitemShut {NoStop}%
\bibitem [{\citenamefont {Zhang}\ \emph {et~al.}(2009)\citenamefont {Zhang},
  \citenamefont {Sknepnek}, \citenamefont {Fernandes},\ and\ \citenamefont
  {Schmalian}}]{Zhang2009}%
  \BibitemOpen
  \bibfield  {author} {\bibinfo {author} {\bibfnamefont {J.}~\bibnamefont
  {Zhang}}, \bibinfo {author} {\bibfnamefont {R.}~\bibnamefont {Sknepnek}},
  \bibinfo {author} {\bibfnamefont {R.~M.}\ \bibnamefont {Fernandes}},\ and\
  \bibinfo {author} {\bibfnamefont {J.}~\bibnamefont {Schmalian}},\ }\bibfield
  {title} {\bibinfo {title} {Orbital coupling and superconductivity in the iron
  pnictides},\ }\href {https://doi.org/10.1103/physrevb.79.220502} {\bibfield
  {journal} {\bibinfo  {journal} {Phys. Rev. B}\ }\textbf {\bibinfo {volume}
  {79}},\ \bibinfo {pages} {220502} (\bibinfo {year} {2009})},\ \Eprint
  {https://arxiv.org/abs/0903.4473} {arXiv:0903.4473} \BibitemShut {NoStop}%
\bibitem [{\citenamefont {Moriya}\ and\ \citenamefont
  {Ueda}(2003)}]{Moriya2003}%
  \BibitemOpen
  \bibfield  {author} {\bibinfo {author} {\bibfnamefont {T.}~\bibnamefont
  {Moriya}}\ and\ \bibinfo {author} {\bibfnamefont {K.}~\bibnamefont {Ueda}},\
  }\bibfield  {title} {\bibinfo {title} {Antiferromagnetic spin fluctuation and
  superconductivity},\ }\href {https://doi.org/10.1088/0034-4885/66/8/202}
  {\bibfield  {journal} {\bibinfo  {journal} {Rep. Prog. Phys.}\ }\textbf
  {\bibinfo {volume} {66}},\ \bibinfo {pages} {1299} (\bibinfo {year}
  {2003})}\BibitemShut {NoStop}%
\bibitem [{\citenamefont {Scalapino}(2012)}]{Scalapino2012}%
  \BibitemOpen
  \bibfield  {author} {\bibinfo {author} {\bibfnamefont {D.~J.}\ \bibnamefont
  {Scalapino}},\ }\bibfield  {title} {\bibinfo {title} {A common thread: {T}he
  pairing interaction for unconventional superconductors},\ }\href
  {https://doi.org/10.1103/revmodphys.84.1383} {\bibfield  {journal} {\bibinfo
  {journal} {Rev. Mod. Phys.}\ }\textbf {\bibinfo {volume} {84}},\ \bibinfo
  {pages} {1383} (\bibinfo {year} {2012})},\ \Eprint
  {https://arxiv.org/abs/1207.4093} {arXiv:1207.4093} \BibitemShut {NoStop}%
\bibitem [{\citenamefont {Ying}\ and\ \citenamefont {Wessel}(2018)}]{Ying2018}%
  \BibitemOpen
  \bibfield  {author} {\bibinfo {author} {\bibfnamefont {T.}~\bibnamefont
  {Ying}}\ and\ \bibinfo {author} {\bibfnamefont {S.}~\bibnamefont {Wessel}},\
  }\bibfield  {title} {\bibinfo {title} {Pairing and chiral spin density wave
  instabilities on the honeycomb lattice: {A} comparative quantum {M}onte
  {C}arlo study},\ }\href {https://doi.org/10.1103/physrevb.97.075127}
  {\bibfield  {journal} {\bibinfo  {journal} {Phys. Rev. B}\ }\textbf {\bibinfo
  {volume} {97}},\ \bibinfo {pages} {075127} (\bibinfo {year}
  {2018})}\BibitemShut {NoStop}%
\bibitem [{\citenamefont {McMillan}(1968)}]{PhysRev.167.331}%
  \BibitemOpen
  \bibfield  {author} {\bibinfo {author} {\bibfnamefont {W.~L.}\ \bibnamefont
  {McMillan}},\ }\bibfield  {title} {\bibinfo {title} {Transition {T}emperature
  of {S}trong-{C}oupled {S}uperconductors},\ }\href
  {https://doi.org/10.1103/PhysRevB.12.905} {\bibfield  {journal} {\bibinfo
  {journal} {Phys. Rev.}\ }\textbf {\bibinfo {volume} {167}},\ \bibinfo {pages}
  {331} (\bibinfo {year} {1968})}\BibitemShut {NoStop}%
\bibitem [{\citenamefont {Dynes}(1972)}]{DYNES1972615}%
  \BibitemOpen
  \bibfield  {author} {\bibinfo {author} {\bibfnamefont {R.}~\bibnamefont
  {Dynes}},\ }\bibfield  {title} {\bibinfo {title} {{McMillan's equation and
  the {$T_c$} of superconductors}},\ }\href
  {https://doi.org/10.1016/0038-1098(72)90603-5} {\bibfield  {journal}
  {\bibinfo  {journal} {Solid State Commun.}\ }\textbf {\bibinfo {volume}
  {10}},\ \bibinfo {pages} {615} (\bibinfo {year} {1972})}\BibitemShut
  {NoStop}%
\bibitem [{\citenamefont {Tolmachev}(1961)}]{Tolmachev1961}%
  \BibitemOpen
  \bibfield  {author} {\bibinfo {author} {\bibfnamefont {V.~V.}\ \bibnamefont
  {Tolmachev}},\ }\bibfield  {title} {\bibinfo {title} {Logarithmic criterion
  for superconductivity},\ }\href {http://mi.mathnet.ru/eng/dan25552}
  {\bibfield  {journal} {\bibinfo  {journal} {Dokl. Akad. Nauk SSSR}\ }\textbf
  {\bibinfo {volume} {140}},\ \bibinfo {pages} {563} (\bibinfo {year}
  {1961})}\BibitemShut {NoStop}%
\bibitem [{\citenamefont {Morel}\ and\ \citenamefont
  {Anderson}(1962)}]{Morel1962}%
  \BibitemOpen
  \bibfield  {author} {\bibinfo {author} {\bibfnamefont {P.}~\bibnamefont
  {Morel}}\ and\ \bibinfo {author} {\bibfnamefont {P.~W.}\ \bibnamefont
  {Anderson}},\ }\bibfield  {title} {\bibinfo {title} {Calculation of the
  {S}uperconducting {S}tate {P}arameters with {R}etarded {E}lectron-{P}honon
  {I}nteraction},\ }\href {https://doi.org/10.1103/physrev.125.1263} {\bibfield
   {journal} {\bibinfo  {journal} {Phys. Rev.}\ }\textbf {\bibinfo {volume}
  {125}},\ \bibinfo {pages} {1263} (\bibinfo {year} {1962})}\BibitemShut
  {NoStop}%
\bibitem [{\citenamefont {Wu}\ \emph {et~al.}(2018)\citenamefont {Wu},
  \citenamefont {MacDonald},\ and\ \citenamefont {Martin}}]{Wu2018}%
  \BibitemOpen
  \bibfield  {author} {\bibinfo {author} {\bibfnamefont {F.}~\bibnamefont
  {Wu}}, \bibinfo {author} {\bibfnamefont {A.}~\bibnamefont {MacDonald}},\ and\
  \bibinfo {author} {\bibfnamefont {I.}~\bibnamefont {Martin}},\ }\bibfield
  {title} {\bibinfo {title} {Theory of {P}honon-{M}ediated {S}uperconductivity
  in {T}wisted {B}ilayer {G}raphene},\ }\href
  {https://doi.org/10.1103/physrevlett.121.257001} {\bibfield  {journal}
  {\bibinfo  {journal} {Phys. Rev. Lett.}\ }\textbf {\bibinfo {volume} {121}},\
  \bibinfo {pages} {257001} (\bibinfo {year} {2018})}\BibitemShut {NoStop}%
\bibitem [{\citenamefont {Lin}\ \emph {et~al.}(2018)\citenamefont {Lin},
  \citenamefont {Tan}, \citenamefont {Wu}, \citenamefont {Chen}, \citenamefont
  {Wang}, \citenamefont {Pan}, \citenamefont {Zhang}, \citenamefont {Cong},
  \citenamefont {Zhang}, \citenamefont {Ji}, \citenamefont {Hu}, \citenamefont
  {Liu},\ and\ \citenamefont {Tan}}]{Lin2018}%
  \BibitemOpen
  \bibfield  {author} {\bibinfo {author} {\bibfnamefont {M.-L.}\ \bibnamefont
  {Lin}}, \bibinfo {author} {\bibfnamefont {Q.-H.}\ \bibnamefont {Tan}},
  \bibinfo {author} {\bibfnamefont {J.-B.}\ \bibnamefont {Wu}}, \bibinfo
  {author} {\bibfnamefont {X.-S.}\ \bibnamefont {Chen}}, \bibinfo {author}
  {\bibfnamefont {J.-H.}\ \bibnamefont {Wang}}, \bibinfo {author}
  {\bibfnamefont {Y.-H.}\ \bibnamefont {Pan}}, \bibinfo {author} {\bibfnamefont
  {X.}~\bibnamefont {Zhang}}, \bibinfo {author} {\bibfnamefont
  {X.}~\bibnamefont {Cong}}, \bibinfo {author} {\bibfnamefont {J.}~\bibnamefont
  {Zhang}}, \bibinfo {author} {\bibfnamefont {W.}~\bibnamefont {Ji}}, \bibinfo
  {author} {\bibfnamefont {P.-A.}\ \bibnamefont {Hu}}, \bibinfo {author}
  {\bibfnamefont {K.-H.}\ \bibnamefont {Liu}},\ and\ \bibinfo {author}
  {\bibfnamefont {P.-H.}\ \bibnamefont {Tan}},\ }\bibfield  {title} {\bibinfo
  {title} {Moir{\'{e}} {P}honons in {T}wisted {B}ilayer {MoS}{$_2$}},\ }\href
  {https://doi.org/10.1021/acsnano.8b05006} {\bibfield  {journal} {\bibinfo
  {journal} {{ACS} Nano}\ }\textbf {\bibinfo {volume} {12}},\ \bibinfo {pages}
  {8770} (\bibinfo {year} {2018})}\BibitemShut {NoStop}%
\bibitem [{\citenamefont {Choi}\ and\ \citenamefont {Choi}(2018)}]{Choi2018}%
  \BibitemOpen
  \bibfield  {author} {\bibinfo {author} {\bibfnamefont {Y.~W.}\ \bibnamefont
  {Choi}}\ and\ \bibinfo {author} {\bibfnamefont {H.~J.}\ \bibnamefont
  {Choi}},\ }\bibfield  {title} {\bibinfo {title} {Strong electron-phonon
  coupling, electron-hole asymmetry, and nonadiabaticity in magic-angle twisted
  bilayer graphene},\ }\href {https://doi.org/10.1103/physrevb.98.241412}
  {\bibfield  {journal} {\bibinfo  {journal} {Phys. Rev. B}\ }\textbf {\bibinfo
  {volume} {98}},\ \bibinfo {pages} {241412} (\bibinfo {year}
  {2018})}\BibitemShut {NoStop}%
\bibitem [{\citenamefont {Lian}\ \emph {et~al.}(2019)\citenamefont {Lian},
  \citenamefont {Wang},\ and\ \citenamefont {Bernevig}}]{Lian2019}%
  \BibitemOpen
  \bibfield  {author} {\bibinfo {author} {\bibfnamefont {B.}~\bibnamefont
  {Lian}}, \bibinfo {author} {\bibfnamefont {Z.}~\bibnamefont {Wang}},\ and\
  \bibinfo {author} {\bibfnamefont {B.~A.}\ \bibnamefont {Bernevig}},\
  }\bibfield  {title} {\bibinfo {title} {Twisted {B}ilayer {G}raphene: {A}
  {P}honon-{D}riven {S}uperconductor},\ }\href
  {https://doi.org/10.1103/physrevlett.122.257002} {\bibfield  {journal}
  {\bibinfo  {journal} {Phys. Rev. Lett.}\ }\textbf {\bibinfo {volume} {122}},\
  \bibinfo {pages} {257002} (\bibinfo {year} {2019})}\BibitemShut {NoStop}%
\bibitem [{\citenamefont {Debnath}\ \emph {et~al.}(2020)\citenamefont
  {Debnath}, \citenamefont {Maity}, \citenamefont {Biswas}, \citenamefont
  {Raghunathan}, \citenamefont {Jain},\ and\ \citenamefont
  {Ghosh}}]{Debnath2020}%
  \BibitemOpen
  \bibfield  {author} {\bibinfo {author} {\bibfnamefont {R.}~\bibnamefont
  {Debnath}}, \bibinfo {author} {\bibfnamefont {I.}~\bibnamefont {Maity}},
  \bibinfo {author} {\bibfnamefont {R.}~\bibnamefont {Biswas}}, \bibinfo
  {author} {\bibfnamefont {V.}~\bibnamefont {Raghunathan}}, \bibinfo {author}
  {\bibfnamefont {M.}~\bibnamefont {Jain}},\ and\ \bibinfo {author}
  {\bibfnamefont {A.}~\bibnamefont {Ghosh}},\ }\bibfield  {title} {\bibinfo
  {title} {Evolution of high-frequency {R}aman modes and their doping
  dependence in twisted bilayer {MoS}{$_2$}},\ }\href
  {https://doi.org/10.1039/c9nr09897f} {\bibfield  {journal} {\bibinfo
  {journal} {Nanoscale}\ }\textbf {\bibinfo {volume} {12}},\ \bibinfo {pages}
  {17272} (\bibinfo {year} {2020})}\BibitemShut {NoStop}%
\bibitem [{\citenamefont {Lin}\ \emph {et~al.}(2021)\citenamefont {Lin},
  \citenamefont {Holler}, \citenamefont {Bauer}, \citenamefont {Parzefall},
  \citenamefont {Scheuck}, \citenamefont {Peng}, \citenamefont {Korn},
  \citenamefont {Bange}, \citenamefont {Lupton},\ and\ \citenamefont
  {Schüller}}]{Lin2021}%
  \BibitemOpen
  \bibfield  {author} {\bibinfo {author} {\bibfnamefont {K.-Q.}\ \bibnamefont
  {Lin}}, \bibinfo {author} {\bibfnamefont {J.}~\bibnamefont {Holler}},
  \bibinfo {author} {\bibfnamefont {J.~M.}\ \bibnamefont {Bauer}}, \bibinfo
  {author} {\bibfnamefont {P.}~\bibnamefont {Parzefall}}, \bibinfo {author}
  {\bibfnamefont {M.}~\bibnamefont {Scheuck}}, \bibinfo {author} {\bibfnamefont
  {B.}~\bibnamefont {Peng}}, \bibinfo {author} {\bibfnamefont {T.}~\bibnamefont
  {Korn}}, \bibinfo {author} {\bibfnamefont {S.}~\bibnamefont {Bange}},
  \bibinfo {author} {\bibfnamefont {J.~M.}\ \bibnamefont {Lupton}},\ and\
  \bibinfo {author} {\bibfnamefont {C.}~\bibnamefont {Schüller}},\ }\bibfield
  {title} {\bibinfo {title} {Large-{S}cale {M}apping of {M}oir{\'{e}}
  {S}uperlattices by {H}yperspectral {R}aman {I}maging},\ }\href
  {https://doi.org/10.1002/adma.202008333} {\bibfield  {journal} {\bibinfo
  {journal} {Adv. Mater.}\ ,\ \bibinfo {pages} {2008333}} (\bibinfo {year}
  {2021})}\BibitemShut {NoStop}%
\bibitem [{\citenamefont {Parzefall}\ \emph {et~al.}(2021)\citenamefont
  {Parzefall}, \citenamefont {Holler}, \citenamefont {Scheuck}, \citenamefont
  {Beer}, \citenamefont {Lin}, \citenamefont {Peng}, \citenamefont {Monserrat},
  \citenamefont {Nagler}, \citenamefont {Kempf}, \citenamefont {Korn},\ and\
  \citenamefont {Schüller}}]{Parzefall2021}%
  \BibitemOpen
  \bibfield  {author} {\bibinfo {author} {\bibfnamefont {P.}~\bibnamefont
  {Parzefall}}, \bibinfo {author} {\bibfnamefont {J.}~\bibnamefont {Holler}},
  \bibinfo {author} {\bibfnamefont {M.}~\bibnamefont {Scheuck}}, \bibinfo
  {author} {\bibfnamefont {A.}~\bibnamefont {Beer}}, \bibinfo {author}
  {\bibfnamefont {K.-Q.}\ \bibnamefont {Lin}}, \bibinfo {author} {\bibfnamefont
  {B.}~\bibnamefont {Peng}}, \bibinfo {author} {\bibfnamefont {B.}~\bibnamefont
  {Monserrat}}, \bibinfo {author} {\bibfnamefont {P.}~\bibnamefont {Nagler}},
  \bibinfo {author} {\bibfnamefont {M.}~\bibnamefont {Kempf}}, \bibinfo
  {author} {\bibfnamefont {T.}~\bibnamefont {Korn}},\ and\ \bibinfo {author}
  {\bibfnamefont {C.}~\bibnamefont {Schüller}},\ }\bibfield  {title} {\bibinfo
  {title} {Moir{\'{e}} phonons in twisted {MoSe}{$_2$}{\textendash}{WSe}{$_2$}
  heterobilayers and their correlation with interlayer excitons},\ }\href
  {https://doi.org/10.1088/2053-1583/abf98e} {\bibfield  {journal} {\bibinfo
  {journal} {2D Mater.}\ }\textbf {\bibinfo {volume} {8}},\ \bibinfo {pages}
  {035030} (\bibinfo {year} {2021})}\BibitemShut {NoStop}%
\bibitem [{\citenamefont {Quan}\ \emph {et~al.}(2021)\citenamefont {Quan},
  \citenamefont {Linhart}, \citenamefont {Lin}, \citenamefont {Lee},
  \citenamefont {Zhu}, \citenamefont {Wang}, \citenamefont {Hsu}, \citenamefont
  {Choi}, \citenamefont {Embley}, \citenamefont {Young}, \citenamefont
  {Taniguchi}, \citenamefont {Watanabe}, \citenamefont {Shih}, \citenamefont
  {Lai}, \citenamefont {MacDonald}, \citenamefont {Tan}, \citenamefont
  {Libisch},\ and\ \citenamefont {Li}}]{Quan2021}%
  \BibitemOpen
  \bibfield  {author} {\bibinfo {author} {\bibfnamefont {J.}~\bibnamefont
  {Quan}}, \bibinfo {author} {\bibfnamefont {L.}~\bibnamefont {Linhart}},
  \bibinfo {author} {\bibfnamefont {M.-L.}\ \bibnamefont {Lin}}, \bibinfo
  {author} {\bibfnamefont {D.}~\bibnamefont {Lee}}, \bibinfo {author}
  {\bibfnamefont {J.}~\bibnamefont {Zhu}}, \bibinfo {author} {\bibfnamefont
  {C.-Y.}\ \bibnamefont {Wang}}, \bibinfo {author} {\bibfnamefont {W.-T.}\
  \bibnamefont {Hsu}}, \bibinfo {author} {\bibfnamefont {J.}~\bibnamefont
  {Choi}}, \bibinfo {author} {\bibfnamefont {J.}~\bibnamefont {Embley}},
  \bibinfo {author} {\bibfnamefont {C.}~\bibnamefont {Young}}, \bibinfo
  {author} {\bibfnamefont {T.}~\bibnamefont {Taniguchi}}, \bibinfo {author}
  {\bibfnamefont {K.}~\bibnamefont {Watanabe}}, \bibinfo {author}
  {\bibfnamefont {C.-K.}\ \bibnamefont {Shih}}, \bibinfo {author}
  {\bibfnamefont {K.}~\bibnamefont {Lai}}, \bibinfo {author} {\bibfnamefont
  {A.~H.}\ \bibnamefont {MacDonald}}, \bibinfo {author} {\bibfnamefont {P.-H.}\
  \bibnamefont {Tan}}, \bibinfo {author} {\bibfnamefont {F.}~\bibnamefont
  {Libisch}},\ and\ \bibinfo {author} {\bibfnamefont {X.}~\bibnamefont {Li}},\
  }\bibfield  {title} {\bibinfo {title} {Phonon renormalization in
  reconstructed {MoS}2 moir{\'{e}} superlattices},\ }\href
  {https://doi.org/10.1038/s41563-021-00960-1} {\bibfield  {journal} {\bibinfo
  {journal} {Nat. Mater.}\ }\textbf {\bibinfo {volume} {20}},\ \bibinfo {pages}
  {1100} (\bibinfo {year} {2021})}\BibitemShut {NoStop}%
\bibitem [{\citenamefont {Rice}\ and\ \citenamefont {Scott}(1975)}]{Rice1975}%
  \BibitemOpen
  \bibfield  {author} {\bibinfo {author} {\bibfnamefont {T.~M.}\ \bibnamefont
  {Rice}}\ and\ \bibinfo {author} {\bibfnamefont {G.~K.}\ \bibnamefont
  {Scott}},\ }\bibfield  {title} {\bibinfo {title} {New {M}echanism for a
  {C}harge-{D}ensity-{W}ave {I}nstability},\ }\href
  {https://doi.org/10.1103/physrevlett.35.120} {\bibfield  {journal} {\bibinfo
  {journal} {Phys. Rev. Lett.}\ }\textbf {\bibinfo {volume} {35}},\ \bibinfo
  {pages} {120} (\bibinfo {year} {1975})}\BibitemShut {NoStop}%
\bibitem [{\citenamefont {Jiang}\ \emph {et~al.}(2014)\citenamefont {Jiang},
  \citenamefont {Mesaros},\ and\ \citenamefont {Ran}}]{Jiang2014}%
  \BibitemOpen
  \bibfield  {author} {\bibinfo {author} {\bibfnamefont {S.}~\bibnamefont
  {Jiang}}, \bibinfo {author} {\bibfnamefont {A.}~\bibnamefont {Mesaros}},\
  and\ \bibinfo {author} {\bibfnamefont {Y.}~\bibnamefont {Ran}},\ }\bibfield
  {title} {\bibinfo {title} {Chiral {S}pin-{D}ensity {W}ave,
  {S}pin-{C}harge-{C}hern {L}iquid, and {$d+id$} {S}uperconductivity in
  {1/4}-{D}oped {C}orrelated {E}lectronic {S}ystems on the {H}oneycomb
  {L}attice},\ }\href {https://doi.org/10.1103/physrevx.4.031040} {\bibfield
  {journal} {\bibinfo  {journal} {Phys. Rev. X}\ }\textbf {\bibinfo {volume}
  {4}},\ \bibinfo {pages} {031040} (\bibinfo {year} {2014})}\BibitemShut
  {NoStop}%
\bibitem [{\citenamefont {Berges}\ \emph {et~al.}(2020)\citenamefont {Berges},
  \citenamefont {van Loon}, \citenamefont {Schobert}, \citenamefont {Rösner},\
  and\ \citenamefont {Wehling}}]{Berges2020}%
  \BibitemOpen
  \bibfield  {author} {\bibinfo {author} {\bibfnamefont {J.}~\bibnamefont
  {Berges}}, \bibinfo {author} {\bibfnamefont {E.~G. C.~P.}\ \bibnamefont {van
  Loon}}, \bibinfo {author} {\bibfnamefont {A.}~\bibnamefont {Schobert}},
  \bibinfo {author} {\bibfnamefont {M.}~\bibnamefont {Rösner}},\ and\ \bibinfo
  {author} {\bibfnamefont {T.~O.}\ \bibnamefont {Wehling}},\ }\bibfield
  {title} {\bibinfo {title} {Ab initio phonon self-energies and fluctuation
  diagnostics of phonon anomalies: {L}attice instabilities from {D}irac
  pseudospin physics in transition metal dichalcogenides},\ }\href
  {https://doi.org/10.1103/physrevb.101.155107} {\bibfield  {journal} {\bibinfo
   {journal} {Phys. Rev. B}\ }\textbf {\bibinfo {volume} {101}},\ \bibinfo
  {pages} {155107} (\bibinfo {year} {2020})}\BibitemShut {NoStop}%
\bibitem [{\citenamefont {Keimer}\ \emph {et~al.}(2015)\citenamefont {Keimer},
  \citenamefont {Kivelson}, \citenamefont {Norman}, \citenamefont {Uchida},\
  and\ \citenamefont {Zaanen}}]{KeiN5182015}%
  \BibitemOpen
  \bibfield  {author} {\bibinfo {author} {\bibfnamefont {B.}~\bibnamefont
  {Keimer}}, \bibinfo {author} {\bibfnamefont {S.~A.}\ \bibnamefont
  {Kivelson}}, \bibinfo {author} {\bibfnamefont {M.~R.}\ \bibnamefont
  {Norman}}, \bibinfo {author} {\bibfnamefont {S.}~\bibnamefont {Uchida}},\
  and\ \bibinfo {author} {\bibfnamefont {J.}~\bibnamefont {Zaanen}},\
  }\bibfield  {title} {\bibinfo {title} {From quantum matter to
  high-temperature superconductivity in copper oxides},\ }\href
  {https://doi.org/https://doi.org/10.1038/nature14165} {\bibfield  {journal}
  {\bibinfo  {journal} {Nature}\ }\textbf {\bibinfo {volume} {518}},\ \bibinfo
  {pages} {179} (\bibinfo {year} {2015})}\BibitemShut {NoStop}%
\end{thebibliography}%


\begin{thebibliography}{56}%
\makeatletter
\providecommand \@ifxundefined [1]{%
 \@ifx{#1\undefined}
}%
\providecommand \@ifnum [1]{%
 \ifnum #1\expandafter \@firstoftwo
 \else \expandafter \@secondoftwo
 \fi
}%
\providecommand \@ifx [1]{%
 \ifx #1\expandafter \@firstoftwo
 \else \expandafter \@secondoftwo
 \fi
}%
\providecommand \natexlab [1]{#1}%
\providecommand \enquote  [1]{``#1''}%
\providecommand \bibnamefont  [1]{#1}%
\providecommand \bibfnamefont [1]{#1}%
\providecommand \citenamefont [1]{#1}%
\providecommand \href@noop [0]{\@secondoftwo}%
\providecommand \href [0]{\begingroup \@sanitize@url \@href}%
\providecommand \@href[1]{\@@startlink{#1}\@@href}%
\providecommand \@@href[1]{\endgroup#1\@@endlink}%
\providecommand \@sanitize@url [0]{\catcode `\\12\catcode `\$12\catcode
  `\&12\catcode `\#12\catcode `\^12\catcode `\_12\catcode `\%12\relax}%
\providecommand \@@startlink[1]{}%
\providecommand \@@endlink[0]{}%
\providecommand \url  [0]{\begingroup\@sanitize@url \@url }%
\providecommand \@url [1]{\endgroup\@href {#1}{\urlprefix }}%
\providecommand \urlprefix  [0]{URL }%
\providecommand \Eprint [0]{\href }%
\providecommand \doibase [0]{https://doi.org/}%
\providecommand \selectlanguage [0]{\@gobble}%
\providecommand \bibinfo  [0]{\@secondoftwo}%
\providecommand \bibfield  [0]{\@secondoftwo}%
\providecommand \translation [1]{[#1]}%
\providecommand \BibitemOpen [0]{}%
\providecommand \bibitemStop [0]{}%
\providecommand \bibitemNoStop [0]{.\EOS\space}%
\providecommand \EOS [0]{\spacefactor3000\relax}%
\providecommand \BibitemShut  [1]{\csname bibitem#1\endcsname}%
\let\auto@bib@innerbib\@empty
%</preamble>
\bibitem [{\citenamefont {{Angeli}}\ and\ \citenamefont
  {{MacDonald}}(2021)}]{gammaTMDCsMacdonald2021}%
  \BibitemOpen
  \bibfield  {author} {\bibinfo {author} {\bibfnamefont {M.}~\bibnamefont
  {{Angeli}}}\ and\ \bibinfo {author} {\bibfnamefont {A.~H.}\ \bibnamefont
  {{MacDonald}}},\ }\bibfield  {title} {\bibinfo {title}
  {{{\ensuremath{\Gamma}} valley transition metal dichalcogenide moir{\'e}
  bands}},\ }\href {https://doi.org/10.1073/pnas.2021826118} {\bibfield
  {journal} {\bibinfo  {journal} {Proc. Natl. Acad. Sci.}\ }\textbf {\bibinfo
  {volume} {118}},\ \bibinfo {pages} {e2021826118} (\bibinfo {year} {2021})},\
  \Eprint {https://arxiv.org/abs/2008.01735} {arXiv:2008.01735} \BibitemShut
  {NoStop}%
\bibitem [{\citenamefont {Cea}\ \emph {et~al.}(2019)\citenamefont {Cea},
  \citenamefont {Walet},\ and\ \citenamefont {Guinea}}]{PhysRevB.100.205113}%
  \BibitemOpen
  \bibfield  {author} {\bibinfo {author} {\bibfnamefont {T.}~\bibnamefont
  {Cea}}, \bibinfo {author} {\bibfnamefont {N.~R.}\ \bibnamefont {Walet}},\
  and\ \bibinfo {author} {\bibfnamefont {F.}~\bibnamefont {Guinea}},\
  }\bibfield  {title} {\bibinfo {title} {Electronic band structure and pinning
  of {F}ermi energy to {V}an {H}ove singularities in twisted bilayer graphene:
  {A} self-consistent approach},\ }\href
  {https://doi.org/10.1103/PhysRevB.100.205113} {\bibfield  {journal} {\bibinfo
   {journal} {Phys. Rev. B}\ }\textbf {\bibinfo {volume} {100}},\ \bibinfo
  {pages} {205113} (\bibinfo {year} {2019})}\BibitemShut {NoStop}%
\bibitem [{\citenamefont {Cea}\ and\ \citenamefont
  {Guinea}(2020)}]{PhysRevB.102.045107}%
  \BibitemOpen
  \bibfield  {author} {\bibinfo {author} {\bibfnamefont {T.}~\bibnamefont
  {Cea}}\ and\ \bibinfo {author} {\bibfnamefont {F.}~\bibnamefont {Guinea}},\
  }\bibfield  {title} {\bibinfo {title} {Band structure and insulating states
  driven by {C}oulomb interaction in twisted bilayer graphene},\ }\href
  {https://doi.org/10.1103/PhysRevB.102.045107} {\bibfield  {journal} {\bibinfo
   {journal} {Phys. Rev. B}\ }\textbf {\bibinfo {volume} {102}},\ \bibinfo
  {pages} {045107} (\bibinfo {year} {2020})}\BibitemShut {NoStop}%
\bibitem [{\citenamefont {Bickers}\ \emph {et~al.}(1989)\citenamefont
  {Bickers}, \citenamefont {Scalapino},\ and\ \citenamefont
  {White}}]{Bickers1989a}%
  \BibitemOpen
  \bibfield  {author} {\bibinfo {author} {\bibfnamefont {N.~E.}\ \bibnamefont
  {Bickers}}, \bibinfo {author} {\bibfnamefont {D.~J.}\ \bibnamefont
  {Scalapino}},\ and\ \bibinfo {author} {\bibfnamefont {S.~R.}\ \bibnamefont
  {White}},\ }\bibfield  {title} {\bibinfo {title} {Conserving {A}pproximations
  for {S}trongly {C}orrelated {E}lectron {S}ystems: {B}ethe-{S}alpeter
  {E}quation and {D}ynamics for the {T}wo-{D}imensional {H}ubbard {M}odel},\
  }\href {https://doi.org/10.1103/physrevlett.62.961} {\bibfield  {journal}
  {\bibinfo  {journal} {Phys. Rev. Lett.}\ }\textbf {\bibinfo {volume} {62}},\
  \bibinfo {pages} {961} (\bibinfo {year} {1989})}\BibitemShut {NoStop}%
\bibitem [{\citenamefont {Bickers}\ and\ \citenamefont
  {Scalapino}(1989)}]{Bickers1989b}%
  \BibitemOpen
  \bibfield  {author} {\bibinfo {author} {\bibfnamefont {N.~E.}\ \bibnamefont
  {Bickers}}\ and\ \bibinfo {author} {\bibfnamefont {D.~J.}\ \bibnamefont
  {Scalapino}},\ }\bibfield  {title} {\bibinfo {title} {Conserving
  approximations for strongly fluctuating electron systems. {I}. {F}ormalism
  and calculational approach},\ }\href
  {https://doi.org/10.1016/0003-4916(89)90359-x} {\bibfield  {journal}
  {\bibinfo  {journal} {Ann. Phys. (NY)}\ }\textbf {\bibinfo {volume} {193}},\
  \bibinfo {pages} {206} (\bibinfo {year} {1989})}\BibitemShut {NoStop}%
\bibitem [{\citenamefont {Xiao}\ \emph {et~al.}(2012)\citenamefont {Xiao},
  \citenamefont {Liu}, \citenamefont {Feng}, \citenamefont {Xu},\ and\
  \citenamefont {Yao}}]{PhysRevLett.108.196802}%
  \BibitemOpen
  \bibfield  {author} {\bibinfo {author} {\bibfnamefont {D.}~\bibnamefont
  {Xiao}}, \bibinfo {author} {\bibfnamefont {G.-B.}\ \bibnamefont {Liu}},
  \bibinfo {author} {\bibfnamefont {W.}~\bibnamefont {Feng}}, \bibinfo {author}
  {\bibfnamefont {X.}~\bibnamefont {Xu}},\ and\ \bibinfo {author}
  {\bibfnamefont {W.}~\bibnamefont {Yao}},\ }\bibfield  {title} {\bibinfo
  {title} {{Coupled Spin and Valley Physics in Monolayers of
  ${\mathrm{MoS}}_{2}$ and Other Group-VI Dichalcogenides}},\ }\href
  {https://doi.org/10.1103/PhysRevLett.108.196802} {\bibfield  {journal}
  {\bibinfo  {journal} {Phys. Rev. Lett.}\ }\textbf {\bibinfo {volume} {108}},\
  \bibinfo {pages} {196802} (\bibinfo {year} {2012})}\BibitemShut {NoStop}%
\bibitem [{\citenamefont {Venkateswarlu}\ \emph {et~al.}(2020)\citenamefont
  {Venkateswarlu}, \citenamefont {Honecker},\ and\ \citenamefont {Trambly~de
  Laissardi\`ere}}]{PhysRevB.102.081103}%
  \BibitemOpen
  \bibfield  {author} {\bibinfo {author} {\bibfnamefont {S.}~\bibnamefont
  {Venkateswarlu}}, \bibinfo {author} {\bibfnamefont {A.}~\bibnamefont
  {Honecker}},\ and\ \bibinfo {author} {\bibfnamefont {G.}~\bibnamefont
  {Trambly~de Laissardi\`ere}},\ }\bibfield  {title} {\bibinfo {title}
  {Electronic localization in twisted bilayer ${\mathrm{mos}}_{2}$ with small
  rotation angle},\ }\href {https://doi.org/10.1103/PhysRevB.102.081103}
  {\bibfield  {journal} {\bibinfo  {journal} {Phys. Rev. B}\ }\textbf {\bibinfo
  {volume} {102}},\ \bibinfo {pages} {081103} (\bibinfo {year}
  {2020})}\BibitemShut {NoStop}%
\bibitem [{\citenamefont {Marzari}\ \emph {et~al.}(2012)\citenamefont
  {Marzari}, \citenamefont {Mostofi}, \citenamefont {Yates}, \citenamefont
  {Souza},\ and\ \citenamefont {Vanderbilt}}]{RevModPhys.84.1419}%
  \BibitemOpen
  \bibfield  {author} {\bibinfo {author} {\bibfnamefont {N.}~\bibnamefont
  {Marzari}}, \bibinfo {author} {\bibfnamefont {A.~A.}\ \bibnamefont
  {Mostofi}}, \bibinfo {author} {\bibfnamefont {J.~R.}\ \bibnamefont {Yates}},
  \bibinfo {author} {\bibfnamefont {I.}~\bibnamefont {Souza}},\ and\ \bibinfo
  {author} {\bibfnamefont {D.}~\bibnamefont {Vanderbilt}},\ }\bibfield  {title}
  {\bibinfo {title} {{Maximally localized Wannier functions: Theory and
  applications}},\ }\href {https://doi.org/10.1103/RevModPhys.84.1419}
  {\bibfield  {journal} {\bibinfo  {journal} {Rev. Mod. Phys.}\ }\textbf
  {\bibinfo {volume} {84}},\ \bibinfo {pages} {1419} (\bibinfo {year}
  {2012})}\BibitemShut {NoStop}%
\bibitem [{\citenamefont {Yuan}\ \emph {et~al.}(2019)\citenamefont {Yuan},
  \citenamefont {Isobe},\ and\ \citenamefont {Fu}}]{Yuan2019}%
  \BibitemOpen
  \bibfield  {author} {\bibinfo {author} {\bibfnamefont {N.~F.~Q.}\
  \bibnamefont {Yuan}}, \bibinfo {author} {\bibfnamefont {H.}~\bibnamefont
  {Isobe}},\ and\ \bibinfo {author} {\bibfnamefont {L.}~\bibnamefont {Fu}},\
  }\bibfield  {title} {\bibinfo {title} {Magic of high-order van {H}ove
  singularity},\ }\href {https://doi.org/10.1038/s41467-019-13670-9} {\bibfield
   {journal} {\bibinfo  {journal} {Nat Commun.}\ }\textbf {\bibinfo {volume}
  {10}},\ \bibinfo {pages} {5769} (\bibinfo {year} {2019})}\BibitemShut
  {NoStop}%
\bibitem [{\citenamefont {Classen}\ \emph {et~al.}(2020)\citenamefont
  {Classen}, \citenamefont {Chubukov}, \citenamefont {Honerkamp},\ and\
  \citenamefont {Scherer}}]{Classen2020}%
  \BibitemOpen
  \bibfield  {author} {\bibinfo {author} {\bibfnamefont {L.}~\bibnamefont
  {Classen}}, \bibinfo {author} {\bibfnamefont {A.~V.}\ \bibnamefont
  {Chubukov}}, \bibinfo {author} {\bibfnamefont {C.}~\bibnamefont
  {Honerkamp}},\ and\ \bibinfo {author} {\bibfnamefont {M.~M.}\ \bibnamefont
  {Scherer}},\ }\bibfield  {title} {\bibinfo {title} {Competing orders at
  higher-order {V}an {H}ove points},\ }\href
  {https://doi.org/10.1103/physrevb.102.125141} {\bibfield  {journal} {\bibinfo
   {journal} {Phys. Rev. B}\ }\textbf {\bibinfo {volume} {102}},\ \bibinfo
  {pages} {125141} (\bibinfo {year} {2020})}\BibitemShut {NoStop}%
\bibitem [{\citenamefont {Guinea}\ and\ \citenamefont
  {Walet}(2018)}]{Guinea2018}%
  \BibitemOpen
  \bibfield  {author} {\bibinfo {author} {\bibfnamefont {F.}~\bibnamefont
  {Guinea}}\ and\ \bibinfo {author} {\bibfnamefont {N.~R.}\ \bibnamefont
  {Walet}},\ }\bibfield  {title} {\bibinfo {title} {Electrostatic effects, band
  distortions, and superconductivity in twisted graphene bilayers},\ }\href
  {https://doi.org/10.1073/pnas.1810947115} {\bibfield  {journal} {\bibinfo
  {journal} {Proc. Natl. Acad. Sci.}\ }\textbf {\bibinfo {volume} {115}},\
  \bibinfo {pages} {13174} (\bibinfo {year} {2018})}\BibitemShut {NoStop}%
\bibitem [{\citenamefont {Calder\'on}\ and\ \citenamefont
  {Bascones}(2020)}]{PhysRevB.102.155149}%
  \BibitemOpen
  \bibfield  {author} {\bibinfo {author} {\bibfnamefont {M.~J.}\ \bibnamefont
  {Calder\'on}}\ and\ \bibinfo {author} {\bibfnamefont {E.}~\bibnamefont
  {Bascones}},\ }\bibfield  {title} {\bibinfo {title} {Interactions in the
  8-orbital model for twisted bilayer graphene},\ }\href
  {https://doi.org/10.1103/PhysRevB.102.155149} {\bibfield  {journal} {\bibinfo
   {journal} {Phys. Rev. B}\ }\textbf {\bibinfo {volume} {102}},\ \bibinfo
  {pages} {155149} (\bibinfo {year} {2020})}\BibitemShut {NoStop}%
\bibitem [{\citenamefont {Fischer}\ \emph {et~al.}(2022)\citenamefont
  {Fischer}, \citenamefont {Goodwin}, \citenamefont {Mostofi}, \citenamefont
  {Lischner}, \citenamefont {Kennes},\ and\ \citenamefont
  {Klebl}}]{Fischer2022}%
  \BibitemOpen
  \bibfield  {author} {\bibinfo {author} {\bibfnamefont {A.}~\bibnamefont
  {Fischer}}, \bibinfo {author} {\bibfnamefont {Z.~A.~H.}\ \bibnamefont
  {Goodwin}}, \bibinfo {author} {\bibfnamefont {A.~A.}\ \bibnamefont
  {Mostofi}}, \bibinfo {author} {\bibfnamefont {J.}~\bibnamefont {Lischner}},
  \bibinfo {author} {\bibfnamefont {D.~M.}\ \bibnamefont {Kennes}},\ and\
  \bibinfo {author} {\bibfnamefont {L.}~\bibnamefont {Klebl}},\ }\bibfield
  {title} {\bibinfo {title} {Unconventional superconductivity in magic-angle
  twisted trilayer graphene},\ }\href
  {https://doi.org/10.1038/s41535-021-00410-w} {\bibfield  {journal} {\bibinfo
  {journal} {npj Quantum Mater.}\ }\textbf {\bibinfo {volume} {7}},\ \bibinfo
  {pages} {5} (\bibinfo {year} {2022})}\BibitemShut {NoStop}%
\bibitem [{\citenamefont {Kerker}(1981)}]{PhysRevB.23.3082}%
  \BibitemOpen
  \bibfield  {author} {\bibinfo {author} {\bibfnamefont {G.~P.}\ \bibnamefont
  {Kerker}},\ }\bibfield  {title} {\bibinfo {title} {Efficient iteration scheme
  for self-consistent pseudopotential calculations},\ }\href
  {https://doi.org/10.1103/PhysRevB.23.3082} {\bibfield  {journal} {\bibinfo
  {journal} {Phys. Rev. B}\ }\textbf {\bibinfo {volume} {23}},\ \bibinfo
  {pages} {3082} (\bibinfo {year} {1981})}\BibitemShut {NoStop}%
\bibitem [{\citenamefont {van Loon}\ \emph {et~al.}(2016)\citenamefont {van
  Loon}, \citenamefont {Sch\"uler}, \citenamefont {Katsnelson},\ and\
  \citenamefont {Wehling}}]{PhysRevB.94.165141}%
  \BibitemOpen
  \bibfield  {author} {\bibinfo {author} {\bibfnamefont {E.~G. C.~P.}\
  \bibnamefont {van Loon}}, \bibinfo {author} {\bibfnamefont {M.}~\bibnamefont
  {Sch\"uler}}, \bibinfo {author} {\bibfnamefont {M.~I.}\ \bibnamefont
  {Katsnelson}},\ and\ \bibinfo {author} {\bibfnamefont {T.~O.}\ \bibnamefont
  {Wehling}},\ }\bibfield  {title} {\bibinfo {title} {Capturing nonlocal
  interaction effects in the {H}ubbard model: {O}ptimal mappings and limits of
  applicability},\ }\href {https://doi.org/10.1103/PhysRevB.94.165141}
  {\bibfield  {journal} {\bibinfo  {journal} {Phys. Rev. B}\ }\textbf {\bibinfo
  {volume} {94}},\ \bibinfo {pages} {165141} (\bibinfo {year}
  {2016})}\BibitemShut {NoStop}%
\bibitem [{\citenamefont {Pizarro}\ \emph {et~al.}(2019)\citenamefont
  {Pizarro}, \citenamefont {Rösner}, \citenamefont {Thomale}, \citenamefont
  {Valent{\'{\i}}},\ and\ \citenamefont {Wehling}}]{Pizarro2019}%
  \BibitemOpen
  \bibfield  {author} {\bibinfo {author} {\bibfnamefont {J.~M.}\ \bibnamefont
  {Pizarro}}, \bibinfo {author} {\bibfnamefont {M.}~\bibnamefont {Rösner}},
  \bibinfo {author} {\bibfnamefont {R.}~\bibnamefont {Thomale}}, \bibinfo
  {author} {\bibfnamefont {R.}~\bibnamefont {Valent{\'{\i}}}},\ and\ \bibinfo
  {author} {\bibfnamefont {T.~O.}\ \bibnamefont {Wehling}},\ }\bibfield
  {title} {\bibinfo {title} {Internal screening and dielectric engineering in
  magic-angle twisted bilayer graphene},\ }\href
  {https://doi.org/10.1103/physrevb.100.161102} {\bibfield  {journal} {\bibinfo
   {journal} {Phys. Rev. B}\ }\textbf {\bibinfo {volume} {100}},\ \bibinfo
  {pages} {161102} (\bibinfo {year} {2019})}\BibitemShut {NoStop}%
\bibitem [{\citenamefont {Ohno}(1964)}]{Ohno1964}%
  \BibitemOpen
  \bibfield  {author} {\bibinfo {author} {\bibfnamefont {K.}~\bibnamefont
  {Ohno}},\ }\bibfield  {title} {\bibinfo {title} {Some remarks on the
  {Pariser}-{Parr}-{Pople} method},\ }\href
  {https://doi.org/https://doi.org/10.1007/BF00528281} {\bibfield  {journal}
  {\bibinfo  {journal} {Theoretica chimica acta}\ }\textbf {\bibinfo {volume}
  {2}},\ \bibinfo {pages} {219} (\bibinfo {year} {1964})}\BibitemShut {NoStop}%
\bibitem [{\citenamefont {Cea}\ and\ \citenamefont {Guinea}(2021)}]{Cea2021}%
  \BibitemOpen
  \bibfield  {author} {\bibinfo {author} {\bibfnamefont {T.}~\bibnamefont
  {Cea}}\ and\ \bibinfo {author} {\bibfnamefont {F.}~\bibnamefont {Guinea}},\
  }\bibfield  {title} {\bibinfo {title} {Coulomb interaction, phonons, and
  superconductivity in twisted bilayer graphene},\ }\href
  {https://doi.org/10.1073/pnas.2107874118} {\bibfield  {journal} {\bibinfo
  {journal} {Proc. Natl. Acad. Sci.}\ }\textbf {\bibinfo {volume} {118}},\
  \bibinfo {pages} {e2107874118} (\bibinfo {year} {2021})}\BibitemShut
  {NoStop}%
\bibitem [{\citenamefont {Laturia}\ \emph {et~al.}(2018)\citenamefont
  {Laturia}, \citenamefont {Van~de Put},\ and\ \citenamefont
  {Vandenberghe}}]{Laturia2018}%
  \BibitemOpen
  \bibfield  {author} {\bibinfo {author} {\bibfnamefont {A.}~\bibnamefont
  {Laturia}}, \bibinfo {author} {\bibfnamefont {M.~L.}\ \bibnamefont {Van~de
  Put}},\ and\ \bibinfo {author} {\bibfnamefont {W.~G.}\ \bibnamefont
  {Vandenberghe}},\ }\bibfield  {title} {\bibinfo {title} {Dielectric
  properties of hexagonal boron nitride and transition metal dichalcogenides:
  from monolayer to bulk},\ }\href {https://doi.org/10.1038/s41699-018-0050-x}
  {\bibfield  {journal} {\bibinfo  {journal} {npj 2D Mater. Appl.}\ }\textbf
  {\bibinfo {volume} {2}},\ \bibinfo {pages} {6} (\bibinfo {year}
  {2018})}\BibitemShut {NoStop}%
\bibitem [{\citenamefont {Weston}\ \emph {et~al.}(2020)\citenamefont {Weston},
  \citenamefont {Zou}, \citenamefont {Enaldiev}, \citenamefont {Summerfield},
  \citenamefont {Clark}, \citenamefont {Z{\'o}lyomi}, \citenamefont {Graham},
  \citenamefont {Yelgel}, \citenamefont {Magorrian}, \citenamefont {Zhou},
  \citenamefont {Zultak}, \citenamefont {Hopkinson}, \citenamefont {Barinov},
  \citenamefont {Bointon}, \citenamefont {Kretinin}, \citenamefont {Wilson},
  \citenamefont {Beton}, \citenamefont {Fal'ko}, \citenamefont {Haigh},\ and\
  \citenamefont {Gorbachev}}]{Weston2020}%
  \BibitemOpen
  \bibfield  {author} {\bibinfo {author} {\bibfnamefont {A.}~\bibnamefont
  {Weston}}, \bibinfo {author} {\bibfnamefont {Y.}~\bibnamefont {Zou}},
  \bibinfo {author} {\bibfnamefont {V.}~\bibnamefont {Enaldiev}}, \bibinfo
  {author} {\bibfnamefont {A.}~\bibnamefont {Summerfield}}, \bibinfo {author}
  {\bibfnamefont {N.}~\bibnamefont {Clark}}, \bibinfo {author} {\bibfnamefont
  {V.}~\bibnamefont {Z{\'o}lyomi}}, \bibinfo {author} {\bibfnamefont
  {A.}~\bibnamefont {Graham}}, \bibinfo {author} {\bibfnamefont
  {C.}~\bibnamefont {Yelgel}}, \bibinfo {author} {\bibfnamefont
  {S.}~\bibnamefont {Magorrian}}, \bibinfo {author} {\bibfnamefont
  {M.}~\bibnamefont {Zhou}}, \bibinfo {author} {\bibfnamefont {J.}~\bibnamefont
  {Zultak}}, \bibinfo {author} {\bibfnamefont {D.}~\bibnamefont {Hopkinson}},
  \bibinfo {author} {\bibfnamefont {A.}~\bibnamefont {Barinov}}, \bibinfo
  {author} {\bibfnamefont {T.~H.}\ \bibnamefont {Bointon}}, \bibinfo {author}
  {\bibfnamefont {A.}~\bibnamefont {Kretinin}}, \bibinfo {author}
  {\bibfnamefont {N.~R.}\ \bibnamefont {Wilson}}, \bibinfo {author}
  {\bibfnamefont {P.~H.}\ \bibnamefont {Beton}}, \bibinfo {author}
  {\bibfnamefont {V.~I.}\ \bibnamefont {Fal'ko}}, \bibinfo {author}
  {\bibfnamefont {S.~J.}\ \bibnamefont {Haigh}},\ and\ \bibinfo {author}
  {\bibfnamefont {R.}~\bibnamefont {Gorbachev}},\ }\bibfield  {title} {\bibinfo
  {title} {Atomic reconstruction in twisted bilayers of transition metal
  dichalcogenides},\ }\href {https://doi.org/10.1038/s41565-020-0682-9}
  {\bibfield  {journal} {\bibinfo  {journal} {Nat. Nanotechnol.}\ }\textbf
  {\bibinfo {volume} {15}},\ \bibinfo {pages} {592} (\bibinfo {year}
  {2020})}\BibitemShut {NoStop}%
\bibitem [{\citenamefont {Goodwin}\ \emph {et~al.}(2019)\citenamefont
  {Goodwin}, \citenamefont {Corsetti}, \citenamefont {Mostofi},\ and\
  \citenamefont {Lischner}}]{Goodwin2019}%
  \BibitemOpen
  \bibfield  {author} {\bibinfo {author} {\bibfnamefont {Z.~A.~H.}\
  \bibnamefont {Goodwin}}, \bibinfo {author} {\bibfnamefont {F.}~\bibnamefont
  {Corsetti}}, \bibinfo {author} {\bibfnamefont {A.~A.}\ \bibnamefont
  {Mostofi}},\ and\ \bibinfo {author} {\bibfnamefont {J.}~\bibnamefont
  {Lischner}},\ }\bibfield  {title} {\bibinfo {title} {Twist-angle sensitivity
  of electron correlations in moir{\'{e}} graphene bilayers},\ }\href
  {https://doi.org/10.1103/physrevb.100.121106} {\bibfield  {journal} {\bibinfo
   {journal} {Phys. Rev. B}\ }\textbf {\bibinfo {volume} {100}},\ \bibinfo
  {pages} {121106} (\bibinfo {year} {2019})}\BibitemShut {NoStop}%
\bibitem [{\citenamefont {Stepanov}\ \emph {et~al.}(2020)\citenamefont
  {Stepanov}, \citenamefont {Das}, \citenamefont {Lu}, \citenamefont
  {Fahimniya}, \citenamefont {Watanabe}, \citenamefont {Taniguchi},
  \citenamefont {Koppens}, \citenamefont {Lischner}, \citenamefont {Levitov},\
  and\ \citenamefont {Efetov}}]{Stepanov2020}%
  \BibitemOpen
  \bibfield  {author} {\bibinfo {author} {\bibfnamefont {P.}~\bibnamefont
  {Stepanov}}, \bibinfo {author} {\bibfnamefont {I.}~\bibnamefont {Das}},
  \bibinfo {author} {\bibfnamefont {X.}~\bibnamefont {Lu}}, \bibinfo {author}
  {\bibfnamefont {A.}~\bibnamefont {Fahimniya}}, \bibinfo {author}
  {\bibfnamefont {K.}~\bibnamefont {Watanabe}}, \bibinfo {author}
  {\bibfnamefont {T.}~\bibnamefont {Taniguchi}}, \bibinfo {author}
  {\bibfnamefont {F.~H.~L.}\ \bibnamefont {Koppens}}, \bibinfo {author}
  {\bibfnamefont {J.}~\bibnamefont {Lischner}}, \bibinfo {author}
  {\bibfnamefont {L.}~\bibnamefont {Levitov}},\ and\ \bibinfo {author}
  {\bibfnamefont {D.~K.}\ \bibnamefont {Efetov}},\ }\bibfield  {title}
  {\bibinfo {title} {Untying the insulating and superconducting orders in
  magic-angle graphene},\ }\href {https://doi.org/10.1038/s41586-020-2459-6}
  {\bibfield  {journal} {\bibinfo  {journal} {Nature}\ }\textbf {\bibinfo
  {volume} {583}},\ \bibinfo {pages} {375} (\bibinfo {year}
  {2020})}\BibitemShut {NoStop}%
\bibitem [{\citenamefont {Saito}\ \emph {et~al.}(2020)\citenamefont {Saito},
  \citenamefont {Ge}, \citenamefont {Watanabe}, \citenamefont {Taniguchi},\
  and\ \citenamefont {Young}}]{Saito2020}%
  \BibitemOpen
  \bibfield  {author} {\bibinfo {author} {\bibfnamefont {Y.}~\bibnamefont
  {Saito}}, \bibinfo {author} {\bibfnamefont {J.}~\bibnamefont {Ge}}, \bibinfo
  {author} {\bibfnamefont {K.}~\bibnamefont {Watanabe}}, \bibinfo {author}
  {\bibfnamefont {T.}~\bibnamefont {Taniguchi}},\ and\ \bibinfo {author}
  {\bibfnamefont {A.~F.}\ \bibnamefont {Young}},\ }\bibfield  {title} {\bibinfo
  {title} {Independent superconductors and correlated insulators in twisted
  bilayer graphene},\ }\href {https://doi.org/10.1038/s41567-020-0928-3}
  {\bibfield  {journal} {\bibinfo  {journal} {Nat. Phys.}\ }\textbf {\bibinfo
  {volume} {16}},\ \bibinfo {pages} {926} (\bibinfo {year} {2020})}\BibitemShut
  {NoStop}%
\bibitem [{\citenamefont {Arora}\ \emph {et~al.}(2020)\citenamefont {Arora},
  \citenamefont {Polski}, \citenamefont {Zhang}, \citenamefont {Thomson},
  \citenamefont {Choi}, \citenamefont {Kim}, \citenamefont {Lin}, \citenamefont
  {Wilson}, \citenamefont {Xu}, \citenamefont {Chu}, \citenamefont {Watanabe},
  \citenamefont {Taniguchi}, \citenamefont {Alicea},\ and\ \citenamefont
  {Nadj-Perge}}]{Arora2020}%
  \BibitemOpen
  \bibfield  {author} {\bibinfo {author} {\bibfnamefont {H.~S.}\ \bibnamefont
  {Arora}}, \bibinfo {author} {\bibfnamefont {R.}~\bibnamefont {Polski}},
  \bibinfo {author} {\bibfnamefont {Y.}~\bibnamefont {Zhang}}, \bibinfo
  {author} {\bibfnamefont {A.}~\bibnamefont {Thomson}}, \bibinfo {author}
  {\bibfnamefont {Y.}~\bibnamefont {Choi}}, \bibinfo {author} {\bibfnamefont
  {H.}~\bibnamefont {Kim}}, \bibinfo {author} {\bibfnamefont {Z.}~\bibnamefont
  {Lin}}, \bibinfo {author} {\bibfnamefont {I.~Z.}\ \bibnamefont {Wilson}},
  \bibinfo {author} {\bibfnamefont {X.}~\bibnamefont {Xu}}, \bibinfo {author}
  {\bibfnamefont {J.-H.}\ \bibnamefont {Chu}}, \bibinfo {author} {\bibfnamefont
  {K.}~\bibnamefont {Watanabe}}, \bibinfo {author} {\bibfnamefont
  {T.}~\bibnamefont {Taniguchi}}, \bibinfo {author} {\bibfnamefont
  {J.}~\bibnamefont {Alicea}},\ and\ \bibinfo {author} {\bibfnamefont
  {S.}~\bibnamefont {Nadj-Perge}},\ }\bibfield  {title} {\bibinfo {title}
  {Superconductivity in metallic twisted bilayer graphene stabilized by
  {WSe}{$_2$}},\ }\href {https://doi.org/10.1038/s41586-020-2473-8} {\bibfield
  {journal} {\bibinfo  {journal} {Nature}\ }\textbf {\bibinfo {volume} {583}},\
  \bibinfo {pages} {379} (\bibinfo {year} {2020})}\BibitemShut {NoStop}%
\bibitem [{\citenamefont {Liu}\ \emph {et~al.}(2021)\citenamefont {Liu},
  \citenamefont {Wang}, \citenamefont {Watanabe}, \citenamefont {Taniguchi},
  \citenamefont {Vafek},\ and\ \citenamefont {Li}}]{Liu2021}%
  \BibitemOpen
  \bibfield  {author} {\bibinfo {author} {\bibfnamefont {X.}~\bibnamefont
  {Liu}}, \bibinfo {author} {\bibfnamefont {Z.}~\bibnamefont {Wang}}, \bibinfo
  {author} {\bibfnamefont {K.}~\bibnamefont {Watanabe}}, \bibinfo {author}
  {\bibfnamefont {T.}~\bibnamefont {Taniguchi}}, \bibinfo {author}
  {\bibfnamefont {O.}~\bibnamefont {Vafek}},\ and\ \bibinfo {author}
  {\bibfnamefont {J.~I.~A.}\ \bibnamefont {Li}},\ }\bibfield  {title} {\bibinfo
  {title} {Tuning electron correlation in magic-angle twisted bilayer graphene
  using {C}oulomb screening},\ }\href {https://doi.org/10.1126/science.abb8754}
  {\bibfield  {journal} {\bibinfo  {journal} {Science}\ }\textbf {\bibinfo
  {volume} {371}},\ \bibinfo {pages} {1261} (\bibinfo {year}
  {2021})}\BibitemShut {NoStop}%
\bibitem [{\citenamefont {Kontani}\ and\ \citenamefont
  {Ueda}(1998)}]{Kontani1998}%
  \BibitemOpen
  \bibfield  {author} {\bibinfo {author} {\bibfnamefont {H.}~\bibnamefont
  {Kontani}}\ and\ \bibinfo {author} {\bibfnamefont {K.}~\bibnamefont {Ueda}},\
  }\bibfield  {title} {\bibinfo {title} {Electronic {P}roperties of the
  {T}rellis-{L}attice {H}ubbard {M}odel: {P}seudogap and {S}uperconductivity},\
  }\href {https://doi.org/10.1103/physrevlett.80.5619} {\bibfield  {journal}
  {\bibinfo  {journal} {Phys. Rev. Lett.}\ }\textbf {\bibinfo {volume} {80}},\
  \bibinfo {pages} {5619} (\bibinfo {year} {1998})}\BibitemShut {NoStop}%
\bibitem [{\citenamefont {Koikegami}\ \emph {et~al.}(1997)\citenamefont
  {Koikegami}, \citenamefont {Fujimoto},\ and\ \citenamefont
  {Yamada}}]{Koikegami1997}%
  \BibitemOpen
  \bibfield  {author} {\bibinfo {author} {\bibfnamefont {S.}~\bibnamefont
  {Koikegami}}, \bibinfo {author} {\bibfnamefont {S.}~\bibnamefont
  {Fujimoto}},\ and\ \bibinfo {author} {\bibfnamefont {K.}~\bibnamefont
  {Yamada}},\ }\bibfield  {title} {\bibinfo {title} {Electronic {S}tructure and
  {T}ransition {T}emperature of the {$d$}-{$p$} {M}odel},\ }\href
  {https://doi.org/10.1143/jpsj.66.1438} {\bibfield  {journal} {\bibinfo
  {journal} {J. Phys. Soc. Jpn.}\ }\textbf {\bibinfo {volume} {66}},\ \bibinfo
  {pages} {1438} (\bibinfo {year} {1997})}\BibitemShut {NoStop}%
\bibitem [{\citenamefont {Li}\ \emph {et~al.}(2020)\citenamefont {Li},
  \citenamefont {Wallerberger}, \citenamefont {Chikano}, \citenamefont {Yeh},
  \citenamefont {Gull},\ and\ \citenamefont {Shinaoka}}]{Li2020}%
  \BibitemOpen
  \bibfield  {author} {\bibinfo {author} {\bibfnamefont {J.}~\bibnamefont
  {Li}}, \bibinfo {author} {\bibfnamefont {M.}~\bibnamefont {Wallerberger}},
  \bibinfo {author} {\bibfnamefont {N.}~\bibnamefont {Chikano}}, \bibinfo
  {author} {\bibfnamefont {C.-N.}\ \bibnamefont {Yeh}}, \bibinfo {author}
  {\bibfnamefont {E.}~\bibnamefont {Gull}},\ and\ \bibinfo {author}
  {\bibfnamefont {H.}~\bibnamefont {Shinaoka}},\ }\bibfield  {title} {\bibinfo
  {title} {Sparse sampling approach to efficient ab initio calculations at
  finite temperature},\ }\href {https://doi.org/10.1103/physrevb.101.035144}
  {\bibfield  {journal} {\bibinfo  {journal} {Phys. Rev. B}\ }\textbf {\bibinfo
  {volume} {101}},\ \bibinfo {pages} {035144} (\bibinfo {year} {2020})},\
  \Eprint {https://arxiv.org/abs/1908.07575} {arxiv:1908.07575} \BibitemShut
  {NoStop}%
\bibitem [{\citenamefont {Witt}\ \emph {et~al.}(2021)\citenamefont {Witt},
  \citenamefont {van Loon}, \citenamefont {Nomoto}, \citenamefont {Arita},\
  and\ \citenamefont {Wehling}}]{Witt2021}%
  \BibitemOpen
  \bibfield  {author} {\bibinfo {author} {\bibfnamefont {N.}~\bibnamefont
  {Witt}}, \bibinfo {author} {\bibfnamefont {E.~G. C.~P.}\ \bibnamefont {van
  Loon}}, \bibinfo {author} {\bibfnamefont {T.}~\bibnamefont {Nomoto}},
  \bibinfo {author} {\bibfnamefont {R.}~\bibnamefont {Arita}},\ and\ \bibinfo
  {author} {\bibfnamefont {T.~O.}\ \bibnamefont {Wehling}},\ }\bibfield
  {title} {\bibinfo {title} {Efficient fluctuation-exchange approach to
  low-temperature spin fluctuations and superconductivity: {F}rom the {H}ubbard
  model to {N}a{$_x$}{C}o{O}{$_2$}{$\cdotp$}{$y$}{H}{$_2$}{O}},\ }\href
  {https://doi.org/10.1103/physrevb.103.205148} {\bibfield  {journal} {\bibinfo
   {journal} {Phys. Rev. B}\ }\textbf {\bibinfo {volume} {103}},\ \bibinfo
  {pages} {205148} (\bibinfo {year} {2021})},\ \Eprint
  {https://arxiv.org/abs/2012.04562} {arXiv:2012.04562} \BibitemShut {NoStop}%
\bibitem [{\citenamefont {Shinaoka}\ \emph {et~al.}()\citenamefont {Shinaoka},
  \citenamefont {Chikano}, \citenamefont {Gull}, \citenamefont {Li},
  \citenamefont {Nomoto}, \citenamefont {Otsuki}, \citenamefont {Wallerberger},
  \citenamefont {Wang},\ and\ \citenamefont {Yoshimi}}]{Shinaoka2021}%
  \BibitemOpen
  \bibfield  {author} {\bibinfo {author} {\bibfnamefont {H.}~\bibnamefont
  {Shinaoka}}, \bibinfo {author} {\bibfnamefont {N.}~\bibnamefont {Chikano}},
  \bibinfo {author} {\bibfnamefont {E.}~\bibnamefont {Gull}}, \bibinfo {author}
  {\bibfnamefont {J.}~\bibnamefont {Li}}, \bibinfo {author} {\bibfnamefont
  {T.}~\bibnamefont {Nomoto}}, \bibinfo {author} {\bibfnamefont
  {J.}~\bibnamefont {Otsuki}}, \bibinfo {author} {\bibfnamefont
  {M.}~\bibnamefont {Wallerberger}}, \bibinfo {author} {\bibfnamefont
  {T.}~\bibnamefont {Wang}},\ and\ \bibinfo {author} {\bibfnamefont
  {K.}~\bibnamefont {Yoshimi}},\ }\bibfield  {title} {\bibinfo {title}
  {Efficient ab initio many-body calculations based on sparse modeling of
  {M}atsubara {G}reen's function},\ }\href@noop {} {\ }\Eprint
  {https://arxiv.org/abs/2106.12685} {2106.12685} \BibitemShut {NoStop}%
\bibitem [{\citenamefont {Shinaoka}\ \emph {et~al.}(2017)\citenamefont
  {Shinaoka}, \citenamefont {Otsuki}, \citenamefont {Ohzeki},\ and\
  \citenamefont {Yoshimi}}]{Shinaoka2017}%
  \BibitemOpen
  \bibfield  {author} {\bibinfo {author} {\bibfnamefont {H.}~\bibnamefont
  {Shinaoka}}, \bibinfo {author} {\bibfnamefont {J.}~\bibnamefont {Otsuki}},
  \bibinfo {author} {\bibfnamefont {M.}~\bibnamefont {Ohzeki}},\ and\ \bibinfo
  {author} {\bibfnamefont {K.}~\bibnamefont {Yoshimi}},\ }\bibfield  {title}
  {\bibinfo {title} {Compressing {G}reen's function using intermediate
  representation between imaginary-time and real-frequency domains},\ }\href
  {https://doi.org/10.1103/physrevb.96.035147} {\bibfield  {journal} {\bibinfo
  {journal} {Phys. Rev. B}\ }\textbf {\bibinfo {volume} {96}},\ \bibinfo
  {pages} {035147} (\bibinfo {year} {2017})},\ \Eprint
  {https://arxiv.org/abs/1702.03054} {arxiv:1702.03054} \BibitemShut {NoStop}%
\bibitem [{\citenamefont {Chikano}\ \emph {et~al.}(2019)\citenamefont
  {Chikano}, \citenamefont {Yoshimi}, \citenamefont {Otsuki},\ and\
  \citenamefont {Shinaoka}}]{Chikano2019}%
  \BibitemOpen
  \bibfield  {author} {\bibinfo {author} {\bibfnamefont {N.}~\bibnamefont
  {Chikano}}, \bibinfo {author} {\bibfnamefont {K.}~\bibnamefont {Yoshimi}},
  \bibinfo {author} {\bibfnamefont {J.}~\bibnamefont {Otsuki}},\ and\ \bibinfo
  {author} {\bibfnamefont {H.}~\bibnamefont {Shinaoka}},\ }\bibfield  {title}
  {\bibinfo {title} {irbasis: {O}pen-source database and software for
  intermediate-representation basis functions of imaginary-time {G}reen's
  function},\ }\href {https://doi.org/10.1016/j.cpc.2019.02.006} {\bibfield
  {journal} {\bibinfo  {journal} {Comput. Phys. Commun.}\ }\textbf {\bibinfo
  {volume} {240}},\ \bibinfo {pages} {181} (\bibinfo {year} {2019})},\ \Eprint
  {https://arxiv.org/abs/1807.05237} {arXiv:1807.05237} \BibitemShut {NoStop}%
\bibitem [{\citenamefont {Kuroki}\ and\ \citenamefont
  {Arita}(2001)}]{Kuroki2001}%
  \BibitemOpen
  \bibfield  {author} {\bibinfo {author} {\bibfnamefont {K.}~\bibnamefont
  {Kuroki}}\ and\ \bibinfo {author} {\bibfnamefont {R.}~\bibnamefont {Arita}},\
  }\bibfield  {title} {\bibinfo {title} {Spin-triplet superconductivity in
  repulsive {H}ubbard models with disconnected {F}ermi surfaces: {A} case study
  on triangular and honeycomb lattices},\ }\href
  {https://doi.org/10.1103/physrevb.63.174507} {\bibfield  {journal} {\bibinfo
  {journal} {Phys. Rev. B}\ }\textbf {\bibinfo {volume} {63}},\ \bibinfo
  {pages} {174507} (\bibinfo {year} {2001})}\BibitemShut {NoStop}%
\bibitem [{\citenamefont {Onari}\ \emph {et~al.}(2002)\citenamefont {Onari},
  \citenamefont {Kuroki}, \citenamefont {Arita},\ and\ \citenamefont
  {Aoki}}]{Onari2002}%
  \BibitemOpen
  \bibfield  {author} {\bibinfo {author} {\bibfnamefont {S.}~\bibnamefont
  {Onari}}, \bibinfo {author} {\bibfnamefont {K.}~\bibnamefont {Kuroki}},
  \bibinfo {author} {\bibfnamefont {R.}~\bibnamefont {Arita}},\ and\ \bibinfo
  {author} {\bibfnamefont {H.}~\bibnamefont {Aoki}},\ }\bibfield  {title}
  {\bibinfo {title} {Superconductivity induced by interband nesting in the
  three-dimensional honeycomb lattice},\ }\href
  {https://doi.org/10.1103/physrevb.65.184525} {\bibfield  {journal} {\bibinfo
  {journal} {Phys. Rev. B}\ }\textbf {\bibinfo {volume} {65}},\ \bibinfo
  {pages} {184525} (\bibinfo {year} {2002})}\BibitemShut {NoStop}%
\bibitem [{\citenamefont {Kuroki}(2010)}]{Kuroki2010}%
  \BibitemOpen
  \bibfield  {author} {\bibinfo {author} {\bibfnamefont {K.}~\bibnamefont
  {Kuroki}},\ }\bibfield  {title} {\bibinfo {title} {Spin-fluctuation-mediated
  {$d+id^{\prime}$} pairing mechanism in doped
  {$\beta$}-{MNCl}({M}={H}f,{Z}r)superconductors},\ }\href
  {https://doi.org/10.1103/physrevb.81.104502} {\bibfield  {journal} {\bibinfo
  {journal} {Phys. Rev. B}\ }\textbf {\bibinfo {volume} {81}},\ \bibinfo
  {pages} {104502} (\bibinfo {year} {2010})}\BibitemShut {NoStop}%
\bibitem [{\citenamefont {Sigrist}\ and\ \citenamefont
  {Ueda}(1991)}]{Sigrist1991}%
  \BibitemOpen
  \bibfield  {author} {\bibinfo {author} {\bibfnamefont {M.}~\bibnamefont
  {Sigrist}}\ and\ \bibinfo {author} {\bibfnamefont {K.}~\bibnamefont {Ueda}},\
  }\bibfield  {title} {\bibinfo {title} {Phenomenological theory of
  unconventional superconductivity},\ }\href
  {https://doi.org/10.1103/revmodphys.63.239} {\bibfield  {journal} {\bibinfo
  {journal} {Rev. Mod. Phys.}\ }\textbf {\bibinfo {volume} {63}},\ \bibinfo
  {pages} {239} (\bibinfo {year} {1991})}\BibitemShut {NoStop}%
\bibitem [{\citenamefont {Mermin}\ and\ \citenamefont
  {Wagner}(1966)}]{Mermin1966}%
  \BibitemOpen
  \bibfield  {author} {\bibinfo {author} {\bibfnamefont {N.~D.}\ \bibnamefont
  {Mermin}}\ and\ \bibinfo {author} {\bibfnamefont {H.}~\bibnamefont
  {Wagner}},\ }\bibfield  {title} {\bibinfo {title} {Absence of
  {F}erromagnetism or {A}ntiferromagnetism in {O}ne- or {T}wo-{D}imensional
  {I}sotropic {H}eisenberg {M}odels},\ }\href
  {https://doi.org/10.1103/physrevlett.17.1307} {\bibfield  {journal} {\bibinfo
   {journal} {Phys. Rev. Lett.}\ }\textbf {\bibinfo {volume} {17}},\ \bibinfo
  {pages} {1307} (\bibinfo {year} {1966})}\BibitemShut {NoStop}%
\bibitem [{\citenamefont {Kontani}\ and\ \citenamefont
  {Ohno}(2006)}]{Kontani2006}%
  \BibitemOpen
  \bibfield  {author} {\bibinfo {author} {\bibfnamefont {H.}~\bibnamefont
  {Kontani}}\ and\ \bibinfo {author} {\bibfnamefont {M.}~\bibnamefont {Ohno}},\
  }\bibfield  {title} {\bibinfo {title} {Effect of a nonmagnetic impurity in a
  nearly antiferromagnetic {F}ermi liquid: {M}agnetic correlations and
  transport phenomena},\ }\href {https://doi.org/10.1103/physrevb.74.014406}
  {\bibfield  {journal} {\bibinfo  {journal} {Phys. Rev. B}\ }\textbf {\bibinfo
  {volume} {74}},\ \bibinfo {pages} {014406} (\bibinfo {year}
  {2006})}\BibitemShut {NoStop}%
\bibitem [{\citenamefont {Kino}\ and\ \citenamefont
  {Kontani}(1998)}]{Kino1998}%
  \BibitemOpen
  \bibfield  {author} {\bibinfo {author} {\bibfnamefont {H.}~\bibnamefont
  {Kino}}\ and\ \bibinfo {author} {\bibfnamefont {H.}~\bibnamefont {Kontani}},\
  }\bibfield  {title} {\bibinfo {title} {Phase {D}iagram of {S}uperconductivity
  on the {A}nisotropic {T}riangular {L}attice {H}ubbard {M}odel: {A}n
  {E}ffective {M}odel of {$\kappa$}-({BEDT}-{TTF}) {S}alts},\ }\href
  {https://doi.org/10.1143/jpsj.67.3691} {\bibfield  {journal} {\bibinfo
  {journal} {J. Phys. Soc. Jpn.}\ }\textbf {\bibinfo {volume} {67}},\ \bibinfo
  {pages} {3691} (\bibinfo {year} {1998})},\ \Eprint
  {https://arxiv.org/abs/cond-mat/9807147} {arxiv:cond-mat/9807147}
  \BibitemShut {NoStop}%
\bibitem [{\citenamefont {Kitatani}\ \emph {et~al.}(2015)\citenamefont
  {Kitatani}, \citenamefont {Tsuji},\ and\ \citenamefont
  {Aoki}}]{Kitatani2015}%
  \BibitemOpen
  \bibfield  {author} {\bibinfo {author} {\bibfnamefont {M.}~\bibnamefont
  {Kitatani}}, \bibinfo {author} {\bibfnamefont {N.}~\bibnamefont {Tsuji}},\
  and\ \bibinfo {author} {\bibfnamefont {H.}~\bibnamefont {Aoki}},\ }\bibfield
  {title} {\bibinfo {title} {{FLEX}+{DMFT} approach to the {$d$}-wave
  superconducting phase diagram of the two-dimensional {H}ubbard model},\
  }\href {https://doi.org/10.1103/physrevb.92.085104} {\bibfield  {journal}
  {\bibinfo  {journal} {Phys. Rev. B}\ }\textbf {\bibinfo {volume} {92}},\
  \bibinfo {pages} {085104} (\bibinfo {year} {2015})},\ \Eprint
  {https://arxiv.org/abs/1505.04865} {arxiv:1505.04865} \BibitemShut {NoStop}%
\bibitem [{\citenamefont {Wang}\ \emph {et~al.}(2012)\citenamefont {Wang},
  \citenamefont {Xiang}, \citenamefont {Wang}, \citenamefont {Wang},
  \citenamefont {Yang},\ and\ \citenamefont {Lee}}]{Wang2012}%
  \BibitemOpen
  \bibfield  {author} {\bibinfo {author} {\bibfnamefont {W.-S.}\ \bibnamefont
  {Wang}}, \bibinfo {author} {\bibfnamefont {Y.-Y.}\ \bibnamefont {Xiang}},
  \bibinfo {author} {\bibfnamefont {Q.-H.}\ \bibnamefont {Wang}}, \bibinfo
  {author} {\bibfnamefont {F.}~\bibnamefont {Wang}}, \bibinfo {author}
  {\bibfnamefont {F.}~\bibnamefont {Yang}},\ and\ \bibinfo {author}
  {\bibfnamefont {D.-H.}\ \bibnamefont {Lee}},\ }\bibfield  {title} {\bibinfo
  {title} {Functional renormalization group and variational {M}onte {C}arlo
  studies of the electronic instabilities in graphene near {$\frac{1}{4}$}
  doping},\ }\href {https://doi.org/10.1103/physrevb.85.035414} {\bibfield
  {journal} {\bibinfo  {journal} {Phys. Rev. B}\ }\textbf {\bibinfo {volume}
  {85}},\ \bibinfo {pages} {035414} (\bibinfo {year} {2012})}\BibitemShut
  {NoStop}%
\bibitem [{\citenamefont {Jiang}\ \emph {et~al.}(2014)\citenamefont {Jiang},
  \citenamefont {Mesaros},\ and\ \citenamefont {Ran}}]{Jiang2014}%
  \BibitemOpen
  \bibfield  {author} {\bibinfo {author} {\bibfnamefont {S.}~\bibnamefont
  {Jiang}}, \bibinfo {author} {\bibfnamefont {A.}~\bibnamefont {Mesaros}},\
  and\ \bibinfo {author} {\bibfnamefont {Y.}~\bibnamefont {Ran}},\ }\bibfield
  {title} {\bibinfo {title} {Chiral {S}pin-{D}ensity {W}ave,
  {S}pin-{C}harge-{C}hern {L}iquid, and {$d+id$} {S}uperconductivity in
  {1/4}-{D}oped {C}orrelated {E}lectronic {S}ystems on the {H}oneycomb
  {L}attice},\ }\href {https://doi.org/10.1103/physrevx.4.031040} {\bibfield
  {journal} {\bibinfo  {journal} {Phys. Rev. X}\ }\textbf {\bibinfo {volume}
  {4}},\ \bibinfo {pages} {031040} (\bibinfo {year} {2014})}\BibitemShut
  {NoStop}%
\bibitem [{\citenamefont {Kuznetsova}\ and\ \citenamefont
  {Barzykin}(2005)}]{Kuznetsova2005}%
  \BibitemOpen
  \bibfield  {author} {\bibinfo {author} {\bibfnamefont {Z.}~\bibnamefont
  {Kuznetsova}}\ and\ \bibinfo {author} {\bibfnamefont {V.}~\bibnamefont
  {Barzykin}},\ }\bibfield  {title} {\bibinfo {title} {Pairing state in
  multicomponent superconductors},\ }\href
  {https://doi.org/10.1209/epl/i2005-10242-8} {\bibfield  {journal} {\bibinfo
  {journal} {Europhys. Lett.}\ }\textbf {\bibinfo {volume} {72}},\ \bibinfo
  {pages} {437} (\bibinfo {year} {2005})}\BibitemShut {NoStop}%
\bibitem [{\citenamefont {Nandkishore}\ \emph {et~al.}(2012)\citenamefont
  {Nandkishore}, \citenamefont {Levitov},\ and\ \citenamefont
  {Chubukov}}]{Nandkishore2012}%
  \BibitemOpen
  \bibfield  {author} {\bibinfo {author} {\bibfnamefont {R.}~\bibnamefont
  {Nandkishore}}, \bibinfo {author} {\bibfnamefont {L.~S.}\ \bibnamefont
  {Levitov}},\ and\ \bibinfo {author} {\bibfnamefont {A.~V.}\ \bibnamefont
  {Chubukov}},\ }\bibfield  {title} {\bibinfo {title} {Chiral superconductivity
  from repulsive interactions in doped graphene},\ }\href
  {https://doi.org/10.1038/nphys2208} {\bibfield  {journal} {\bibinfo
  {journal} {Nat. Phys.}\ }\textbf {\bibinfo {volume} {8}},\ \bibinfo {pages}
  {158} (\bibinfo {year} {2012})}\BibitemShut {NoStop}%
\bibitem [{\citenamefont {Black-Schaffer}\ and\ \citenamefont
  {Honerkamp}(2014)}]{BlackSchaffer2014}%
  \BibitemOpen
  \bibfield  {author} {\bibinfo {author} {\bibfnamefont {A.~M.}\ \bibnamefont
  {Black-Schaffer}}\ and\ \bibinfo {author} {\bibfnamefont {C.}~\bibnamefont
  {Honerkamp}},\ }\bibfield  {title} {\bibinfo {title} {Chiral {$d$}-wave
  superconductivity in doped graphene},\ }\href
  {https://doi.org/10.1088/0953-8984/26/42/423201} {\bibfield  {journal}
  {\bibinfo  {journal} {J. Phys.: Condens. Matter}\ }\textbf {\bibinfo {volume}
  {26}},\ \bibinfo {pages} {423201} (\bibinfo {year} {2014})},\ \Eprint
  {https://arxiv.org/abs/1406.0101} {arxiv:1406.0101} \BibitemShut {NoStop}%
\bibitem [{\citenamefont {Harrison}(2004)}]{Harrison2004}%
  \BibitemOpen
  \bibfield  {author} {\bibinfo {author} {\bibfnamefont {W.~A.}\ \bibnamefont
  {Harrison}},\ }\href {https://doi.org/10.1142/5432} {\emph {\bibinfo {title}
  {Elementary {E}lectronic {S}tructure}}},\ \bibinfo {edition} {rev.}\ ed.\
  (\bibinfo  {publisher} {World Scientific},\ \bibinfo {address} {Singapore},\
  \bibinfo {year} {2004})\BibitemShut {NoStop}%
\bibitem [{\citenamefont {Allen}\ and\ \citenamefont
  {Dynes}(1975)}]{PhysRevB.12.905}%
  \BibitemOpen
  \bibfield  {author} {\bibinfo {author} {\bibfnamefont {P.~B.}\ \bibnamefont
  {Allen}}\ and\ \bibinfo {author} {\bibfnamefont {R.~C.}\ \bibnamefont
  {Dynes}},\ }\bibfield  {title} {\bibinfo {title} {Transition temperature of
  strong-coupled superconductors reanalyzed},\ }\href
  {https://doi.org/10.1103/PhysRevB.12.905} {\bibfield  {journal} {\bibinfo
  {journal} {Phys. Rev. B}\ }\textbf {\bibinfo {volume} {12}},\ \bibinfo
  {pages} {905} (\bibinfo {year} {1975})}\BibitemShut {NoStop}%
\bibitem [{\citenamefont {McMillan}(1968)}]{PhysRev.167.331}%
  \BibitemOpen
  \bibfield  {author} {\bibinfo {author} {\bibfnamefont {W.~L.}\ \bibnamefont
  {McMillan}},\ }\bibfield  {title} {\bibinfo {title} {Transition {T}emperature
  of {S}trong-{C}oupled {S}uperconductors},\ }\href
  {https://doi.org/10.1103/PhysRevB.12.905} {\bibfield  {journal} {\bibinfo
  {journal} {Phys. Rev.}\ }\textbf {\bibinfo {volume} {167}},\ \bibinfo {pages}
  {331} (\bibinfo {year} {1968})}\BibitemShut {NoStop}%
\bibitem [{\citenamefont {Rösner}\ \emph {et~al.}(2014)\citenamefont
  {Rösner}, \citenamefont {Haas},\ and\ \citenamefont
  {Wehling}}]{Roesner2014}%
  \BibitemOpen
  \bibfield  {author} {\bibinfo {author} {\bibfnamefont {M.}~\bibnamefont
  {Rösner}}, \bibinfo {author} {\bibfnamefont {S.}~\bibnamefont {Haas}},\ and\
  \bibinfo {author} {\bibfnamefont {T.~O.}\ \bibnamefont {Wehling}},\
  }\bibfield  {title} {\bibinfo {title} {Phase diagram of electron-doped
  dichalcogenides},\ }\href {https://doi.org/10.1103/physrevb.90.245105}
  {\bibfield  {journal} {\bibinfo  {journal} {Phys. Rev. B}\ }\textbf {\bibinfo
  {volume} {90}},\ \bibinfo {pages} {245105} (\bibinfo {year}
  {2014})}\BibitemShut {NoStop}%
\bibitem [{\citenamefont {Schönhoff}\ \emph {et~al.}(2016)\citenamefont
  {Schönhoff}, \citenamefont {Rösner}, \citenamefont {Groenewald},
  \citenamefont {Haas},\ and\ \citenamefont {Wehling}}]{Schoenhoff2016}%
  \BibitemOpen
  \bibfield  {author} {\bibinfo {author} {\bibfnamefont {G.}~\bibnamefont
  {Schönhoff}}, \bibinfo {author} {\bibfnamefont {M.}~\bibnamefont {Rösner}},
  \bibinfo {author} {\bibfnamefont {R.~E.}\ \bibnamefont {Groenewald}},
  \bibinfo {author} {\bibfnamefont {S.}~\bibnamefont {Haas}},\ and\ \bibinfo
  {author} {\bibfnamefont {T.~O.}\ \bibnamefont {Wehling}},\ }\bibfield
  {title} {\bibinfo {title} {Interplay of screening and superconductivity in
  low-dimensional materials},\ }\href
  {https://doi.org/10.1103/physrevb.94.134504} {\bibfield  {journal} {\bibinfo
  {journal} {Phys. Rev. B}\ }\textbf {\bibinfo {volume} {94}},\ \bibinfo
  {pages} {134504} (\bibinfo {year} {2016})}\BibitemShut {NoStop}%
\bibitem [{\citenamefont {Mahan}(2000)}]{Mahan2000}%
  \BibitemOpen
  \bibfield  {author} {\bibinfo {author} {\bibfnamefont {G.~D.}\ \bibnamefont
  {Mahan}},\ }\href {https://doi.org/10.1007/978-1-4757-5714-9} {\emph
  {\bibinfo {title} {Many-{P}article {P}hysics}}},\ \bibinfo {edition} {3rd}\
  ed.\ (\bibinfo  {publisher} {Springer {US}},\ \bibinfo {year}
  {2000})\BibitemShut {NoStop}%
\bibitem [{\citenamefont {Feldman}(1976)}]{Feldman1976}%
  \BibitemOpen
  \bibfield  {author} {\bibinfo {author} {\bibfnamefont {J.~L.}\ \bibnamefont
  {Feldman}},\ }\bibfield  {title} {\bibinfo {title} {Elastic constants of
  2{H}-{MoS}{$_2$} and 2{H}-{NbSe}{$_2$} extracted from measured dispersion
  curves and linear compressibilities},\ }\href
  {https://doi.org/10.1016/0022-3697(76)90143-8} {\bibfield  {journal}
  {\bibinfo  {journal} {J. Phys. Chem. Solids}\ }\textbf {\bibinfo {volume}
  {37}},\ \bibinfo {pages} {1141} (\bibinfo {year} {1976})}\BibitemShut
  {NoStop}%
\bibitem [{\citenamefont {Zhao}\ \emph {et~al.}(2013)\citenamefont {Zhao},
  \citenamefont {Luo}, \citenamefont {Li}, \citenamefont {Zhang}, \citenamefont
  {Araujo}, \citenamefont {Gan}, \citenamefont {Wu}, \citenamefont {Zhang},
  \citenamefont {Quek}, \citenamefont {Dresselhaus},\ and\ \citenamefont
  {Xiong}}]{Zhao2013}%
  \BibitemOpen
  \bibfield  {author} {\bibinfo {author} {\bibfnamefont {Y.}~\bibnamefont
  {Zhao}}, \bibinfo {author} {\bibfnamefont {X.}~\bibnamefont {Luo}}, \bibinfo
  {author} {\bibfnamefont {H.}~\bibnamefont {Li}}, \bibinfo {author}
  {\bibfnamefont {J.}~\bibnamefont {Zhang}}, \bibinfo {author} {\bibfnamefont
  {P.~T.}\ \bibnamefont {Araujo}}, \bibinfo {author} {\bibfnamefont {C.~K.}\
  \bibnamefont {Gan}}, \bibinfo {author} {\bibfnamefont {J.}~\bibnamefont
  {Wu}}, \bibinfo {author} {\bibfnamefont {H.}~\bibnamefont {Zhang}}, \bibinfo
  {author} {\bibfnamefont {S.~Y.}\ \bibnamefont {Quek}}, \bibinfo {author}
  {\bibfnamefont {M.~S.}\ \bibnamefont {Dresselhaus}},\ and\ \bibinfo {author}
  {\bibfnamefont {Q.}~\bibnamefont {Xiong}},\ }\bibfield  {title} {\bibinfo
  {title} {Interlayer {B}reathing and {S}hear {M}odes in {F}ew-{T}rilayer
  {MoS}{$_2$} and {WSe}{$_2$}},\ }\href {https://doi.org/10.1021/nl304169w}
  {\bibfield  {journal} {\bibinfo  {journal} {Nano Lett.}\ }\textbf {\bibinfo
  {volume} {13}},\ \bibinfo {pages} {1007} (\bibinfo {year}
  {2013})}\BibitemShut {NoStop}%
\bibitem [{\citenamefont {Landau}\ and\ \citenamefont
  {Lifshitz}(1970)}]{Landau1970}%
  \BibitemOpen
  \bibfield  {author} {\bibinfo {author} {\bibfnamefont {L.~D.}\ \bibnamefont
  {Landau}}\ and\ \bibinfo {author} {\bibfnamefont {E.~M.}\ \bibnamefont
  {Lifshitz}},\ }\href@noop {} {\emph {\bibinfo {title} {Theory of
  {E}lasticity}}},\ \bibinfo {edition} {2nd}\ ed.,\ Vol.~\bibinfo {volume} {7}\
  (\bibinfo  {publisher} {Oxford: Pergamon},\ \bibinfo {year}
  {1970})\BibitemShut {NoStop}%
\bibitem [{\citenamefont {{\c{C}}ak{\i}r}\ \emph {et~al.}(2014)\citenamefont
  {{\c{C}}ak{\i}r}, \citenamefont {Peeters},\ and\ \citenamefont
  {Sevik}}]{Cakir2014}%
  \BibitemOpen
  \bibfield  {author} {\bibinfo {author} {\bibfnamefont {D.}~\bibnamefont
  {{\c{C}}ak{\i}r}}, \bibinfo {author} {\bibfnamefont {F.~M.}\ \bibnamefont
  {Peeters}},\ and\ \bibinfo {author} {\bibfnamefont {C.}~\bibnamefont
  {Sevik}},\ }\bibfield  {title} {\bibinfo {title} {Mechanical and thermal
  properties of h-{MX}{$_2$} ({M}={C}r, {M}o, {W}; {X}={O}, {S}, {S}e, {T}e)
  monolayers: {A} comparative study},\ }\href
  {https://doi.org/10.1063/1.4879543} {\bibfield  {journal} {\bibinfo
  {journal} {Appl. Phys. Lett.}\ }\textbf {\bibinfo {volume} {104}},\ \bibinfo
  {pages} {203110} (\bibinfo {year} {2014})}\BibitemShut {NoStop}%
\bibitem [{\citenamefont {Liu}\ \emph {et~al.}(2014)\citenamefont {Liu},
  \citenamefont {Yan}, \citenamefont {Chen}, \citenamefont {Fan}, \citenamefont
  {Sun}, \citenamefont {Suh}, \citenamefont {Fu}, \citenamefont {Lee},
  \citenamefont {Zhou}, \citenamefont {Tongay}, \citenamefont {Ji},
  \citenamefont {Neaton},\ and\ \citenamefont {Wu}}]{Liu2014}%
  \BibitemOpen
  \bibfield  {author} {\bibinfo {author} {\bibfnamefont {K.}~\bibnamefont
  {Liu}}, \bibinfo {author} {\bibfnamefont {Q.}~\bibnamefont {Yan}}, \bibinfo
  {author} {\bibfnamefont {M.}~\bibnamefont {Chen}}, \bibinfo {author}
  {\bibfnamefont {W.}~\bibnamefont {Fan}}, \bibinfo {author} {\bibfnamefont
  {Y.}~\bibnamefont {Sun}}, \bibinfo {author} {\bibfnamefont {J.}~\bibnamefont
  {Suh}}, \bibinfo {author} {\bibfnamefont {D.}~\bibnamefont {Fu}}, \bibinfo
  {author} {\bibfnamefont {S.}~\bibnamefont {Lee}}, \bibinfo {author}
  {\bibfnamefont {J.}~\bibnamefont {Zhou}}, \bibinfo {author} {\bibfnamefont
  {S.}~\bibnamefont {Tongay}}, \bibinfo {author} {\bibfnamefont
  {J.}~\bibnamefont {Ji}}, \bibinfo {author} {\bibfnamefont {J.~B.}\
  \bibnamefont {Neaton}},\ and\ \bibinfo {author} {\bibfnamefont
  {J.}~\bibnamefont {Wu}},\ }\bibfield  {title} {\bibinfo {title} {Elastic
  {P}roperties of {C}hemical-{V}apor-{D}eposited {M}onolayer {MoS}{$_2$},
  {WS}{$_2$}, and {T}heir {B}ilayer {H}eterostructures},\ }\href
  {https://doi.org/10.1021/nl501793a} {\bibfield  {journal} {\bibinfo
  {journal} {Nano Lett.}\ }\textbf {\bibinfo {volume} {14}},\ \bibinfo {pages}
  {5097} (\bibinfo {year} {2014})}\BibitemShut {NoStop}%
\end{thebibliography}%
\end{document}

\title{Supplemental Material: Doping fingerprints of spin and lattice fluctuations in moir\'{e} superlattice systems}

\author{Niklas Witt}
\email{niklas.witt@physik.uni-hamburg.de}
\affiliation{Institute of Theoretical Physics, Bremen Center for Computational Materials Science,
	and MAPEX Center for Materials and Processes, University of Bremen, Otto-Hahn-Allee 1, 28359 Bremen, Germany}
\affiliation{I. Institute of Theoretical Physics, University of Hamburg, Notkestraße 9, 22607 Hamburg, Germany}

\author{José M. Pizarro}
\email{jose.pizarro@mpsd.mpg.de}
\affiliation{Institute of Theoretical Physics, Bremen Center for Computational Materials Science,
	and MAPEX Center for Materials and Processes, University of Bremen, Otto-Hahn-Allee 1, 28359 Bremen, Germany}
\affiliation{Max Planck Institute for the Structure and Dynamics of Matter, Luruper Chaussee 149, 22671 Hamburg, Germany}

\author{Jan Berges}
\affiliation{Institute of Theoretical Physics, Bremen Center for Computational Materials Science,
	and MAPEX Center for Materials and Processes, University of Bremen, Otto-Hahn-Allee 1, 28359 Bremen, Germany}

\author{Takuya Nomoto}
\affiliation{Department of Applied Physics, The University of Tokyo, 7-3-1 Hongo, Bunkyo-ku, Tokyo 113-8656, Japan}

\author{Ryotaro Arita}
\affiliation{Department of Applied Physics, The University of Tokyo, 7-3-1 Hongo, Bunkyo-ku, Tokyo 113-8656, Japan}
\affiliation{RIKEN Center for Emergent Matter Science, 2-1 Hirosawa, Wako, Saitama 351-0198, Japan}

\author{Tim O. Wehling}
%\email{twehling@uni-bremen.de}
\affiliation{Institute of Theoretical Physics, Bremen Center for Computational Materials Science,
	and MAPEX Center for Materials and Processes, University of Bremen, Otto-Hahn-Allee 1, 28359 Bremen, Germany}
\affiliation{I. Institute of Theoretical Physics, University of Hamburg, Notkestraße 9, 22607 Hamburg, Germany}

\maketitle

\onecolumngrid
\setcounter{table}{0}
\setcounter{section}{0}
\setcounter{figure}{0}
\setcounter{equation}{0}
\renewcommand{\thepage}{\Roman{page}}
\renewcommand{\thesection}{S\arabic{section}}
\renewcommand{\thetable}{S\arabic{table}}
\renewcommand{\thefigure}{S\arabic{figure}}
\renewcommand{\theequation}{S\arabic{equation}}
%\cleardoublepage
%\vfill\eject
%\thispagestyle{empty}
%\phantom{hi}\vfill\eject
%\section{Supplemental Material}

\renewcommand*{\citenumfont}[1]{S#1}
\renewcommand*{\bibnumfmt}[1]{[S#1]}

In Section \ref{sec:continuum_model} of this Supplemental Material, we give a short description of the low-energy continuum model developed in Ref.~\cite{gammaTMDCsMacdonald2021} for $\Gamma$-valley twisted transition metal dichalcogenides (TMDCs) which we employed here. We show the band structure for all $\Gamma$-valley twisted TMDCs, WS$_2$, MoS$_2$, and MoSe$_2$, at different twist angles $\theta$. 
In Section \ref{sec:wannier_projection}, we present the Wannier construction for the two top-most superlattice valence bands in a twist angle range of $1^{\circ} < \theta < 5^{\circ}$. In Section \ref{sec:long_range_Coulomb}, we discuss the effect of the long-range Coulomb interactions on the low-energy flat bands in presence of doping in the Hartree approximation. We follow the procedure from Refs. \cite{PhysRevB.100.205113,PhysRevB.102.045107} for magic-angle twisted bilayer graphene (MATBG). Section \ref{sec:Hubbard_interaction} gives estimations of the of the electronic Coulomb interaction strength, in particular the on-site and nearest-neighbor interaction parameters $U$ and $V$.
In Section \ref{sec:FLEX_details}, we explain the calculations in the fluctuation exchange (FLEX) approximation \cite{Bickers1989a,Bickers1989b}.
In Section \ref{sec:spin_fluctuation}, we discuss the nature of magnetic ordering, analyze the momentum dependence of the static spin susceptibility at different dopings $\delta$ and temperatures $T/t$, and show its real space profile.
In Section \ref{sec:sc_instability}, we investigate the leading superconducting order parameter and possible pairing symmetries in the honeycomb Hubbard model at various dopings and temperatures.
Section \ref{sec:asymmetry_influence_SC} discusses the influence of longer-ranged hopping terms on spin fluctuations and superconductivity. In Section \ref{sec:nonlocal} we investigate the influence of non-local electron-phonon coupling as well as dispersive phonon frequencies on the doping dependence of phonon-mediated superconductivity.
In Section \ref{sec:phonon_estimation} we provide an estimation of the effective electron-phonon interaction $U_{\mathrm{eff}}$ used for the Holstein model calculations in the main text and for the non-local Peierls coupling employed in Section \ref{sec:nonlocal}.

\section{Low-energy continuum model for $\Gamma$-valley twisted TMDCs}\label{sec:continuum_model}
We outline here the low-energy continuum model which we used for the description of the moir\'e valence band structure of $\Gamma$-valley twisted TMDCs. The model was introduced in Ref.~\cite{gammaTMDCsMacdonald2021}, from where we outline here the main points. In this continuum model, only the valence antibonding state at the $\Gamma$ point is considered, which is isolated from other bands by hundreds of meV because of the interlayer coupling. Since the bands around the $\Gamma$-point are mainly of transition metal $d_{z^2}$-character, spin-orbit coupling effects are small and can be neglected. Because of this, the description of these $\Gamma$-valley twisted TMDCs is easier than other TMDC systems like homobilayer WSe$_2$, where the valence band maximum is at the $K$-point with strong spin-orbit coupling \cite{PhysRevLett.108.196802}. The low-energy Hamiltonian of the continuum model can be written as
\begin{equation}
H = -\frac{\hbar^2 k^2}{2 m^{*}} + V_{\mathrm{M}} (\bold{r}),
\label{eqSM1}
\end{equation}
where $m^{*}$ is the effective mass and $V_{\mathrm{M}} (\bold{r})$ is the moir\'{e} potential felt by the holes at the valence band maximum in $\Gamma$. The moir\'{e} potential has the following expression in real space:
\begin{equation}
V_{\mathrm{M}} (\bold{r}) = \sum_{s=1}^3 \sum_{j=1}^{6} V^s \mathrm{e}^{i (\bold{g}_{j}^s \cdot \bold{r} + \phi^s)}.
\label{eqSM2}
\end{equation}
Here, $s$ is the $s$-th shell of six moir\'{e} reciprocal lattice vectors  $\bold{g}_j^s = \mathcal{R}_{(j-1)\pi/3} \bold{G}^s$ (\(j=1,\ldots,6\)) with $\mathcal{R}_{\alpha}$ being the two-dimensional (2D) rotation matrix about an angle $\alpha$. We choose reciprocal lattice vectors pointing to the $s$-th shell as $\bold{G}^1=\bold{G}_2^{\mathrm{M}}$, $\bold{G}^2=\bold{G}_1^{\mathrm{M}}+\bold{G}_2^{\mathrm{M}}$, and $\bold{G}^3=2\bold{G}_2^{\mathrm{M}}$, where $\bold{G}_{1,2}^{\mathrm{M}}$ span the reciprocal moir\'{e} lattice. The phase factors $\phi^s$ are constrained by the $C_{6z}$ symmetry of the moir\'{e} lattice to be either $0$ or $\pi$.

\begin{table}
\label{tableSM1}
\begin{center}
\caption{Continuum model parameters for $\Gamma$-valley twisted TMDCs. $a_0$ is the lattice constant in $\text{\AA}$, $m^{*}$ is the effective  mass in bare electron mass units, and $V^s$ are in meV. Data taken from Ref.~\cite{gammaTMDCsMacdonald2021}.}
%\begin{ruledtabular}
\begin{tabular}{ p{4em}*{3}{p{3em}} } 
	\toprule
   & WS$_2$ & MoS$_2$ & MoSe$_2$ \\ 
 \colrule
 $a_0$ & 3.18 & 3.182 & 3.295 \\ 
 $m^{*}$ & 0.87 & 0.9 & 1.17 \\ 
 $V^{1}$ & 33.5 & 39.45 & 36.8 \\ 
 $V^{2}$ & 4.0 & 6.5 & 8.4 \\ 
 $V^{3}$ & 5.5 & 10.0 & 10.2 \\ 
 $\phi^{1,2,3}$ & $\pi$ & $\pi$ & $\pi$ \\ 
 \botrule
\end{tabular}
%\end{ruledtabular}
\end{center}

\end{table}

The continuum model parameters $(m^{*},V^s,\phi^s)$ were obtained from the \emph{ab initio} calculation of the fully relaxed twisted bilayers, and they are given in Table \ref{tableSM1} for the different $\Gamma$-valley twisted TMDCs, WS$_2$, MoS$_2$, and MoSe$_2$ \cite{gammaTMDCsMacdonald2021}. The maximum of the moir\'{e} potential of Eq.~(\ref{eqSM2}) felt by the holes in the valence band maximum is found in the AB/BA regions (see Fig.~1B of Ref.~[\onlinecite{gammaTMDCsMacdonald2021}]), so that the low-energy physics of the $\Gamma$-valley twisted TMDCs is controlled by orbitals sitting in the honeycomb AB/BA regions.

\begin{figure}
	\centering
	\includegraphics[width=0.95\textwidth]{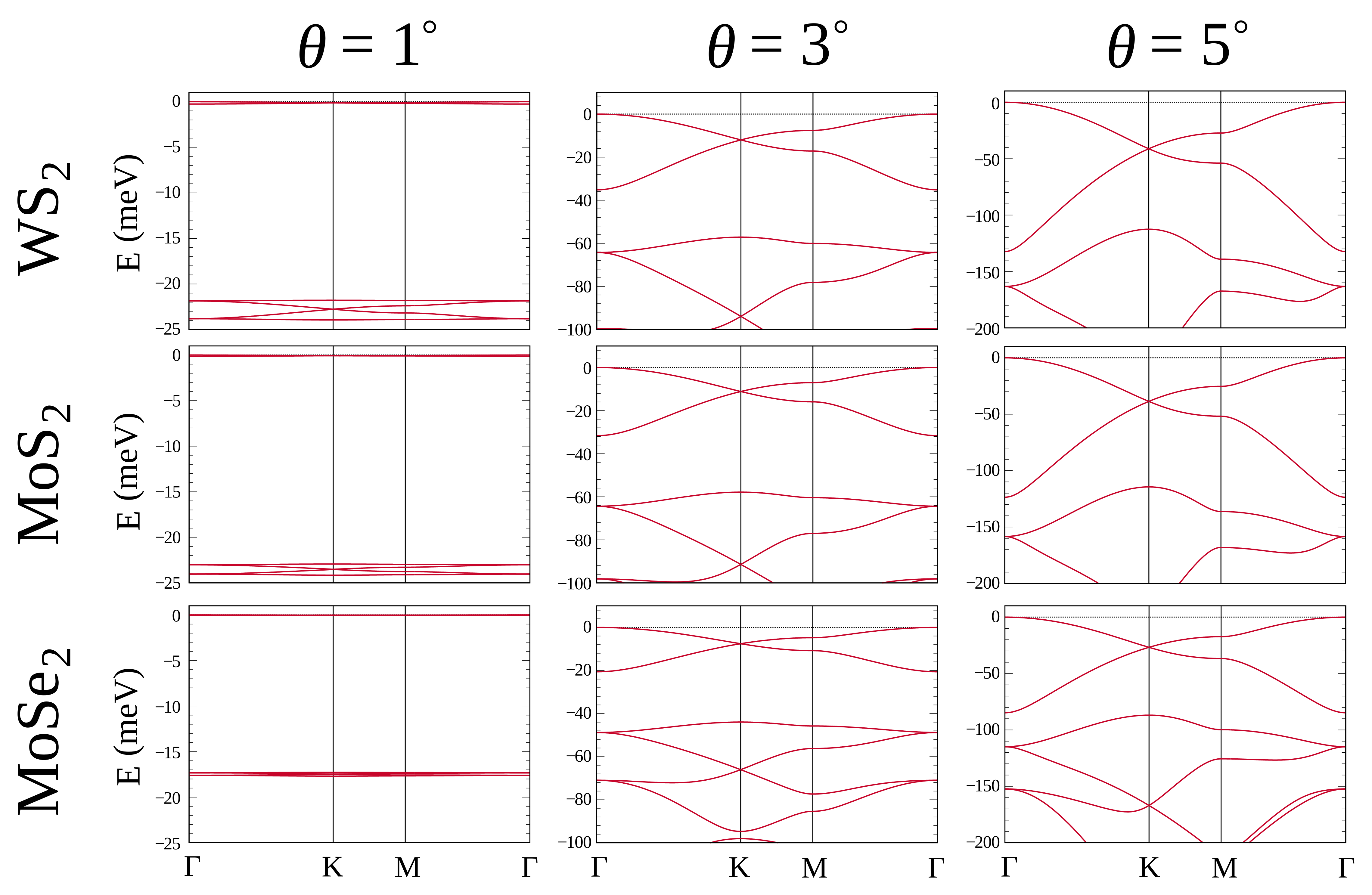}
	\caption{(Color online) Band structures for the $\Gamma$-valley twisted TMDCs, WS$_2$ (top row), MoS$_2$ (middle row), and MoSe$_2$ (bottom row). We show the low-energy band evolution for the twist angle range $1^{\circ}<\theta<5^{\circ}$.}
	\label{figSM1} 
\end{figure}

The diagonalization of $H$ is performed in reciprocal space, where the Hamiltonian of Eq.~(\ref{eqSM1}) is given by (see Eq.~(1) in the main text)
\begin{equation}
H = -\frac{\hbar^2 | \bold{k} + \bold{G} |^2}{2m^{*}} \delta_{\bold{G},\bold{G'}} + V_{\mathrm{M}}(\bold{G} - \bold{G'}).
\label{eq1rep}
\end{equation}
$\bold{k}$ and $\bold{G}$ denote moir\'{e} crystal momentum and vectors from the  moir\'{e} reciprocal lattice, respectively. $V_{\mathrm{M}}(\bold{G} - \bold{G'})$ is the Fourier transformation of $V_{\mathrm{M}}(\bold{r})$. This Hamiltonian is expanded up to a plane-wave cutoff $G_{\mathrm{c}}$ for a given twist angle $\theta$. In Fig.~\ref{figSM1} we show the band structures for $\Gamma$-valley twisted TMDCs at twist angles in the range $1^{\circ}<\theta<5^{\circ}$. The zero energy is defined as the top of the valence band. For this twist angle range it is sufficient to use $G_c=4$\,--\,5~$G^{\mathrm{M}}$, where $G^{\mathrm{M}}=|\bold{G}_{1,2}^{\mathrm{M}}|$. 

From the valence band edge, two flat bands emerge which touch at a Dirac point in the corner of the Brillouin zone and at a certain negative energy. We refer to these bands as "flat Dirac bands" for brevity. These bands are well isolated from other higher energy bands for $\theta<5^{\circ}$. The bandwidth of the flat Dirac bands continuously increases approximately quadratically with the twist angle. 

Another approach to calculate band structures is the derivation of atomistic tight-binding models, which, for instance, has been done in Ref.~\cite{PhysRevB.102.081103} for MoS$_2$. This approach yields the same results and reproduces ab-initio calculations consistently.

\section{Wannier projection of flat Dirac bands}\label{sec:wannier_projection}
\begin{figure}
	\centering
	\includegraphics[width=0.95\textwidth]{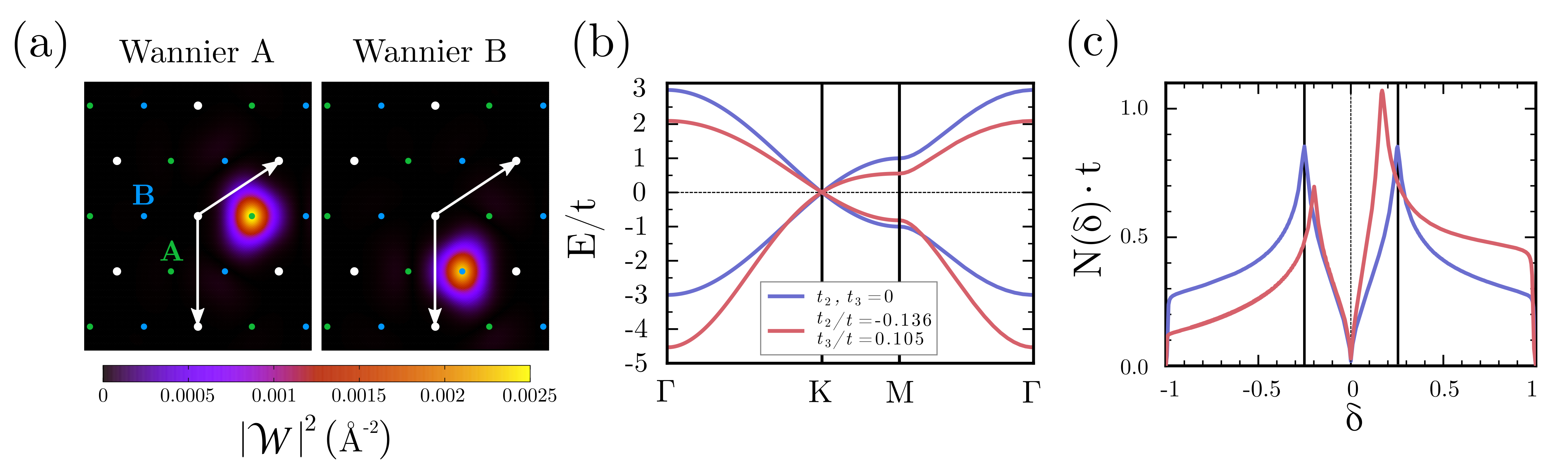}
	\caption{(Color online) Wannier tight-binding model for twisted WS$_2$ at 3.5$^\circ$. (a) Wannier densities $|\mathcal{W}_m|^2$ for each orbital centered in the sublattices A (green dots) and B (blue dots). White dots denote the Bravais lattice $\bold{R}$ and white arrows are the moir\'{e} lattice vectors $\bold{L}_{1,2}^{\mathrm{M}}$. (b) Band structure and (c) density of states $N(\delta)$ per spin and unit cell of the honeycomb tight-binding model with nearest-neighbor hopping only ($t_2,t_3=0$, blue line) and including also longer-ranged hopping terms ($t_2/t_1 = -0.136,\, t_3/t_1=0.105$, red line). The doping $\delta$ is counted relative to the Dirac point. Van Hove singularities (VHS) appear at the $M$-points in the Brillouin zone, corresponding to a doping of $\delta=\pm 0.25$ ($\delta\approx-0.2,0.17$) in the particle-hole symmetric (asymmetric) case.}
	\label{figSM2} 
\end{figure}

Based on our observations in the previous section, we construct a  tight-binding Hamiltonian for the isolated flat Dirac bands via Wannier projection with one orbital per sublattice site. The AB and BA regions play the role of the A and B sublattice sites in a honeycomb lattice. The eigenstates of the low-energy continuum model are
\begin{equation}
\Phi_{\bold{k}}^\alpha (\bold{r}) \equiv \lvert \Phi_{\bold{k}}^\alpha \rangle = \sum_{\bold{G}} c_{\bold{k}\bold{G}}^{\alpha} \mathrm{e}^{i(\bold{k}+\bold{G})\cdot \bold{r}},
\label{eqSM3}
\end{equation}
where $\alpha$ is the band index and $c_{\bold{k}\bold{G}}^{\alpha}$ are the plane-wave coefficients obtained from the diagonalization of the Hamiltonian from Eq.~(\ref{eq1rep}). We set $G_c=5G^{\mathrm{M}}$ and use a k-mesh of $15\times 15$. We consider Gaussian functions centered on A and B sites as the trial orbitals $\left\lvert g_\bold{k}^m \right\rangle$, whose plane-wave expansion coefficients are given by
\begin{equation}
g_{\bold{kG}}^m = \mathrm{e}^{-(\Delta \bold{K})^2/2} \mathrm{e}^{-i \bold{K} \cdot \bold{l}_m}\;.
\label{eqSM4}
\end{equation}
Here, $m\in\{\mathrm{A,B}\}$ is the sublattice (orbital) index, $\bold{K} = \bold{k} + \bold{G}$, and $\bold{l}_{\mathrm{A}}=\bold{L}^{\mathrm{M}}_1/3 + 2\bold{L}^{\mathrm{M}}_2/3$ and $\bold{l}_{\mathrm{B}}=2\bold{L}^{\mathrm{M}}_1/3 + \bold{L}^{\mathrm{M}}_2/3$ are vectors pointing from the moir\'{e} unit cell origin to A and B sites, respectively. $\bold{L}_{1,2}^{\mathrm{M}}$ are the moir\'{e} lattice vectors, $\lambda^{\mathrm{M}}=|\bold{L}_{1,2}^{\mathrm{M}}|$, and $\Delta=\lambda^{\mathrm{M}}/3$ is the extent of the trial orbitals. These trial orbitals are then projected onto the eigenstates manifold of the low energy Dirac bands $\left\lvert \phi^m_{\bold{k}} \right\rangle = \sum_{\alpha} \left\langle \Phi^{\alpha}_{\bold{k}} \vert g^m_{\bold{k}} \right\rangle \left\lvert \Phi^{\alpha}_{\bold{k}} \right\rangle$, which yields the corresponding plane-wave expansion coefficients of the state $\left\lvert \phi^m_{\bold{k}} \right\rangle$ \cite{RevModPhys.84.1419}
\begin{equation}
\phi_{\bold{kG}}^m = \sum_\alpha c_{\bold{k}\bold{G}}^{\alpha} P_\bold{k}^{\alpha m}\;.
\label{eqSM5}
\end{equation}
The projection matrix $P_{\bold{k}}^{\alpha m}=\left\langle \Phi^{\alpha}_{\bold{k}} \vert g_\bold{k}^m \right \rangle \equiv \sum_{\bold{G}} (c_{\bold{k}\bold{G}}^{\alpha})^\dagger g_{\bold{kG}}^m$ allows to calculate the overlap matrix as
\begin{equation}
S_{\bold{k}}^{mn} = \left\langle \phi^m_{\bold{k}} \vert \phi^n_{\bold{k}} \right\rangle = (P_\bold{k}^\dagger P_\bold{k})^{mn}\;.
\label{eqSM6}
\end{equation}
Eqs. (\ref{eqSM5}) and (\ref{eqSM6}) are used to calculate the so-called smooth gauge plane-wave expansion coefficients of the smooth gauge Bloch states $\lvert \tilde{\Phi}^m_{\bold{k}} \rangle$
\begin{equation}
\tilde{c}_{\bold{k}\bold{G}}^{m} = \sum_n \phi_{\bold{kG}}^n \cdot (S_\bold{k}^{-1/2})^{nm}\;.
\label{eqSM7}
\end{equation}
The resulting set of well-localized Wannier orbitals can be constructed in real space as
\begin{equation}
\mathcal{W}_{\bold{R}m} (\bold{r}) = \frac{1}{N_\bold{k} \sqrt{A_{\mathrm{M}}}} \sum_{\bold{k}} \sum_{\bold{G}} \tilde{c}_{\bold{k}\bold{G}}^{m} \mathrm{e}^{i \bold{K} \cdot (\bold{r} - \bold{R})}\;,
\label{eqSM8}
\end{equation}
where $A_{\mathrm{M}}=\sqrt{3}\lambda^{\mathrm{M}}/2$ is the moir\'{e} unit cell size, $\bold{r}$ denotes the real space coordinates, and $\bold{R}$ describes the Bravais lattice. In Fig.~\ref{figSM2}(a) we show the real-space probability density $|\mathcal{W}|^2$ for the two Wannier orbitals $m$ from the unit cell at the origin of the Bravais lattice obtained from the flat Dirac bands.

Now, the Hamiltonian in the Wannier orbital basis can be calculated by projecting the continuum Hamiltonian onto the smooth gauge Block states $\lvert \tilde{\Phi}^m \rangle$
\begin{equation}
\tilde{H}_\bold{k}^{mn} = \langle \tilde{\Phi}^m_{\bold{k}} \rvert H \lvert \tilde{\Phi}^n_{\bold{k}} \rangle \equiv \sum_{\bold{G},\bold{G}'} \left(\tilde{c}_{\bold {k}\bold{G}}^m\right)^\dagger \tilde{c}_{\bold{k}\bold{G}'}^n H_{\bold{k}\bold{G}\bold{G}'},
\label{eqSM9}
\end{equation}
where $H_{\bold{k}\bold{G}\bold{G}'}$ are the matrix elements of the Hamiltonian in Eq.~(\ref{eq1rep}). Fourier transformation of Eq.~(\ref{eqSM9}) gives the real space Wannier Hamiltonian $\tilde{H}^{mn}_\bold{r}$ whose matrix elements $m$, $n$ are the hopping integrals entering the tight-binding model used in the main text. We find that including up to three nearest-neighbor hoppings is sufficient to describe the band structures found by the continuum model in the twist angle range $1^{\circ}<\theta<5^{\circ}$, as shown in Fig.~1(b) of the main text for WS$_2$ at $\theta=3.5^{\circ}$. The angle dependence of the hopping amplitudes is shown in Fig.~1(c) of the main text.

The resulting honeycomb tight-binding model is
\begin{align}
	\begin{split}
		H_0(\bold{k}) &= \begin{pmatrix}
		H_{\mathrm{AA}}(\bold{k}) & H_{\mathrm{AB}}(\bold{k})\\
		H_{\mathrm{BA}}(\bold{k}) & H_{\mathrm{BB}}(\bold{k})
		\end{pmatrix}\quad\mathrm{with}\\
		H_{\mathrm{AA}}(\bold{k}) = H_{\mathrm{BB}}(\bold{k}) &= 2 t_2\left[\cos(\bold{k}\cdot\bold{L}_1^{\mathrm{M}}) + \cos(\bold{k}\cdot\bold{L}_2^{\mathrm{M}}) + \cos(\bold{k}\cdot(\bold{L}_1^{\mathrm{M}}+\bold{L}_2^{\mathrm{M}})\right]\,,\\
		H_{\mathrm{AB}}(\bold{k}) = H_{\mathrm{BA}}^*(\bold{k}) &= t_1\left[1 + \mathrm{e}^{i\bold{k}\cdot\bold{L}_1^{\mathrm{M}}} +  \mathrm{e}^{i\bold{k}\cdot(\bold{L}_1^{\mathrm{M}}+\bold{L}_2^{\mathrm{M}})}\right]
		+ t_3\left[2\cos(\bold{k}\cdot\bold{L}_2^{\mathrm{M}}) \mathrm{e}^{i\bold{k}\cdot(2\bold{L}_1^{\mathrm{M}} + \bold{L}_2^{\mathrm{M}})}\right]\,.
	\end{split}
	\label{eq:honeycomb_hamiltonian}
\end{align}
The corresponding band structure and density of states (DOS) per spin and unit cell can be found in Figs. \ref{figSM2}(b) and (c), respectively. The DOS is shown as a function of doping $\delta$ that is counted relative to the Dirac point. We show the third nearest-neighbor hopping model with $t_2/t_1 = -0.136$ and $t_3/t_1=0.105$ for WS$_2$ at $\theta=5^\circ$ using $t\equiv t_1$ as unit of energy. It reveals a slight particle-hole asymmetry around the Dirac point. We also show a simplified model which only accounts for nearest-neighbor hopping and which is particle-hole symmetric. In both cases there are Van Hove singularities (VHS) emerging at the $M$ points of the moir\'{e} Brillouin zone.

Since $t_1 \gg t_2, t_3$, the character of the VHS does not change to higher order VHS \cite{Yuan2019,Classen2020} and the qualitative physics occurring in the system are not expected to change significantly (c.f.~Section S8 for an explicit demonstration). Therefore, we neglect $t_2$ and $t_3$ and consider the particle-hole symmetric model with nearest-neighbor hopping $t_1\equiv t$ only in the main text.

%%%%%%%%%%%%%

\section{Long-range Coulomb interactions}\label{sec:long_range_Coulomb}
Several twisted 2D systems are known to show a strong reconstruction of their band structure upon doping caused by the Hartree potential resulting from the long-range Coulomb interaction \cite{Guinea2018,PhysRevB.100.205113,PhysRevB.102.045107,PhysRevB.102.155149,Fischer2022}. Here, we study the effect of the Hartree potential in $\Gamma$-valley twisted TMDCs. We follow the method developed in Ref.~\cite{PhysRevB.100.205113}. The Hartree potential contribution to the total Hamiltonian of Eq.~(\ref{eqSM1}) is given by
\begin{equation}
	V_{\mathrm{H}}(\bold{r}) = \int d^2\bold{r}' V_{\mathrm{C}}(\bold{r}-\bold{r}') \delta \rho(\bold{r}'),
	\label{eqSM12}
\end{equation}
where $V_{\mathrm{C}}(\bold{r}) = \frac{e^2}{\epsilon |\bold{r}|}$ is the Coulomb potential, $\epsilon=4.5$ is the dielectric constant of the environment as produced by hBN, and $\delta \rho (\bold{r})$ is the deviation of the charge density from charge neutrality. $\delta \rho =0$ corresponds to the undoped continuum model, i.e., when the Fermi level is at the top of the flat Dirac bands. We can then write
\begin{equation}
	\delta \rho (\bold{r}) = \frac{1}{A_{\mathrm{M}}} \sum_{\bold{G}} \delta \rho(\bold{G}) \mathrm{e}^{i\bold{G}\cdot\bold{r}}.
	\label{eqSM13}
\end{equation}
The Fourier components $\delta \rho(\bold{G})$ are given by
\begin{equation}
	\delta \rho(\bold{G}) = -\frac{2}{N_\bold{k}} \sum_{\bold{k},\bold{G}'} \sum_{\alpha '} \left(c_{\bold{k}\bold{G}'}^{\alpha '}\right)^\dagger c_{\bold{k} \bold{G}'+\bold{G}}^{\alpha '},
	\label{eqSM14}
\end{equation}
where the sum over $\alpha '$ runs over the unoccupied states in the valence band, so it depends on the doping level $E_{\mathrm{F}}$, the factor $2$ accounts for the spin degeneracy, and the minus sign refers to hole doping. Due to the $D_6$ symmetry of the lattice, $\delta \rho(\bold{G})$ are equally weighted in the same $s$ shell of $\bold{g}_j^s$ vectors, so we can write $\delta \rho_s \equiv \delta \rho(\bold{g}_j^s)$ for any $j$. We also checked that it is enough to consider the first and second shells $s=1,2$ to correctly address the effect of the Hartree potential. Under these assumptions, we can write the Hartree potential as
\begin{equation}
	V_{\mathrm{H}}(\bold{r}) = \sum_s V_0^s \delta \rho_{s} \sum_j \mathrm{e}^{i \bold{g}_j^s \cdot \bold{r}}, 
	\label{eqSM15}
\end{equation}
where $V_0^s = \frac{2\pi e^2}{\epsilon A_{\mathrm{M}} |\bold{G}^s|}$. $\delta \rho_{s}$ are the amplitudes which define the Hartree potential and have to be determined self-consistently. By Fourier transforming Eq.~(\ref{eqSM15}), we can introduce the Hartree potential in Eq.~(\ref{eq1rep}) and solve the total Hamiltonian $H+V_{\mathrm{H}}$ in the reciprocal space.
\begin{figure}
	\centering
	\includegraphics[width=0.95\textwidth]{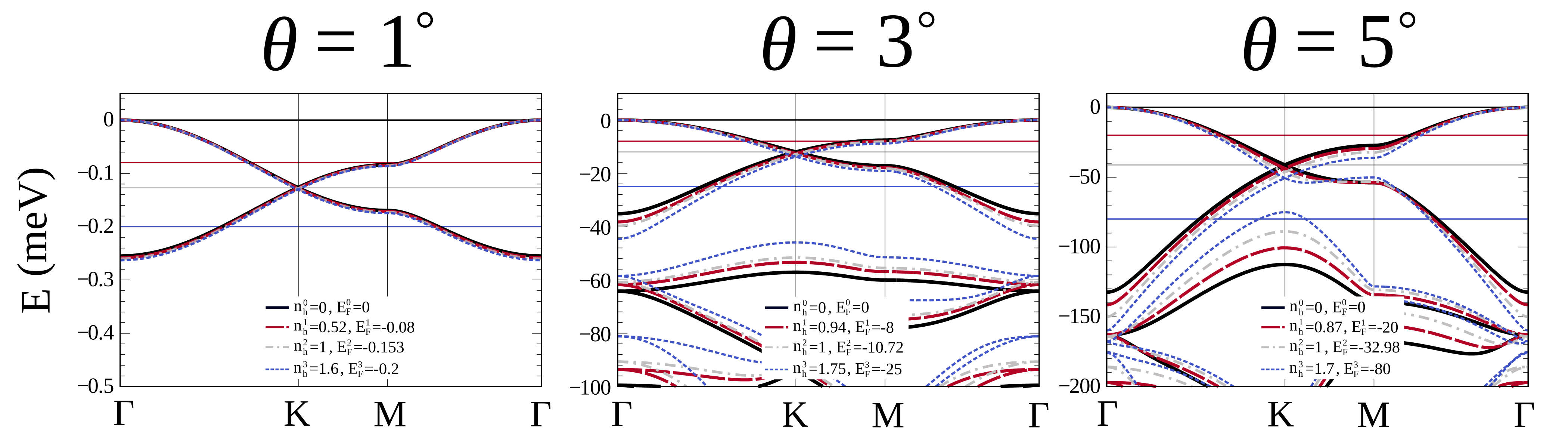}
	\caption{(Color online) Hartree potential effect in the doped band structures of twisted WS$_2$. From left to right, we show results for different twist angles $\theta$. The undoped bands are shown with black solid lines. Band structures corresponding to Fermi levels $E_{\mathrm{F}}^i$ and hole doping $n_h^i$  set between the undoped valence band maximum and the Dirac point (dashed red, $i=1$), at the Dirac point (dashed gray, $i=2$), and between the Dirac point and the bottom of the flat Dirac bands (dashed blue, $i=3$). The solid horizontal lines represent the corresponding Fermi energies. Calculations were performed at $T=0$.}
	\label{figSM3} 
\end{figure}

The self-consistent procedure is as follows:
\begin{itemize}
	\item We consider various doping levels with respect to the top of the valence band $n_h$ (number of holes per spin). Here $n_h=0$ corresponds to the undoped system, $n_h=1$ to the hole doping to the Dirac point, and $n_h=2$ to completely empty flat Dirac bands. We consider $n_h^1$ between the undoped level and the Dirac point, $n_h^2$ at the Dirac point, and $n_h^3$ between the Dirac point and the bottom of the flat Dirac bands. We obtain the plane-wave coefficients $c_{\bold{k}\bold{G}}^{\alpha}$ from diagonalizing $H+V_{\mathrm{H}}$.
	\item Using Eq.~(\ref{eqSM14}), we calculate the new charges $\delta \rho_s^\mathrm{new}$.
	\item In each iteration step, the self-consistent convergence is checked by $|\delta \rho_s^\mathrm{old} - \delta \rho_s^\mathrm{new}| < 10^{-6}$. If the convergence criterion is fulfilled, we finish the code and calculate the new and renormalized band structures.
	\item If the convergence criterion is not fulfilled, then we update $\delta \rho_s$ using a Kerker mixing procedure \cite{PhysRevB.23.3082}, where $\delta \rho_s = \delta \rho_s^\mathrm{old} + \alpha \frac{G_s^2}{G_s^2+\beta^2}\left( \delta \rho_s^\mathrm{new} - \delta \rho_s^\mathrm{old} \right)$. We set $\alpha=0.1$, $\beta=0.9$, and $G_s \equiv \lvert \bold{G}^s \rvert$. A simple straight mixing is obtained if $\beta$ is set to a very small value. We find that the charges are usually converged after less than 30 iterations depending on the chosen twist angle $\theta$ and the doping level $n_h^i$.
\end{itemize}  

We show in Fig.~\ref{figSM3} the effect of the Hartree potential for WS$_2$ at different twist angles $\theta$ and different Fermi energies  $E_{\mathrm{F}}^i$ corresponding to respective hole dopings $n_h^i$. The Hartree potential mainly shifts the bands as a whole and increases the bandwidth. Only for larger twist angles $\theta \geq 5^\circ$ and dopings $n_h^3$, the flat Dirac bands start to be reconstructed, with the higher energy bands below the flat Dirac bands even being partially filled. This is in contrast to graphene-based systems, where the entangled multiorbital nature of the flat bands facilitate strong renormalization of the bands.

\begin{figure}
	\centering
	\includegraphics[width=0.75\textwidth]{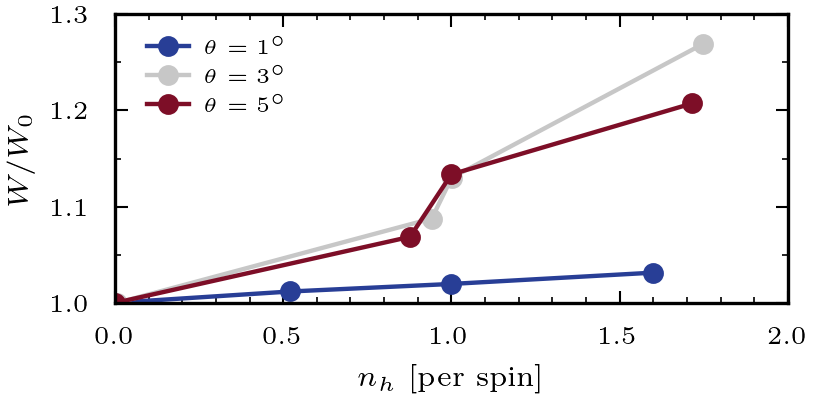}
	\caption{(Color online) Relative change in the bandwidth with respect to the undoped case $W/W_0$ due to the Hartree potential in twisted WS$_2$. The change is larger for large hole dopings and non-monotonous in terms of the twist angle. The largest relative change for $\theta=3^\circ$ at $n_h\approx 1.75$ is approximately 27\,$\%$.}
	\label{figSM3WW0} 
\end{figure}

The bandwidth renormalization can be easily visualized when plotting the relative change of the bandwidth with respect to the undoped case $W/W_0$, see Fig.~\ref{figSM3WW0}, where the change is larger for larger dopings. The change is non-monotonic when changing the twist angle. From these results, we conclude that $\theta \leq 5^\circ$ is the limit of applicability of our calculations. In any case and for the purpose of our FLEX calculations, we can assume that the doping $\delta = 1-n_h$ used in the main text occurs between the valence band maximum and the Dirac point, where the bands are essentially unaffected by the Hartree potential.

%%%%%%%%%%%%%
%% NEW SECTION U,V %%
\section{Estimation of the Coulomb interaction strength}\label{sec:Hubbard_interaction}
\begin{figure}
	\centering
	\includegraphics[width=0.95\textwidth]{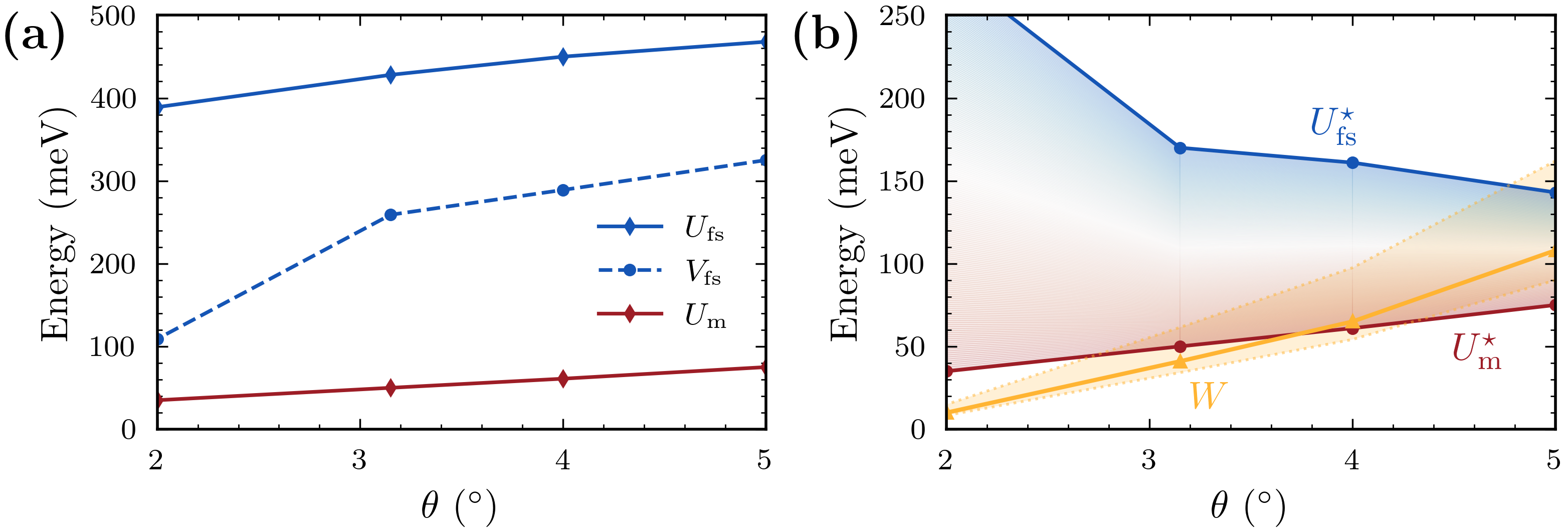}
	\caption{(Color online) Estimated Coulomb interactions in different dielectric environments for WS$_2$. 
		(a) Local and nearest-neighbor Coulomb interaction $U$ and $V$ in two dielectric environments, free-standing twisted bilayer (`fs', red line) and metallic gate in direct contact with the twisted bilayer (`m', blue line). (b) Effective local Hubbard interactions $U^\star=U-V$ and bandwidth $W$ (orange line). The blue-red shaded region describes the possible values that $U^\star$ can take depending on the dielectric environmental setup. The orange-shaded region indicates interaction values we use in the FLEX calculations ($U^\star/t \sim 5$\,--\,9).}
	\label{figSM_HubbardUV} 
\end{figure}

From the definition of the Wannier orbitals in Eq.~(\ref{eqSM8}), we estimate the value of the screened Coulomb interaction matrix elements $W_{\bold{R},mn}$. The local and nearest-neighbor Coulomb interactions can be then calculated as the matrix elements $U\equiv W_{\bold{0},AA}$ and $V\equiv W_{\bold{0},AB}$. The resulting extended Hubbard model can be mapped onto a local Hubbard model by making the assumption $U^\star = U-V$ \cite{PhysRevB.94.165141}. We estimate the upper and lower bounds by projecting an effective interaction $V_{\mathrm{eff}}(\bold{r})$ onto the Wannier functions in two limiting dielectric environment cases: free-standing twisted bilayers, for which the external screening is minimal, and a metallic gate in direct contact with the twisted bilayers, for which the external screening is maximal \cite{Pizarro2019}. The screened Coulomb interaction matrix is given by
\begin{equation}
W_{\bold{R},mn} = \iint d^2\bold{r}\, d^2\bold{r}' \; V_{\mathrm{eff}}(\bold{r}-\bold{r}') \rho_{\bold{R}m}(\bold{r}) \rho_{\bold{0}n}(\bold{r}'),
\label{eqSMc1}
\end{equation}
with $\rho_{\bold{R}m}(\bold{r})=|\mathcal{W}_{\bold{R}m} (\bold{r})|^2$. $V_{\mathrm{eff}}(\bold{r})$ is the Coulomb interaction screened by the TMDC bilayer in its undoped state and the dielectric environment. We start from an Ohno potential \cite{Ohno1964} 
\begin{align}
	V_{\mathrm{Ohno}}(\bold{r}) = \frac{e^2}{\sqrt{r^2+\xi^2}}
\end{align}
that regularizes the bare Coulomb interaction $e^2/r$ at a short wavelength cut-off $\xi=1$~\AA, which is set by the spacial extent of the W $d$-orbitals. The effect of screening is easily included in reciprocal space, so the effective interaction is  $V_{\mathrm{eff}}(\bold{r})$ then calculated from the inverse Fourier transformation of
\begin{equation}
V_{\mathrm{eff}}(\bold{q}) = \frac{V_{\mathrm{Ohno}}(\bold{q})}{\epsilon(\bold{q})} = \frac{2 \pi e^2}{\epsilon(\bold{q}) q} \mathrm{e}^{-q\xi}\;,
\label{Yukawa1}
\end{equation}
where $\epsilon(\bold{q})$ is the dielectric function that encodes the environmental screening effect. For our two limiting cases, free-standing (`fs') and metal in direct contact (`m'), we use the effective dielectric functions \cite{Pizarro2019,Cea2021}:
\begin{align}
	\begin{split}
		\epsilon_{\mathrm{fs}}(\bold{q}) = \kappa \frac{1-\tilde{\kappa}\mathrm{e}^{-qh}}{1+\tilde{\kappa}\mathrm{e}^{-qh}}\;,\\
		\epsilon_{\mathrm{m}}(\bold{q}) = \kappa \coth{\frac{q h}{2}}\;.
	\end{split}
	\label{eq:environmentalscreening}
\end{align}
Here, $\kappa \approx 10$ is the internal screening of the twisted TMDC, $h\approx13$~{\AA} is the bilayer height \cite{Laturia2018,Weston2020}, and $\tilde{\kappa}=(\kappa-1)/(\kappa+1)$.
In Fig.~\ref{figSM_HubbardUV}(a) we plot the on-site and nearest-neighbor interactions $U$ and $V$ at different twist angles $\theta$ for the two limiting cases for WS$_2$. Since the nearest-neighbor interaction with a metal gate contact $V_{\mathrm{m}}$ is on the order of 1~meV, we did not include it in the plot. Fig.~\ref{figSM_HubbardUV}(b) shows the effective local Hubbard interaction  $U^\star$ and the bandwidth $W$. 
%The shaded area between $U_{\mathrm{fs}}^*$ and $U_{\mathrm{m}}^*$ indicates the presence of regions of strongly ($\theta\lesssim 3^\circ$) to intermediately correlated physics ($\theta> 4^\circ$). 
A realistic value for $U^\star$ in $\Gamma$-valley twisted TMDCs will fall inside the shaded regions between the limiting cases $U^\star_{\mathrm{fs}}$ and $U^\star_{\mathrm{m}}$ which depends on the experimental setup and which can be tuned by changing the dielectric environment \cite{Pizarro2019,Goodwin2019,Stepanov2020,Saito2020, Arora2020,Liu2021}. For our FLEX calculations, we use $U^\star/t = 5$\,--\,9 (where we assume an effective Hubbard model with $U\equiv U^\star$) which is indicated by an orange-shaded region, since $W\approx 6t$. These interaction values correspond to experimentally accessible interaction strengths in a twist angle range of $3-5^\circ$.

%%%%%%%%%%%%%%%%%%%%%%%%%%%%%%%%%%%%%%%%%%
%%%%%%%%%%%%%%%%%%%%%%%%%%%%%%%%%%%%%%%%%%

\section{Numerical details of FLEX calculations}\label{sec:FLEX_details}
We summarize the calculation steps performed in the FLEX approximation \cite{Bickers1989a,Bickers1989b} and give details on the numerical parameters used. In the FLEX approximation, one solves the Dyson equation
\begin{align}
	\hat{G}(k)^{-1} = \hat{G}^0(k)^{-1} - \hat{\Sigma}(k)\,,
	\label{eq:Dyson_eq}
\end{align}
with the dressed (bare) Green function \(G\) (\(G^0\)), self-energy \(\Sigma\), and the four-momentum \(k = (i\omega_n,\bold{k})\). $\bold{k}$ is the crystal momentum and  \(\omega_n=(2n+1)\pi k_{\mathrm{B}}T\) are the Matsubara frequencies at a temperature \(T\). In case of the single-orbital honeycomb model, all quantities are given by \(2\times2\) matrices in terms of sublattice indices A and B \cite{Kontani1998,Koikegami1997} which is denoted by a hat \(G_{\alpha\beta}\equiv(\hat{G})_{\alpha\beta}\). The non-interacting Green function is given by
\begin{align}
	\hat{G}^0(k) = \left[i\omega_n\mathds{1} - (\hat{H}_0(\bold{k}) - \mu\mathds{1})\right]^{-1},
\end{align}
where \(H_0\) is the non-interacting Hamiltonian given in Eq.~(\ref{eq:honeycomb_hamiltonian}), \(\mathds{1}\) denotes the \(2\times 2\) identity matrix, and \(\mu\) is the chemical potential of the doping level \(\delta\). The self-energy \(\Sigma\) mainly consists of contributions from spin and charge fluctuations and is calculated from
\begin{align}
	\Sigma_{\alpha\beta}(k) = \frac{T}{N_{\bold{k}}}\sum_{q}G_{\alpha\beta}(k-q)\left\lbrace U^2\left[\frac{3}{2}\hat{\chi}^{\mathrm{s}}(q) + \frac{1}{2}\hat{\chi}^{\mathrm{c}}(q)-\hat{\chi}^{0}(q)\right]+\mathds{1}U\right\rbrace_{\alpha\beta} \, ,
\label{eq:FLEX_two_site_Sigma}
\end{align}
with the number of sites \(N_{\bold{k}}\), and the Hubbard interaction \(U\) as given in Eq.~(2) of the main text. The charge and spin susceptibility entering Eq.~(\ref{eq:FLEX_two_site_Sigma}) are defined by
\begin{align}
	\hat{\chi}^{\mathrm{c,s}}(q) = \hat{\chi}^0(q)\left[\mathds{1} \pm U\hat{\chi}^0(q)\right]^{-1},
\label{eq:Charge_spin_susceptibility}
\end{align}
where the irreducible susceptibility is
\begin{align}
\chi^0_{\alpha\beta}(q) = - \frac{T}{N_{\bold{k}}}\sum_{k} G_{\alpha\beta}(k+q)G_{\beta\alpha}(k)\;.
\label{eq:Irreducible_susceptibility}
\end{align}
Eqs.~(\ref{eq:Dyson_eq}) -- (\ref{eq:Irreducible_susceptibility}) are solved self-consistently. The calculations are initialized using only the bare Green function \(G^0\) with \(\Sigma=0\), i.e., starting from free electrons, and in each iteration step the chemical potential \(\mu\) needs to be adjusted to keep the doping \(\delta\) fixed. We employ a linear mixing \(G = \kappa G^{\mathrm{new}} + (1-\kappa) G^{\mathrm{old}}\) with \(\kappa=0.2\). We then defined self-consistency for a relative difference of \(10^{-4}\) between the self-energy of two iteration steps. In all calculations, we used a \(\bold{k}\)-mesh resolution of \(120\times 120\). For the Matsubara frequencies we used the sparse-sampling approach \cite{Li2020,Witt2021,Shinaoka2021} of the intermediate representation (IR) basis \cite{Shinaoka2017,Chikano2019} with an IR parameter \(\Lambda=10^4\) and a basis cutoff \(\delta_{\mathrm{IR}}=10^{-8}\). Since the numerical cost of FLEX calculations for $T=\mathcal{O}(0.001t)$ is quite expensive, this formalism is crucial. For instance, older works studying honeycomb models \cite{Kuroki2001,Onari2002,Kuroki2010} could not determine the transition temperature $T_{\mathrm{c}}$. Details on the implementation can be found in Ref.~[\onlinecite{Witt2021}] .\\

To study the superconducting phase transition driven by spin fluctuations, we consider the linearized gap equation 
\begin{align}
\lambda \Delta^{S}_{\alpha\beta}(k) = - \frac{T}{N_{\bold{k}}}\sum_{q}\sum_{\alpha\pr,\beta\pr} V^{S}_{\alpha\beta}(q)G_{\alpha\alpha\pr}(k-q)G_{\beta\beta\pr}(q-k)\Delta^{S}_{\alpha\pr\beta\pr}(k-q)\;,
\label{eq:FLEX_multi_orb_Delta_formula}
\end{align}
for the pairing potential or gap function \(\Delta\) on sublattice \(\alpha\) and with spin orientation \(S\). This equation represents an eigenvalue problem for \(\Delta\) where the eigenvalue \(\lambda\) can be understood as the relative pairing strength of a certain pairing channel. The dominant pairing symmetry of the gap function has the largest eigenvalue \(\lambda\)  and the transition temperature is found if \(\lambda\) reaches unity.
Since we do not consider spin-orbit coupling, the linearized gap equation (\ref{eq:FLEX_multi_orb_Delta_formula}) is diagonal in the spin singlet- and triplet-pairing channel (\(S=0,1\)) with the respective interactions due to the exchange of spin and charge fluctuations
\begin{align}
	\hat{V}^{S=0}(q) = \frac{3}{2}U^2\hat{\chi}^{\mathrm{s}}(q) - \frac{1}{2}U^2\hat{\chi}^{\mathrm{c}}(q) + \mathds{1}U\;,\quad
	\hat{V}^{S=1}(q) = -\frac{1}{2}U^2\hat{\chi}^{\mathrm{s}}(q) - \frac{1}{2}U^2\hat{\chi}^{\mathrm{c}}(q)\,.
	\label{eq:FLEX_multi_orb_SC_interaction}
\end{align}
We solve Eqs. (\ref{eq:FLEX_multi_orb_Delta_formula}) and (\ref{eq:FLEX_multi_orb_SC_interaction}) by using the power iteration method with a relative error of \(10^{-4}\) for convergence. As an input serve the converged Green function of the normal state calculations and a trial gap function \(\Delta_0\), which is set up according to the irreducible representations of the \(D_6\) symmetry group \cite{Sigrist1991}.

\section{Magnetic quasiorder and spin fluctuations in the honeycomb Hubbard model}\label{sec:spin_fluctuation}
In two dimensions, the Mermin-Wagner theorem \cite{Mermin1966} prevents the formation of (genuine) long-range order at finite temperature as obeyed by the FLEX approximation \cite{Kontani2006}. However, tendencies towards magnetic quasi-order can be read off from the Stoner enhancement factor \(U\chi^{0}(\bold{q})\), which enters the denominator of the static spin susceptibility $\chi^{\mathrm{s}}(i\nu_0=0,\bold{q})$ (c.f.~Eq.~(\ref{eq:Charge_spin_susceptibility})). Thus, possible formation of spin density waves (SDWs) can be investigated in FLEX by inspecting the instabilities of \(\chi^{\mathrm{s}}(i\nu_0,\bold{q})\). When the Stoner enhancement approaches unity [$U\chi^{0}\sim\mathcal{O}(0.99)$], $\chi^{\mathrm{s}}$ diverges and the transition to a quasi-ordered magnetic state is assumed \cite{Bickers1989a,Kino1998,Kitatani2015}. At this point, the FLEX calculations turn unstable and do not converge anymore. A discussion of the leading Stoner enhancement, indicating regions of strong magnetic fluctuations, is given in the main text.

\begin{figure}
	\centering
	\includegraphics[width=0.95\textwidth]{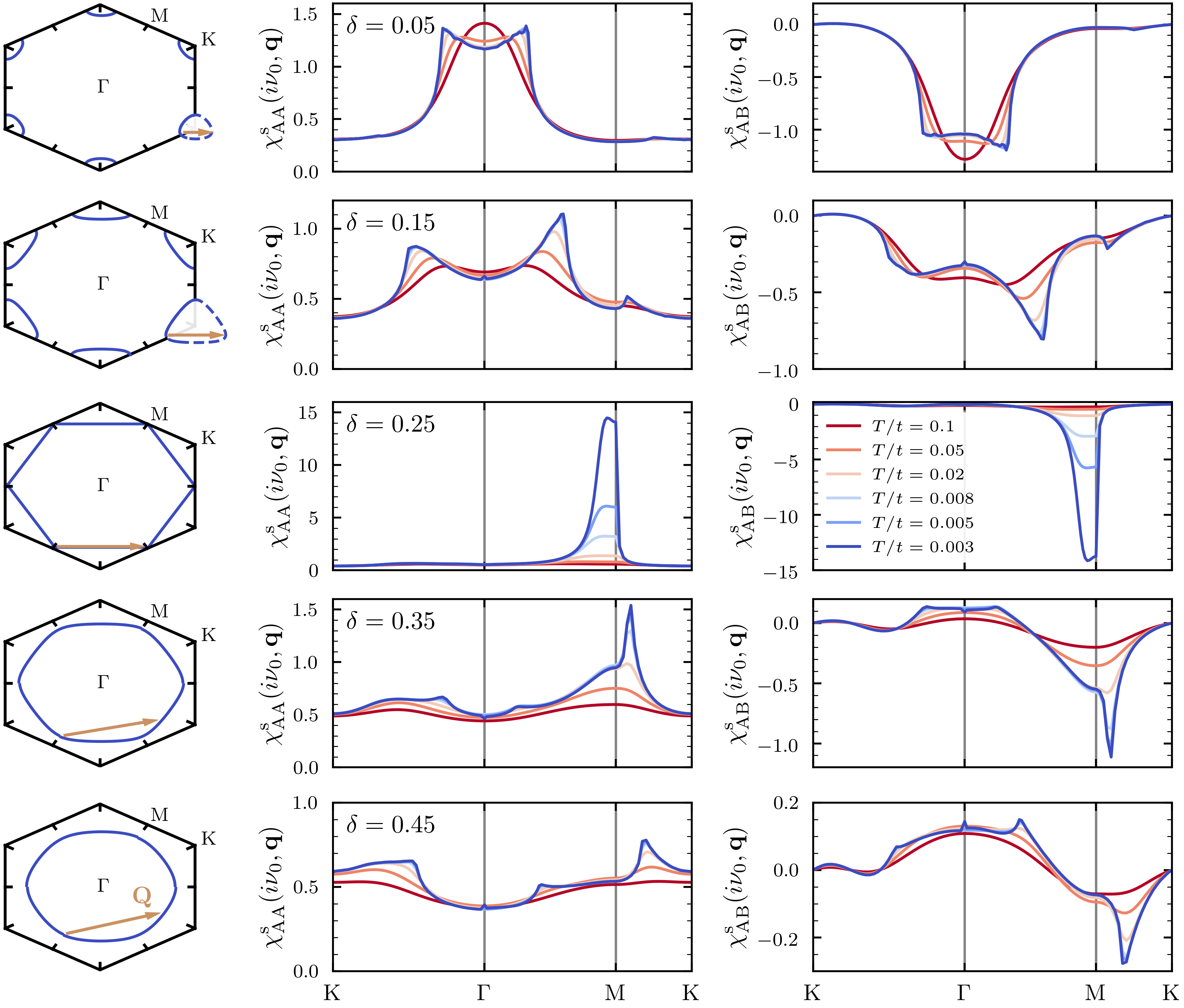}
	\caption{(Color online) Momentum dependent static spin susceptibility calculated in FLEX for different temperatures \(T/t\) and dopings \(\delta\) (rows) at fixed interaction \(U/t=4\). The left column contains the non-interacting Fermi surfaces at the respective doping levels and the dominant nesting vectors \(\bold{Q}\) associated with the largest peak of \(\chi^{\mathrm{s}}(i\nu_0,\bold{q})\). Middle and right column show the momentum resolved real part of the spin susceptibility within (AA-component) and between (AB-component) sublattices, respectively.}
	\label{fig:momentum_susceptibility} 
\end{figure}

While the real-space profile of the magnetic fluctuations is discussed in the main text, Fig.~2(b), further insight into the emerging SDWs can be gained by inspecting the momentum-space structure of \(\chi^{\mathrm{s}}(i\nu_0,\bold{q})\). In Fig.~\ref{fig:momentum_susceptibility}, we show the intra-sublattice (AA) and inter-sublattice (AB) components of \(\chi^{\mathrm{s}}(i\nu_0,\bold{q})\) along high-symmetry paths of the Brillouin zone for different \(\delta\) and \(T\) at an intermediate interaction of \(U/t=4\). Additionally, we included the Fermi surfaces of the non-interacting system associated with each doping level. 

In the doping range between the Dirac point and VHS, the AA- and AB-components of \(\chi^{\mathrm{s}}\) carry predominantly a different sign signaling antiferromagnetic fluctuations with respect to the A and B sublattices. Beyond the VHS, ferromagnetic fluctuations with respect to the sublattice index emerge and the relative fluctuation strength decreases. In each sublattice, the peak structure of \(\chi^{\mathrm{s}}\) changes significantly depending on the Fermi surface shape and becomes more pronounced for lower temperatures. That is, because the spin fluctuations emerge from the nesting conditions of the Fermi surface, i.e., possible intra-pocket electron scattering. To illustrate this, we also draw the nesting vector \(\bold{Q}\) belonging to the dominant peak of \(\chi^{\mathrm{s}}\) between Fermi surface sheets. 

%In this case, the spin orientation is chiral on each sublattice \cite{Wang2012,Jiang2014}. When doping the system beyond the VHS ($\delta > 0.25$), additional ferromagnetic ordering forms. However, the relative spin fluctuation strength rapidly decreases and the system stays paramagnetic. For large interactions \(U/t\gtrsim 5\), spin fluctuations become too strong near the Dirac point and the VHS which prevents the convergence of the FLEX calculation. 

Near Dirac doping (\(\delta=0.05\)), the Fermi surface is formed by small, almost circular pockets around the K point so that long-wavelength SDWs emerge, since  \(\chi^{\mathrm{s}}\) peaks close to the \(\Gamma\) point. This situation corresponds to an almost ferromagnetic ordering in each sublattice, but antiferromagnetic fluctuations between the sublattices. Increasing the doping (\(\delta=0.15\)) deforms the Fermi surface to an equilateral shape whereby the spin fluctuations assume shorter wavelengths, as the peak  in \(\chi^{\mathrm{s}}\) shifts from the \(\Gamma\) point to the M point. At the VHS (\(\delta=0.25\)), the system undergoes a Lifshitz transition and the Fermi surface turns hexagonal with perfect nesting conditions. This causes strong fluctuations with a chiral spin profile on each sublattice \cite{Wang2012,Jiang2014}. Beyond the VHS (\(\delta=0.35,0.45\)), the Fermi surface contracts around the \(\Gamma\) point with decreasing relative fluctuation strength. Increasing the interaction \(U\) enhances the fluctuation strength, but does not affect the general \mbox{structure of \(\chi^{\mathrm{s}}\)}.

\begin{figure}
	\centering
	\includegraphics[width=0.95\textwidth]{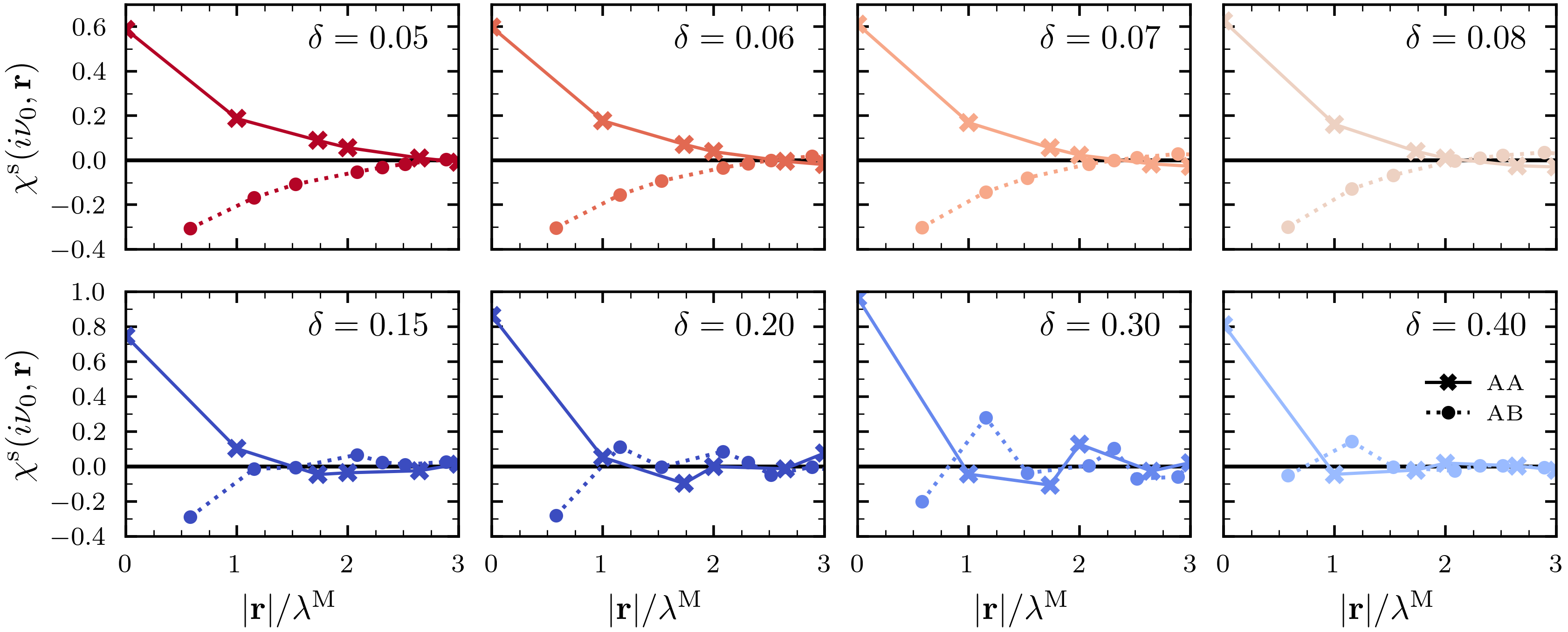}
	\caption{(Color online) Real space profile of the static spin susceptibility calculated in FLEX  for different dopings \(\delta\) at \(T/t= 0.003\) and \(U/t=6\). The distance $\vert\mathbf{r}\vert$ of two lattice sites is given in units of the moir\'{e} length \(\lambda^{\mathrm{M}}\). Correlations between spins on equal (different) sublattice sites are marked by solid lines with crosses (dotted lines with circles), corresponding to the AA (AB) component of \(\chi^{\mathrm{s}}\).}
	\label{fig:rspace_susceptibility}  
\end{figure}

Spin fluctuations can mediate an effective electron-electron interaction, as described by Eq.~(\ref{eq:FLEX_multi_orb_SC_interaction}). This interaction has non-local attractive regions which can pair spatially correlated electrons as they avoid occupying the same site. Thus, the real space profile \(\chi^{\mathrm{s}}(i\nu_0,\bold{r})\) provides information on the pairing potential for electrons. In Fig.~\ref{fig:rspace_susceptibility}, we show \(\chi^{\mathrm{s}}(i\nu_0,\bold{r})\) for different dopings at \(T/t=0.003\) and \(U/t=6\). In accordance with the previous discussion of the momentum space structure, antiferromagnetic correlations between different sublattice sites occur for doping levels in the vicinity of the Dirac point $(\delta\lesssim 0.15)$ which turn ferromagnetic for larger dopings. By increasing the doping, the AA and AB components of \(\chi^{\mathrm{s}}\) change sign on a shorter length scale, so that regions with antiferromagnetic correlations shrink. This reduces the attractive regions ($V^S\sim\chi^{\mathrm{s}}<0$) leading to a less optimal pairing situation since the pair electrons need to move closer while the Coulomb repulsion pushes them apart.

In the main text, we discuss that an optimal pairing condition with maximal transition temperature \(T^{\mathrm{max}}_{\mathrm{c}}\) arises. This can be understood from the structure of \(\chi^{\mathrm{s}}(i\nu_0,\bold{r})\) and DOS. Considering \(U/t=6\), \(T^{\mathrm{max}}_{\mathrm{c}}\) is located around \(\delta_{\mathrm{opt}}\sim0.06\) -- \(0.07\). The top row of Fig.~\ref{fig:rspace_susceptibility} shows that the fourth- and fifth-nearest-neighbor component of \(\chi^{\mathrm{s}}_{\mathrm{AB}}\) change sign in this doping region. Up to this point, \(T_{\mathrm{c}}\) increases with doping driven by the increase in the DOS at the Fermi level (c.f.~ Fig.~\ref{figSM1}(c)). As the attractive region shrinks beyond \(\delta_{\mathrm{opt}}\), pairing conditions deteriorate and \(T_{\mathrm{c}}\) decreases. The optimal situation appears where these two counteracting trends are balanced.

\section{Leading superconducting instability}\label{sec:sc_instability}
The possible pairing symmetries of the superconducting order can be classified according to the irreducible representation of the point group symmetry of the system \cite{Sigrist1991}. The honeycomb lattice is of \(D_6\) symmetry which can possibly host singlet extended \(s\)-wave, or degenerate \(d\)-wave (\(d_{xy},d_{x^2-y^2}\)) as well as triplet degenerate \(p\)-wave (\(p_x,p_y\)), \(f_{x(x^2-3y^2)}\)-wave, or \(f_{y(3x^2-y^2)}\)-wave pairing. The dominant pairing symmetry emerges with the largest eigenvalue \(\lambda\) of the linearized gap equation (\ref{eq:FLEX_multi_orb_Delta_formula}).

\begin{figure}
	\centering
	\includegraphics[width=0.95\textwidth]{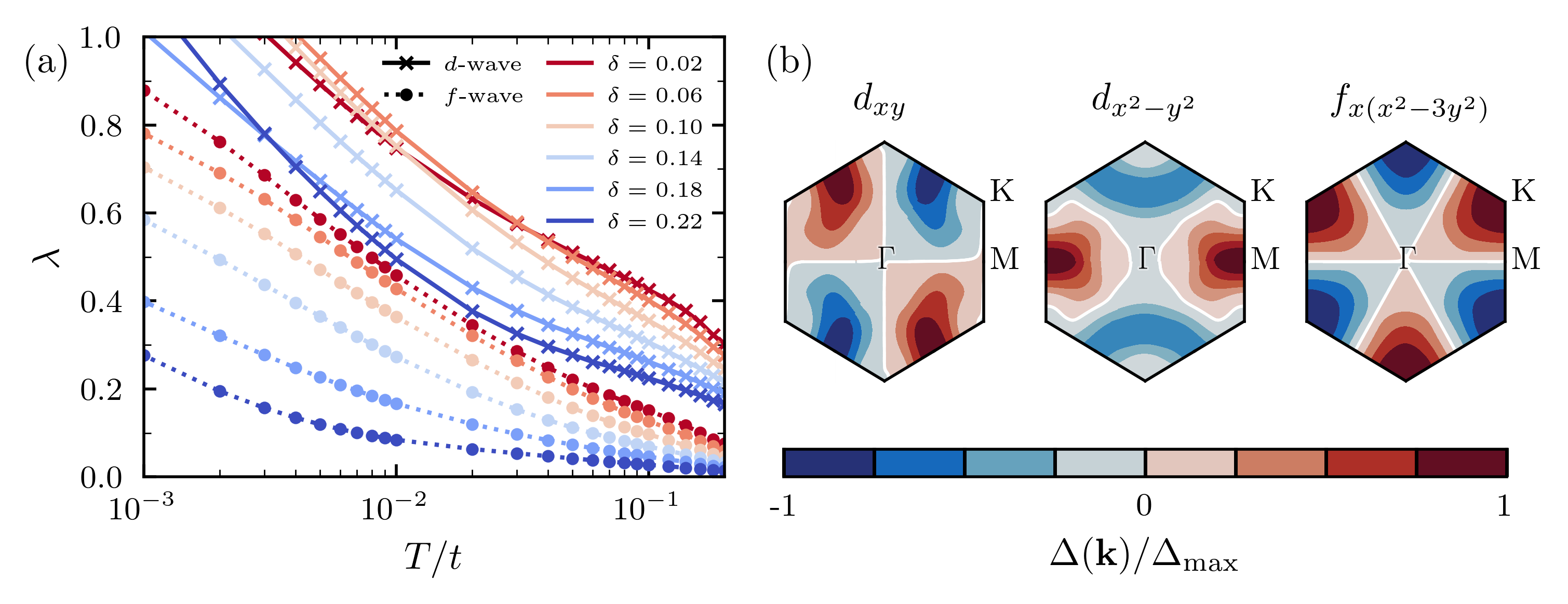}
	\caption{(Color online) Singlet vs. triplet superconductivity in the honeycomb Hubbard model. (a) Eigenvalues \(\lambda\) of the linearized Eliashberg equation for the degenerate \(d\)-wave (solid lines) and the \(f\equiv f_{x(x^2-3y^2)}\)-wave (dotted lines) pairing symmetries. Shown is the temperature dependence of \(\lambda\) for different dopings \(\delta\) at $U/t=6$. Note that the temperature scale is logarithmic. (b) Momentum-space structure of the order character for the $d$-wave and $f$-wave pairing symmetries. Shown are the normalized diagonal elements of the gap function $\Delta_{\mathrm{AA}}$ at lowest Matsubara frequency for converged calculations at $T/t=0.005$, $U/t=6$, and $\delta=0.1$. The nodes of the gap are indicated by white lines.}
	\label{fig:SC_eigenvalues} 
\end{figure}

In Fig.~\ref{fig:SC_eigenvalues}(a), we compare \(\lambda\) of the \(d\)-wave and \(f\equiv f_{x(x^2-3y^2)}\)-wave pairing symmetry for different dopings \(\delta\) between the Dirac point and VHS. These two parings emerge as the dominant pairing symmetries in the singlet and triplet pairing channel, respectively. The momentum dependence of the corresponding intra-sublattice order parameters at lowest Matsubara frequency $\Delta_{\mathrm{AA}}(i\omega_1,\mathbf{k})$ is shown in Fig.~\ref{fig:SC_eigenvalues}(b). By comparing the superconducting eigenvalues, it can be seen that singlet pairing is favored over triplet pairing. This is, in fact, consistent with the observed antiferromagnetic fluctuations as they support singlet pairing. Clearly, the \(d\)-wave pairing is the dominant superconducting instability for which the critical temperature \(T_{\mathrm{c}}\) is read off for \(\lambda_d\to 1\). The values of \(\lambda_f\), on the other hand, do not reach unity in the studied temperature region indicating that a possible transition would occur at considerably lower temperatures. 

For the dominant \(d\)-wave pairing, we find that the pairing mainly takes place between different sublattices since we observe \(|\Delta_{\mathrm{AB}}|>|\Delta_{\mathrm{AA}}|\). This is also in agreement with the antiferromagnetic alignment of the spins between the sublattices. Because of this, the triplet pairing instability can be enhanced and even dominate over the singlet pairing by introducing a staggered potential between the A and B sublattice sites \cite{Kuroki2001}.

Below \(T_{\mathrm{c}}\), a linear combination of the degenerate \(d\)-wave states forms as the superconducting ground state. The exact realization depends on the free energy with the possibility of a chiral or nematic states \cite{Sigrist1991}. For the simple honeycomb lattice, the chiral \(d+id\) state is the preferred solution with the lowest free energy \cite{Kuznetsova2005,Nandkishore2012,BlackSchaffer2014}, as the number of nodes in the quasiparticle spectrum is minimized in this case.

\section{Influence of particle-hole asymmetry on spin-fluctuation-mediated superconductivity}\label{sec:asymmetry_influence_SC}
In Section S1, we discussed the influence of longer-ranged hopping terms on the band structure and DOS of the honeycomb lattice tight-binding model (c.f.~Fig.~\ref{figSM1}(b) and (c)). Here, we assess the change of the spin-fluctuation-mediated superconducting phase transition line due to the resulting particle-hole asymmetry. We use the same parameters $t_2/t = -0.136$ and $t_3/t = 0.105$ as in Section S1. To describe the asymmetry, we need to compare each side of the Dirac point. We calculate the critical temperature $T^{\mathrm{sp}}_{\mathrm{c}}$ for one Hubbard parameter $U/t = 6$.

\begin{figure}
	\centering
	\includegraphics[width=0.88\textwidth]{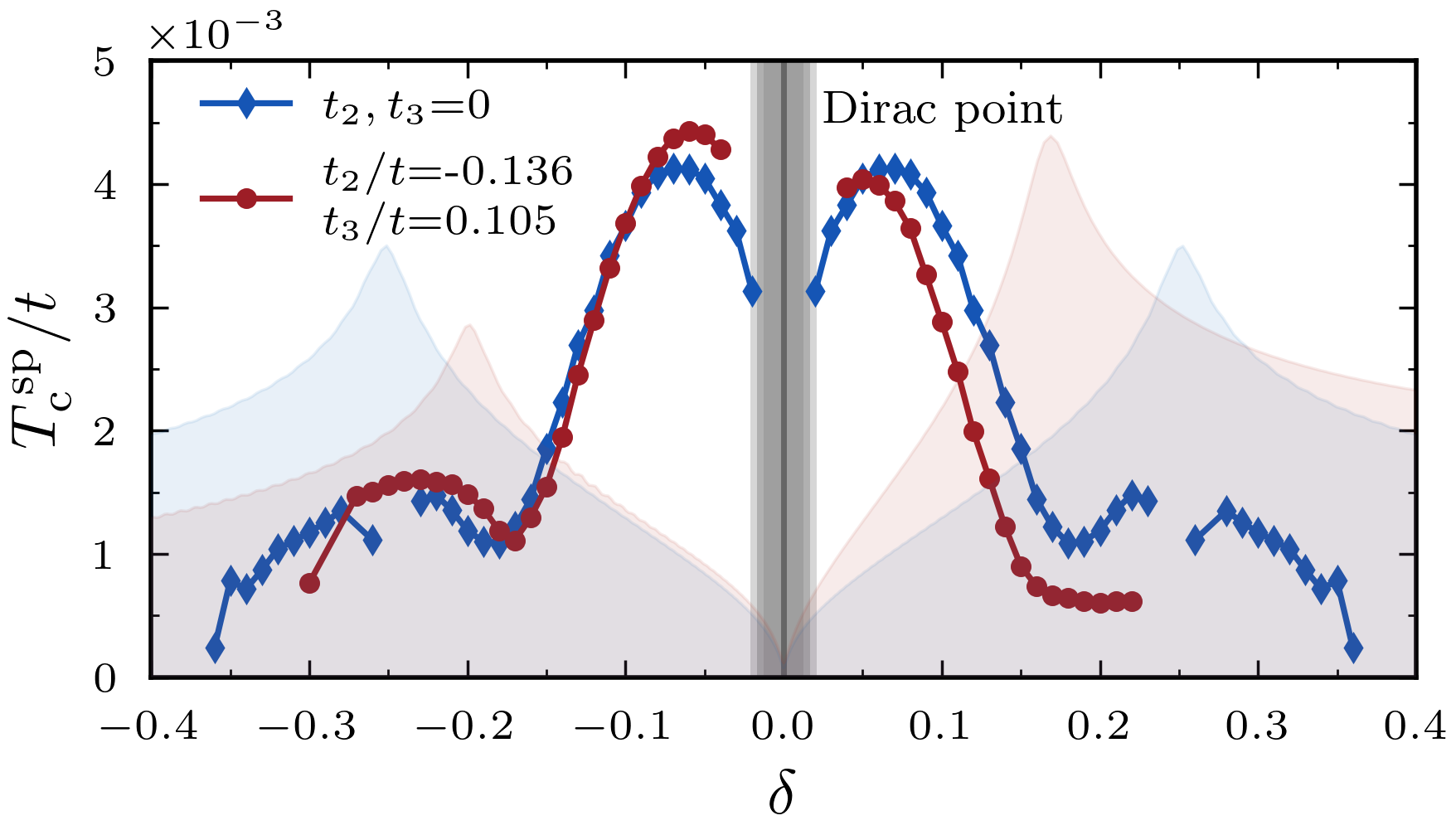}
	\caption{(Color online) Comparison of the doping dependence of the spin-fluctuation-driven superconducting transition line $T^{\mathrm{sp}}_{\mathrm{c}}$ for the particle-hole symmetric ($t_2,t_3=0$, blue line with diamonds) and asymmetric ($t_2/t = -0.136$, $t_3/t = 0.105$, red line with circles) honeycomb lattice model. The density of states for each model is drawn by a shaded area to indicate the position of the Van Hove singularities (VHS) in each case and how the phononic transition line  $T^{\mathrm{ph}}_{\mathrm{c}}$ would differ qualitatively. The Dirac point and its vicinity are indicated by a gray shaded area.}
	\label{fig:SC_asymmetry} 
\end{figure}

A comparison of the doping dependence of $T_{\mathrm{c}}$ for the particle-hole symmetric and asymmetric model is shown in Fig.~\ref{fig:SC_asymmetry}. In accordance with the band structure and DOS asymmetry, an asymmetry in the doping dependence of $T^{\mathrm{sp}}_{\mathrm{c}}$ emerges. On the left side of the Dirac point, superconductivity is slightly enhanced, while it is suppressed on the other side. This might be contrary to expectations, since the enhancement/suppression of the DOS is opposite. The reason for this is a change in the shape of the Fermi surface and hence nesting conditions caused by the additional hopping terms. Near the Dirac point, the triangular parts of the Fermi surface become flatter and the spin fluctuation strength increases due to better nesting. At the VHS, the hexagonal shape of the Fermi surface becomes rounder causing weaker spin fluctuations. The extent to which this happens, is different for each side of the Dirac point resulting in two different curves. For instance, on the left side the DOS of both cases is similar, but  $T^{\mathrm{sp}}_{\mathrm{c}}$ of the asymmetric model is slightly increased due to stronger spin fluctuations. The different nesting conditions also cause the VHS to be less detrimental to the calculations, since the Stoner enhancement does not diverge as strongly.

Even though quantitative aspects of $T^{\mathrm{sp}}_{\mathrm{c}}(\delta)$ change by the presence of long-range contributions to the single-particle dispersion, the general qualitative behavior remains unchanged. A clear maximum of $T^{\mathrm{sp}}_{\mathrm{c}}$ exists on both sides of the Dirac point, while superconductivity is suppressed by doping away from that region. Still ferromagnetic fluctuations form beyond the VHS which lead to the absence of superconductivity.

\section{Superconductivity from non-local electron-phonon interaction}
\label{sec:nonlocal}
In the main text, we have stated that conventional superconductivity driven by the electron--phonon interaction persists over a larger doping range and peaks at different levels than unconventional superconductivity driven by spin fluctuations.
More precisely, using the approximations of a single Einstein phonon mode and a constant Holstein electron--phonon coupling, we have shown that the critical temperature closely follows the electronic DOS.
Here, we will demonstrate that these observations remain valid for more general momentum-dependent phonon frequencies and Peierls electron--phonon coupling.

\begin{figure}
	\centering
	\includegraphics[width=\linewidth]{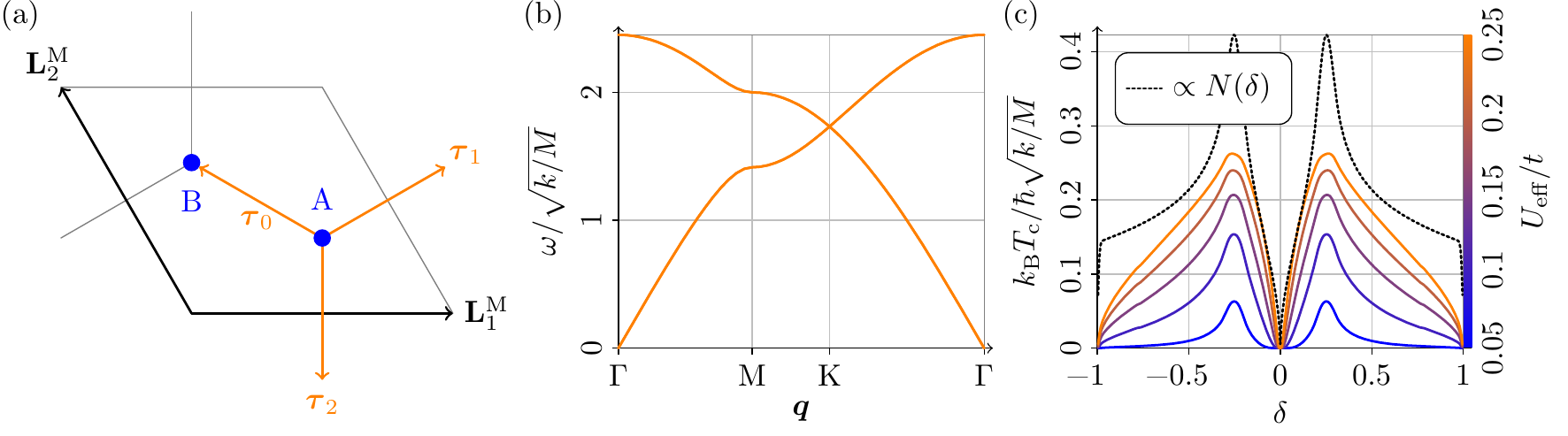}%
	\caption{(Color online) Non-local electron-phonon interactions.
		(a)~Honeycomb lattice with primitive lattice vectors $\vec L_{1, 2}^{\mathrm{M}}$, bond vectors $\vec \tau_{0, 1, 2}$, and sublattices $\mathrm A, \mathrm B$.
		(b)~Phonon dispersion $\omega_{\vec q \pm}$ in units of $\sqrt{k / M}$ with force constant $k$ and mass $M$.
		(c)~Critical temperature $T \sub c$ in units of $\sqrt{k / M}$ and density of states $N$ in arbitrary units as a function of the doping level $\delta$ for different strengths of the effective interaction $U \sub{eff}$ in units of the hopping $t$.}
	\label{fig:nonlocal}
\end{figure}

We describe the electrons and phonons of the moir\'e superlattice using nearest-neighbor-only tight-binding and mass--spring models on a honeycomb lattice.
The tight-binding Hamiltonian is equivalent to the nearest-neighbor part of Eq.~\eqref{eq:honeycomb_hamiltonian}, except that we change the orientation of the two electronic sublattices $\mathrm A, \mathrm B$ and the primitive moir\'e lattice vectors $\vec L_{1, 2}^{\mathrm{M}}$ for the sake of notational simplicity, see Fig.~\ref{fig:nonlocal}\,(a).
Using reciprocal lattice units $k_{1, 2} = \vec k \cdot\vec L_{1, 2}^{\mathrm{M}}$, the tight-binding Hamiltonian can then be defined as
\begin{align}
    H_{\vec k \mathrm A \mathrm B} \pconj
    = t (1 + \E^{\I k_1} + \E^{-\I k_2}),
    \quad
    H_{\vec k \mathrm B \mathrm A} \pconj
    = H_{\vec k \mathrm A \mathrm B} \conj,
    \quad
    H_{\vec k \mathrm A \mathrm A} \pconj
    = H_{\vec k \mathrm B \mathrm B} \pconj = 0,
\end{align}
where $t \equiv t_1$ and the asterisk denotes the complex conjugate.
The corresponding electron dispersion relation (see Fig.~\ref{figSM2}\,(b)) reads
\begin{align}
    E_{\vec k \pm}
    = \pm t \sqrt{3 + 2 \cos(k_1) + 2 \cos(k_2) + 2 \cos(k_1 + k_2)}.
\end{align}
For the phonons, we use a mass--spring model with an isotropic nearest-neighbor force constant.
Using reciprocal lattice units $q_{1, 2} = \vec q \cdot \vec L_{1, 2}^{\mathrm{M}}$, the dynamical matrix can be defined as
\begin{align}
    D_{\vec q \mathrm A \mathrm B} \pconj
    = - \mathds 1 \frac k M (1 + \E^{\I q_1} + \E^{-\I q_2}),
    \quad
    D_{\vec q \mathrm B \mathrm A} \pconj
    = D_{\vec q \mathrm A \mathrm B} \conj,
    \quad
    D_{\vec q \mathrm A \mathrm A} \pconj
    = D_{\vec q \mathrm B \mathrm B} \pconj = \mathds 1 \frac {3 k} M,
\end{align}
where $k$ and $M$ are the effective force constant and a mass, respectively.
The force constant and mass have to be understood as effective quantities related to the moir\'e unit cell and not referring to the primitive unit cell or individual atoms.
$\mathds 1$ denotes the unit matrix in the space of Cartesian displacement directions.

We show in Fig.~\ref{fig:nonlocal}\,(b) the corresponding phonon dispersion which consists of two branches, whose degeneracy is the number of spatial dimensions, and reads $\omega_{\vec q \pm} = \sqrt{k_{\vec q \pm} / M}$ with
\begin{align}
    k_{\vec q \pm}
    = k [3 \pm \sqrt{3 + 2 \cos(q_1) + 2 \cos(q_2) + 2 \cos(q_1 + q_2)}].
    \label{eq:shear_eigenmode_spring}
\end{align}
Finally, modeling the dependence of the hopping $t$ on the bond length $\tau$ as $t / t_0 = (\tau / \tau_0)^{-\beta}$~\cite{Harrison2004} and labeling the sublattices of the ionic displacements as $\mathrm A', \mathrm B'$, the deformation-potential matrix element can be defined as
\begin{align}
    \vec d_{\vec q \mathrm A' \vec k \mathrm A \mathrm B} \pconj
    = \frac {\beta t} \tau (\hat{\vec \tau}_0 + \hat{\vec \tau}_1 \E^{\I k_1} + \hat{\vec \tau}_2 \E^{-\I k_2}),
    \quad
    \vec d_{\vec q \mathrm A' \vec k \mathrm B \mathrm A} \pconj
    = \vec d_{\vec q \mathrm A' \vec k + \vec q \mathrm A \mathrm B} \conj,
    \quad
    \vec d_{\vec q \mathrm B' \vec k i j} \pconj
    = -\vec d_{\vec q \mathrm A' \vec k j i} \conj,
    \label{eq:deformation_potential}
\end{align}
where $\hat{\vec \tau}_{0, 1, 2}$ are the normalized nearest-neighbor bond directions (Fig.~\ref{fig:nonlocal}\,(a)) and $i, j \in \mathrm A, \mathrm B$.
$d_{\vec q x \vec k i j}$ quantifies the scattering of an electron from $\vec k, j$ to $\vec k + \vec q, i$ due to a $\vec q, x$ displacement.
Using the eigenvectors $\vec \psi$ and $\vec e$ of the tight-binding Hamiltonian and the dynamical matrix, the deformation-potential matrix element can be transformed to the band basis via
\begin{align}
    d_{\vec q \nu \vec k m n} \pconj = \sum_{x i j}
    e_{\vphantom{\vec k} \vec q x \nu} \pconj
    \psi_{\vec k + \vec q i m} \conj
    \psi_{\vec k j n} \pconj
    d_{\vec q x \vec k i j} \pconj,
\end{align}
where $\nu$ denotes the phonon branch and $m$, $n$ the electronic band.
The index $x$ combines $\mathrm A', \mathrm B'$ and Cartesian directions.
With this, we have everything needed to calculate the effective electron--phonon coupling strength
\begin{align}
    \lambda(\mu)
    = N(\mu) \frac {
        \sum_{\vec q \nu \vec k m n}
        \delta(\epsilon_{\vec k + \vec q m} - \mu)
        \delta(\epsilon_{\vec k n} - \mu)
        U_{\vec q \nu \vec k m n} \super{eff}
    }{
        \sum_{\vec q \vec k m n}
        \delta(\epsilon_{\vec k + \vec q m} - \mu)
        \delta(\epsilon_{\vec k n} - \mu)
    },
\end{align}
where we have defined the effective attractive interaction $U_{\vec q \nu \vec k m n} \super{eff} = \abs{d_{\vec q \nu \vec k m n}}^2 / k_{\vec q \nu}$, and the logarithmic average of the phonon energy
\begin{align}
    \omega \sub{log} (\mu)
    = \exp \biggl[ \frac {
        \sum_{\vec q \nu \vec k m n}
        \delta(\epsilon_{\vec k + \vec q m} - \mu)
        \delta(\epsilon_{\vec k n} - \mu)
        U_{\vec q \nu \vec k m n} \super{eff}
        \log(\omega_{\vec q \nu})
    }{
        \sum_{\vec q \nu \vec k m n}
        \delta(\epsilon_{\vec k + \vec q m} - \mu)
        \delta(\epsilon_{\vec k n} - \mu)
        U_{\vec q \nu \vec k m n} \super{eff}
    } \biggr]
\end{align}
as a function of the chemical potential $\mu$~\cite{PhysRevB.12.905}.
Here, $N(\mu)$ is the DOS per spin direction and unit cell, see Fig.~\ref{figSM2}(c) for the DOS as a function of the doping level $\delta$.
Both $\lambda$ and $\omega \sub{log}$ are double Fermi-surface averages; the $\delta$ functions ensure that both in- and outgoing states $\vec k, n$ and $\vec k + \vec q, m$ are on the Fermi surface.
Note that the shape of $\lambda$ and $\omega \sub{log}$ as a function of $\mu$ for our model is fixed and their magnitude depends solely on the prefactors $U \sub{eff} / t = \beta^2 t / \tau^2 k$ and $\sqrt{k / M}$, respectively.

We calculate the critical temperature $T \sub c$ using McMillan's formula~\cite{PhysRev.167.331, PhysRevB.12.905} (Eq.~(3) of the main text) for different values of $U \sub{eff}$ covering the entire range from weak to strong coupling, i.e., $0 < \lambda \lessapprox 2$, as a function of the doping $\delta$.
For simplicity, we set the Coulomb pseudopotential $\mu^* = 0$, while finite $\mu^*$ do not change the picture qualitatively.
We sample the Brillouin zone using $96 \times 96 \times 1$ $\vec q$ and $\vec k$ points in combination with a Gaussian broadening of $0.05 t$.
In all cases, $T \sub c$ approximately follows the DOS, see Fig.~\ref{fig:nonlocal}\,(c).
Depending on the value of $U \sub{eff}$, the maxima at the VHS are more or less pronounced.

\section{Estimation of effective electron-phonon interaction parameter $U_{\mathrm{eff}}$}\label{sec:phonon_estimation}
In the main text and in Section \ref{sec:nonlocal} we used McMillan's formula to show that superconductivity arising from electron-phonon coupling reveals generic and robust doping fingerprints by $T_{\mathrm{c}}$ following the DOS. The quantitative details of the superconducting transition are then determined by the material properties. Here, we give an estimation on the order of magnitude for the effective BCS-like interaction $U_{\mathrm{eff}}$ entering the pairing strength $\lambda = U_{\mathrm{eff}} N(E_{\mathrm{F}})$ for $\Gamma$-valley twisted TMDCs. 
%We first do an estimation from extrapolating bulk phonons of the untwisted bilayer and afterwards we estimate the moiré phonon coupling strength associated with the local and nonlocal couplings from continuum mechanics.

The simplest estimation for the pairing strength $\lambda$ for twisted moir\'{e} systems is to extrapolate from calculations for the untwisted material. In homobilayer TMDCs, $\lambda$ can take values up to 8 \cite{Roesner2014} depending on the doping with the DOS varying between 0.4~eV$^{-1}$ and 2~eV$^{-1}$ \cite{Schoenhoff2016}. Hence, the effective interaction strength is $U_{\mathrm{eff}}= \lambda/N(E_{\mathrm{F}}) \approx 4$\,--\,12~eV per unit cell. This value needs to be scaled to the moir\'{e} unit cell which contains approximately $(\lambda^{\mathrm{M}}/a_0)^2  = 1/\sin^2\theta \approx \theta^{-2}$ single unit cells with lattice constant $a_0$ (c.f.~Table \ref{tableSM1}) for small $\theta$, i.e., the effective interaction is twist-angle dependent $U_{\mathrm{eff}}(\theta)\approx 4$--$12\,\theta^2$~eV. Using our observation $t\propto \theta^2$ (c.f.~Fig.~1(c) of the main text), we can express $U_{\mathrm{eff}}$ in units of $t$. For instance, for twisted MoS$_2$ bilayer we can write $t\approx 2$~eV$\cdot \theta^2 = \alpha\theta^2$ with $\theta$ in radians. Thus, we estimate an interaction strength of $U_{\mathrm{eff}}/t = 2$\,--\,6.

%The interaction $U_{\mathrm{eff}}$ is typically expressed by the electron-phonon coupling $g$ and the averaged ("typical") phonon frequency $\langle\omega\rangle$ as $U_{\mathrm{eff}}=2g^2/\hbar\langle\omega\rangle$. However, it can also be expressed by the deformation potential $d = g\sqrt{2M\omega}/\hbar$ with mass $M$ and the effective moir\'{e} force constant $k=\langle\omega\rangle^2 m$ as $U_{\mathrm{eff}}=d^2/k$, showing that it can be interpreted as a classical quantity since all $\hbar$ cancel out.
We also discuss  moir\'{e} phonon modes, where we obtain $U_{\mathrm{eff}}=d^2/k$ from elastic properties of the bilayer TMDCs. Instead of using the microsopic electron-phonon coupling $g$ and the averaged ("typical") phonon frequency $\langle\omega\rangle$, we express $U_{\mathrm{eff}}$ in terms of an effective moir\'{e} deformation potential $d$ and an effective moir\'{e} force constant $k$. They are related by
\begin{align}
	U_{\mathrm{eff}} = \frac{2g}{\hbar \langle\omega\rangle} = \frac{2}{\hbar \langle\omega\rangle} \frac{\hbar d^2}{2M\langle\omega\rangle} = \frac{d^2}{k}\,,
	\label{eq:Ueff_definition}
\end{align}
since $g = \sqrt{\frac{\hbar}{2M\langle\omega\rangle}}d$ \cite{Mahan2000} and $\langle\omega\rangle =\sqrt{k/M}$ with mass $M$. It shows that the attractive phonon-mediated interaction can be interpreted as a classical quantity, as all $\hbar$ cancel out. 

\begin{figure}
	\centering
	\includegraphics[width=\linewidth]{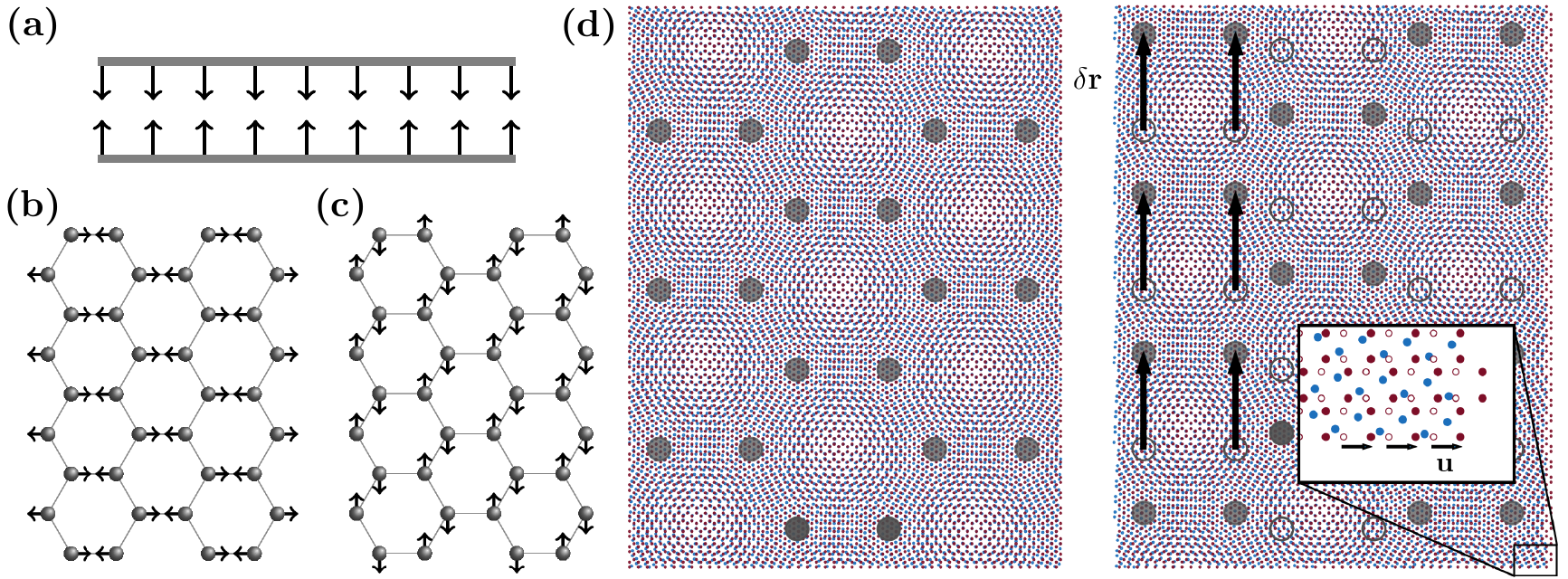}%
	\caption{(Color online) Moir\'{e} phonon modes.
		(a)~Layer-breathing mode (``Holstein'' coupling).
		(b,\,c)~Optical in-plane modes of a honeycomb lattice at $\Gamma$ (``Peierls'' coupling). (d)~Effect of TMDC layer displacement. The left panel shows the unshifted lattice, while in the right panel the red TMDC lattice was displaced to the right by $\bold{u}$, so that the moir\'{e} superlattice is shifted to the top by $\delta\bold{r}$. Empty circles denote the unshifted atom positions.}
	\label{fig:modes}
\end{figure}

First, we consider a case which corresponds to a purely local mode with Holstein-type coupling, which results from an interlayer breathing mode, see Fig.~\ref{fig:modes}(a). In this case, a restoring force $F = k\Delta h$ is induced when the interlayer distance is changed by an amount  $\Delta h$. The response of the system is also encoded in the elastic constant in out-of-plane direction
\begin{align}
	C_{33} = \frac{\sigma}{\varepsilon} = \frac{F/A}{\Delta h/h} = \frac{Fh}{A\Delta h}
\end{align} 
with the tensile stress $\sigma$ of the lifted area $A$ and strain $\varepsilon$ of the equilibrium layer distance $h$. Thus, the force constant can be calculated from 
\begin{align}
	k = \frac{F}{\Delta h} = \frac{A \, C_{33}}{h}\,.
\end{align}
We assume that only a fraction $p<1$ (the AB, BA regions) of the moir\'{e} unit cell needs to be lifted, so that $A = p \frac{\sqrt{3}}{2} (\lambda^{\mathrm{M}})^2 \approx p\frac{\sqrt{3}}{2} a^2_0 \theta^{-2}$. The magnitude of the deformation potential is given in Eq.~(\ref{eq:deformation_potential}) and for a single mode it simplifies to 
\begin{align}
	d = \frac{\beta t_{\perp}}{h}
	\label{eq:Holst_def_pot}
\end{align}
with the interlayer hopping $t_\perp$. The attractive interaction in Eq.~(\ref{eq:Ueff_definition}) then takes the form
\begin{align}
	U_{\mathrm{eff}}^{\mathrm{Holst.}} = \frac{d^2}{k} = \left(\frac{\beta t_{\perp}}{h}\right)^2 \cdot \frac{h}{A\,C_{33}} 
	\approx \frac{\beta^2 t_{\perp}^2}{\frac{\sqrt{3}}{2}p a_0^2\theta^{-2} h C_{33}} 
	 \;.
\end{align}
Expressing $U_{\mathrm{eff}}^{\mathrm{Holst.}}$ in units of the moiré honeycomblattice hopping $t\approx\alpha\theta^2$ yields
\begin{align}
	U_{\mathrm{eff}}^{\mathrm{Holst.}} \approx \frac{2}{\sqrt{3}}\frac{\beta^2 t_{\perp}^2}{\alpha p a_0^2 hC_{33}} \,t\;.
\end{align}
We can estimate $U_{\mathrm{eff}}^{\mathrm{Holst.}}$ to be in the range of $0.05$\,--\,$1.4$~$t$ by assuming $\beta=4$\,--\,5 \cite{Harrison2004}, $a_0 = 3.18$~\AA \cite{Schoenhoff2016}, $C_{33}=52$~GPa \cite{Feldman1976,Zhao2013}, $t_\perp=0.3$\,--\,0.4~eV \cite{gammaTMDCsMacdonald2021}, $h=3$\,--\,6~\AA, $\alpha=2$~eV, and $p=0.167$\,--\,$0.5$. In our simplified approach, we thus get interactions that can induce superconductivity (c.f.~Fig.~S3(b) of the main text) since the pairing strength $\lambda = U_{\mathrm{eff}}/t \cdot N(E_{\mathrm{F}})t$ reaches values up to $\lambda\approx0.5$ (c.f.~Fig.~ \ref{figSM2}(c)).

Now, we consider an interlayer shear mode with the two layers being moved in opposite directions and opposite shearing profile in the AB and BA regions of the moir\'{e}. This effectively modulates the bond lengths in the moir\'{e} honeycomb superlattice, i.e., we estimate the effective interaction arising from the Peierls coupling discussed in Section \ref{sec:nonlocal}. Two equivalent shear modes exist, see Fig.~\ref{fig:modes}(b,c), for which the potential energy is given by the optical $\bold{q}=\bold{0}$ eigenmode (c.f.~Eq.~(\ref{eq:shear_eigenmode_spring})) of the spring model in Section \ref{sec:nonlocal}. The displacement $\delta r_0$ of a Wannier center with respect to the origin at an AB/BA site thus has the elastic energy
\begin{align}
	E_{\mathrm{el}} = \frac{1}{2} k_{\bold{q}=\bold{0},+}\,\delta r_0^2 = 3 k\, \delta r_0^2\;.
	\label{eq:elastic_spring_energy}
\end{align}
On the other hand, we can estimate the equivalent displacement energy \cite{Landau1970}
\begin{align}
	E = \frac{1}{2} \int_{A_{\mathrm{M}}}\!\mathrm{d}^2r \;\sum_{\alpha} \lambda_{\mathrm{L}} u^2_{\alpha\alpha} + \sum_{\alpha\beta} 2\mu_{\mathrm{L}} u_{\alpha\beta}^2
	\label{eq:displacement_energy}
\end{align}
associated with the displacement field $\mathbf{u}$ of a single layer. Here, $u_{\alpha\beta} = \frac{1}{2}(\frac{\partial u_\alpha}{\partial x_\beta} + \frac{\partial u_\beta}{\partial x_\alpha} + \sum_\gamma \frac{\partial u_\gamma}{\partial x_\alpha}\frac{\partial u_\gamma}{\partial x_\beta})$ is the strain tensor and $\lambda_{\mathrm{L}}$, $\mu_{\mathrm{L}}$ are the Lam\'{e} constants which are linked to the Young's modulus $Y$ and Poisson ratio $\nu$ via
\begin{align}
	\lambda_{\mathrm{L}} =  \frac{\nu}{1-2\nu}\frac{1}{1+\nu} Y\;,\quad \mu_{\mathrm{L}} = \frac{1}{2(1+\nu)}Y
	\label{eq:lame_constants}
\end{align}
with $Y\approx 150$~N/m and $\nu\approx0.22$ for TMDC homobilayers \cite{Cakir2014,Liu2014}. The shear displacement $\bold{u}$ of the TMDC layers induces a perpendicular shift $\delta \bold{r}$ of the Wannier center with respect to the origin at an AB site as shown Fig.~\ref{fig:modes}(d). They are linked by
\begin{align}
	\bold{u}_{\pm} = \left[\mathcal{R}_{\pm\theta/2} - \mathds{1}_{2\times2}\right]\delta\bold{r} \approx \begin{pmatrix}
		0 &\pm \frac{\theta}{2}\\
		\mp \frac{\theta}{2} &0
	\end{pmatrix}\delta\bold{r}\;,
	\label{eq:moire_layer_displacement}
\end{align}
for small twist angles where the upper and lower TMDC layer carry a different sign.  For each mode the displacement field $\delta \bold{r}_{(b,c)} = \delta r(\bold{r}) \bold{e}_{x,y} $ can be described by the leading Fourier components
\begin{align}
	\delta r(\bold{r}) = \frac{2\delta r_0}{3\sqrt{3}}\left(\sin(\bold{G}_1^{\mathrm{M}} \bold{r}) + \sin([\bold{G}_2^{\mathrm{M}}-\bold{G}_1^{\mathrm{M}}] \bold{r}) - \sin(\bold{G}_2^{\mathrm{M}} \bold{r})\right)\; .
	\label{eq:moire_mode_displacement}
\end{align} Since the shear modes (and displacement energies) are equivalent, we focus on the mode in Fig.~\ref{fig:modes}(b) in the following. For this mode, we have the stress tensor components
\begin{align}
	\begin{split}
	u_{xx} &= 0\;,\\
	u_{xy} &= u_{yx} = \frac{1}{2} \frac{\partial u_y}{\partial x} = \pm \frac{\theta\delta r_0}{12}G^{\mathrm{M}}\left(\cos(\bold{G}_1^{\mathrm{M}} \bold{r}) + \cos([\bold{G}_2^{\mathrm{M}}-\bold{G}_1^{\mathrm{M}}] \bold{r})\right)\;,\\
	u_{yy} &= \frac{\partial u_y}{\partial y} = \pm \frac{\theta \delta r_0}{6\sqrt{3}}G^{\mathrm{M}}\left(\cos(\bold{G}_1^{\mathrm{M}} \bold{r}) + \cos([\bold{G}_2^{\mathrm{M}}-\bold{G}_1^{\mathrm{M}}] \bold{r}) - 2\cos(\bold{G}_2^{\mathrm{M}} \bold{r}))\right)
	\label{eq:stress_tensor}
	\end{split}
\end{align}
where the higher order terms of the offdiagonal components were neglected for small displacements and $G^{\mathrm{M}} = |\bold{G}^{\mathrm{M}}_{1,2}| = 2\pi/\lambda^{\mathrm{M}}$.
Inserting Eq.~(\ref{eq:stress_tensor}) into Eq.~(\ref{eq:displacement_energy}) and integrating over the moir\'{e} unit cell area, the displacement energy for one layer yields
\begin{align}
	E = \frac{1}{2} \frac{\theta^2\delta r_0^2}{36} (G^{\mathrm{M}})^2A_{\mathrm{M}} (\lambda_{\mathrm{L}} + 3\mu_{\mathrm{L}})\;.
	\label{eq:integrated_displacement_energy}
\end{align} 
We obtain the force constant by equating the displacement energy for both layers with twice (due to two layers) the elastic energy in Eq.~(\ref{eq:elastic_spring_energy}) and using Eq.~(\ref{eq:lame_constants}) as 
\begin{align}
	k = \frac{2}{3\delta r_0^2} E_{\mathrm{el}} = \frac{1}{108} (G^{\mathrm{M}})^2 A_{\mathrm{M}}  (\lambda_{\mathrm{L}} + 3\mu_{\mathrm{L}})\theta^2 = \frac{\sqrt{3}\pi^2}{108} \frac{3-4\nu}{(1-2\nu)(1+\nu)} Y \theta^2 = k_0 \theta^2
\end{align}
with $k_0\approx 5.6$~eV/\AA$^2$. The deformation potential is as in Eq.~(\ref{eq:Holst_def_pot}) with $t_\perp$ and $h$ being replaced by $t$ and $\lambda^{\mathrm{M}}/3$ (Wannier orbital extent), respectively. The effective potential takes the form
\begin{align}
	U_{\mathrm{eff}}^{\mathrm{Peierls}} \approx \frac{\left(\beta t \frac{3}{\lambda^{\mathrm{M}}}\right)^2}{ k_0 \theta^2} \approx \frac{9\beta^2 t^2\left(\frac{\theta}{a_0}\right)^2}{k_0 \theta^2} = \frac{9\beta^2}{a_0^2 k_0} t^2 = \frac{9\alpha\beta^2}{a_0^2 k_0} \theta^2 \cdot t\; .
\end{align}
Since $U_{\mathrm{eff}}^{\mathrm{Peierls}}/t\propto t \propto \theta^2$ with the prefactor $\frac{9\alpha\beta^2}{a_0^2 k_0}\approx 5$\,--\,8, the effective interaction and hence pairing strength is very small. From our estimation we conclude that superconductivity from moir\'{e} Peierls coupling will not be realized in the real material system.

%From Sections \ref{sec:nonlocal} and \ref{sec:phonon_estimation}, we can conclude that independently of the fine details of the electron-phonon coupling profile, the critical temperature $T_c^{ph}$ will follow the density of states. Nevertheless, for a better understanding on the phonon modes profile it is desirable to obtain the effective coupling strength $\lambda$ from the phononic spectral function $\alpha^2F$.

\bibliography{twistFLEX_bib}